

\magnification=\magstephalf

\newbox\SlashedBox
\def\slashed#1{\setbox\SlashedBox=\hbox{#1}
\hbox to 0pt{\hbox to 1\wd\SlashedBox{\hfil/\hfil}\hss}#1}
\def\hboxtosizeof#1#2{\setbox\SlashedBox=\hbox{#1}
\hbox to 1\wd\SlashedBox{#2}}

\def\mathslashed#1{\setbox\SlashedBox=\hbox{$#1$}
\hbox to 0pt{\hbox to 1\wd\SlashedBox{\hfil/\hfil}\hss}#1}

\def\ifsmall{\iffalse}  
\def\titlepagefont{}  

\def\DefineTeXgraphics{%
\special{ps::[global] /TeXgraphics { } def}}  

\def\today{\ifcase\month\or January\or February\or March\or April\or May
\or June\or July\or August\or September\or October\or November\or
December\fi\space\number\day, \number\year}
\def\eatPrefix19{}
\def\Year{\expandafter\eatPrefix\the\year}
\newcount\hours \newcount\minutes
\def\monthname{\ifcase\month\or
January\or February\or March\or April\or May\or June\or July\or
August\or September\or October\or November\or December\fi}
\def\shortmonthname{\ifcase\month\or
Jan\or Feb\or Mar\or Apr\or May\or Jun\or Jul\or
Aug\or Sep\or Oct\or Nov\or Dec\fi}

\def\TimeStamp{\hours\the\time\divide\hours by60%
\minutes -\the\time\divide\minutes by60\multiply\minutes by60%
\advance\minutes by\the\time%
${\rm \shortmonthname}\cdot\if\day<10{}0\fi\the\day\cdot\the\year%
\qquad\the\hours:\if\minutes<10{}0\fi\the\minutes$}




\def\Title#1{%
\vskip 1in{\titlefont\centerline{#1}}\vskip .5in}



\newif\ifdraftmode
\newif\ifleftlabels  

\def\nolabels{\def\wrlabeL##1{}\def\eqlabeL##1{}\def\reflabeL##1{}}
\def\writelabels{\def\wrlabeL##1{\leavevmode\vadjust{\rlap{\smash%
{\line{{\escapechar=` \hfill\rlap{\sevenrm\hskip.03in\string##1}}}}}}}%
\def\eqlabeL##1{{\escapechar-1\rlap{\sevenrm\hskip.05in\string##1}}}%
\def\reflabeL##1{\noexpand\rlap{\noexpand\sevenrm[\string##1]}}}
\def\writeleftlabels{\def\wrlabeL##1{\leavevmode\vadjust{\rlap{\smash%
{\line{{\escapechar=` \hfill\rlap{\sevenrm\hskip.03in\string##1}}}}}}}%
\def\eqlabeL##1{{\escapechar-1%
\rlap{\sixrm\hskip.05in\string##1}%
\llap{\sevenrm\string##1\hskip.03in\hbox to \hsize{}}}}%
\def\reflabeL##1{\noexpand\rlap{\noexpand\sevenrm[\string##1]}}}
\nolabels

\newdimen\fullhsize
\newdimen\hstitle
\hstitle=\hsize 
\newdimen\hsbody
\hsbody=\hsize 
\newdimen\hbodyoffset
\hbodyoffset=\hoffset 
\newbox\leftpage
\def\abstract#1{#1}
\def\rotated{\special{ps: landscape}
\magnification=1000  
\baselineskip=14pt
\global\hstitle=9truein\global\hsbody=4.75truein
\global\vsize=7truein\global\voffset=-.31truein
\global\hoffset=-0.54in\global\hbodyoffset=-.54truein
\global\fullhsize=10truein
\def\DefineTeXgraphics{%
\special{ps::[global]
/TeXgraphics {currentpoint translate 0.7 0.7 scale
              -80 0.72 mul -1000 0.72 mul translate} def}}
\let\lr=L
\def\ifsmall{\iftrue}
\def\titlepagefont{\twelvepoint}
\trueseventeenpoint
\def\almostshipout##1{\if L\lr \count1=1
      \global\setbox\leftpage=##1 \global\let\lr=R
   \else \count1=2
      \shipout\vbox{\hbox to\fullhsize{\box\leftpage\hfil##1}}
      \global\let\lr=L\fi}

\output={\ifnum\count0=1 
 \shipout\vbox{\hbox to \fullhsize{\hfill\pagebody\hfill}}\advancepageno
 \else
 \almostshipout{\leftline{\vbox{\pagebody\makefootline}}}\advancepageno
 \fi}

\def\abstract##1{{\leftskip=1.5in\rightskip=1.5in ##1\par}} }

\def\linemessage#1{\immediate\write16{#1}}

\global\newcount\secno \global\secno=0
\global\newcount\appno \global\appno=0
\global\newcount\meqno \global\meqno=1
\global\newcount\subsecno \global\subsecno=0
\global\newcount\figno \global\figno=0

\newif\ifAnyCounterChanged
\let\terminator=\relax
\def\normalize#1{\ifx#1\terminator\let\next=\relax\else%
\if#1i\aftergroup i\else\if#1v\aftergroup v\else\if#1x\aftergroup x%
\else\if#1l\aftergroup l\else\if#1c\aftergroup c\else%
\if#1m\aftergroup m\else%
\if#1I\aftergroup I\else\if#1V\aftergroup V\else\if#1X\aftergroup X%
\else\if#1L\aftergroup L\else\if#1C\aftergroup C\else%
\if#1M\aftergroup M\else\aftergroup#1\fi\fi\fi\fi\fi\fi\fi\fi\fi\fi\fi\fi%
\let\next=\normalize\fi%
\next}
\def\makeNormal#1#2{\def\doNormalDef{\edef#1}\begingroup%
\aftergroup\doNormalDef\aftergroup{\normalize#2\terminator\aftergroup}%
\endgroup}

\def\warnIfChanged#1#2{%
\ifundef#1
\else\begingroup%
\edef\oldDefinitionOfCounter{#1}\edef\newDefinitionOfCounter{#2}%
\ifx\oldDefinitionOfCounter\newDefinitionOfCounter%
\else%
\linemessage{Warning: definition of \noexpand#1 has changed.}%
\global\AnyCounterChangedtrue\fi\endgroup\fi}

\def\Section#1{\global\advance\secno by1\relax\global\meqno=1%
\global\subsecno=0%
\bigbreak
\centerline{\twelvepoint \bf %
\the\secno. #1}%
\par\nobreak\medskip\nobreak}
\def\tagsection#1{%
\warnIfChanged#1{\the\secno}%
\xdef#1{\the\secno}%
\ifWritingAuxFile\immediate\write\auxfile{\noexpand\xdef\noexpand#1{#1}}\fi%
}
\def\section{\Section}
\def\Subsection#1{\global\advance\subsecno by1\relax\medskip %
\leftline{\bf\the\secno.\the\subsecno\ #1}%
\par\nobreak\nobreak}
\def\tagsubsection#1{%
\warnIfChanged#1{\the\secno.\the\subsecno}%
\xdef#1{\the\secno.\the\subsecno}%
\ifWritingAuxFile\immediate\write\auxfile{\noexpand\xdef\noexpand#1{#1}}\fi%
}

\def\subsection{\Subsection}

\def\romappno{\uppercase\expandafter{\romannumeral\appno}}
\def\makeNormalizedRomappno{%
\expandafter\makeNormal\expandafter\normalizedromappno%
\expandafter{\romannumeral\appno}%
\edef\normalizedromappno{\uppercase{\normalizedromappno}}}
\def\Appendix#1{\global\advance\appno by1\relax\global\meqno=1\global\secno=0%
\global\subsecno=0%
\bigbreak
\goodbreak\bigskip
\centerline{\twelvepoint \bf Appendix %
\romappno. #1}%
\par\nobreak\medskip\nobreak}
\def\tagappendix#1{\makeNormalizedRomappno%
\warnIfChanged#1{\normalizedromappno}%
\xdef#1{\normalizedromappno}%
\ifWritingAuxFile\immediate\write\auxfile{\noexpand\xdef\noexpand#1{#1}}\fi%
}
\def\appendix{\Appendix}
\def\Subappendix#1{\global\advance\subsecno by1\relax\medskip %
\leftline{\bf\romappno.\the\subsecno\ #1}%
\par\nobreak\smallskip\nobreak}
\def\tagsubappendix#1{\makeNormalizedRomappno%
\warnIfChanged#1{\normalizedromappno.\the\subsecno}%
\xdef#1{\normalizedromappno.\the\subsecno}%
\ifWritingAuxFile\immediate\write\auxfile{\noexpand\xdef\noexpand#1{#1}}\fi%
}

\def\subappendix{\Subappendix}

\def\eqn#1{\makeNormalizedRomappno%
\ifnum\secno>0%
  \warnIfChanged#1{\the\secno.\the\meqno}%
  \eqno(\the\secno.\the\meqno)\xdef#1{\the\secno.\the\meqno}%
     \global\advance\meqno by1
\else\ifnum\appno>0%
  \warnIfChanged#1{\normalizedromappno.\the\meqno}%
  \eqno({\rm\romappno}.\the\meqno)%
      \xdef#1{\normalizedromappno.\the\meqno}%
     \global\advance\meqno by1
\else%
  \warnIfChanged#1{\the\meqno}%
  \eqno(\the\meqno)\xdef#1{\the\meqno}%
     \global\advance\meqno by1
\fi\fi%
\eqlabeL#1%
\ifWritingAuxFile\immediate\write\auxfile{\noexpand\xdef\noexpand#1{#1}}\fi%
}
\def\defeqn#1{\makeNormalizedRomappno%
\ifnum\secno>0%
  \warnIfChanged#1{\the\secno.\the\meqno}%
  \xdef#1{\the\secno.\the\meqno}%
     \global\advance\meqno by1
\else\ifnum\appno>0%
  \warnIfChanged#1{\normalizedromappno.\the\meqno}%
  \xdef#1{\normalizedromappno.\the\meqno}%
     \global\advance\meqno by1
\else%
  \warnIfChanged#1{\the\meqno}%
  \xdef#1{\the\meqno}%
     \global\advance\meqno by1
\fi\fi%
\eqlabeL#1%
\ifWritingAuxFile\immediate\write\auxfile{\noexpand\xdef\noexpand#1{#1}}\fi%
}
\def\anoneqn{\makeNormalizedRomappno%
\ifnum\secno>0
  \eqno(\the\secno.\the\meqno)%
     \global\advance\meqno by1
\else\ifnum\appno>0
  \eqno({\rm\normalizedromappno}.\the\meqno)%
     \global\advance\meqno by1
\else
  \eqno(\the\meqno)%
     \global\advance\meqno by1
\fi\fi%
}
\def\mfig#1#2{\global\advance\figno by1%
\relax#1\the\figno%
\warnIfChanged#2{\the\figno}%
\edef#2{\the\figno}%
\reflabeL#2%
\ifWritingAuxFile\immediate\write\auxfile{\noexpand\xdef\noexpand#2{#2}}\fi%
}

\def\fig#1{\mfig{fig.\ ~}#1}

\catcode`@=11 

\newif\ifFiguresInText\FiguresInTexttrue
\newif\if@FigureFileCreated
\newwrite\capfile
\newwrite\figfile

\def\PlaceTextFigure#1#2#3#4{%
\vskip 0.5truein%
#3\hfil\epsfbox{#4}\hfil\break%
\hfil\vbox{Figure #1. #2}\hfil%
\vskip10pt}
\def\PlaceEndFigure#1#2{%
\epsfysize=\vsize\epsfbox{#2}\hfil\break\vfill\centerline{Figure #1.}\eject}

\def\LoadFigure#1#2#3#4{%
\ifundef#1{\phantom{\mfig{}#1}}\fi
\ifWritingAuxFile\immediate\write\auxfile{\noexpand\xdef\noexpand#1{#1}}\fi%
\ifFiguresInText
\PlaceTextFigure{#1}{#2}{#3}{#4}%
\else
\if@FigureFileCreated\else%
\immediate\openout\capfile=\jobname.caps%
\immediate\openout\figfile=\jobname.figs%
\fi%
\immediate\write\capfile{\noexpand\item{Figure \noexpand#1.\ }#2.}%
\immediate\write\figfile{\noexpand\PlaceEndFigure\noexpand#1{\noexpand#4}}%
\fi}

\def\listfigs{\ifFiguresInText\else%
\vfill\eject\immediate\closeout\capfile
\immediate\closeout\figfile%
\centerline{{\bf Figures}}\bigskip\frenchspacing%
\input \jobname.caps\vfill\eject\nonfrenchspacing%
\input\jobname.figs\fi}

\font\ninerm=cmr9
\font\eightrm=cmr8
\font\sixrm=cmr6

\def\loadtrueseventeenpoint{
 \font\seventeenrm=cmr10 at 17.28truept
 \font\seventeeni=cmmi10 at 17.28truept
 \font\seventeenbf=cmbx10 at 17.28truept
 \font\seventeenit=cmti10 at 17.28truept
 \font\seventeensl=cmsl10 at 17.28truept
 \font\seventeensy=cmsy10 at 17.28truept
}
\def\loadfourteenpoint{
\font\fourteenrm=cmr10 at 14.4pt
\font\fourteeni=cmmi10 at 14.4pt
\font\fourteenit=cmti10 at 14.4pt
\font\fourteensl=cmsl10 at 14.4pt
\font\fourteensy=cmsy10 at 14.4pt
\font\fourteenbf=cmbx10 at 14.4pt
}
\def\loadtruetwelvepoint{
\font\twelverm=cmr10 at 12truept
\font\twelvei=cmmi10 at 12truept
\font\twelveit=cmti10 at 12truept
\font\twelvesl=cmsl10 at 12truept
\font\twelvesy=cmsy10 at 12truept
\font\twelvebf=cmbx10 at 12truept
}

\font\ninei=cmmi9
\font\eighti=cmmi8
\font\sixi=cmmi6
\skewchar\ninei='177 \skewchar\eighti='177 \skewchar\sixi='177

\font\ninesy=cmsy9
\font\eightsy=cmsy8
\font\sixsy=cmsy6
\skewchar\ninesy='60 \skewchar\eightsy='60 \skewchar\sixsy='60

\font\ninebf=cmbx9
\font\eightbf=cmbx8
\font\sixbf=cmbx6

\font\ninett=cmtt9
\font\eighttt=cmtt8

\hyphenchar\tentt=-1 
\hyphenchar\ninett=-1
\hyphenchar\eighttt=-1

\font\ninesl=cmsl9
\font\eightsl=cmsl8

\font\nineit=cmti9
\font\eightit=cmti8


\newskip\ttglue
\def\tenpoint{\def\rm{\fam0\tenrm}%
  \textfont0=\tenrm \scriptfont0=\sevenrm \scriptscriptfont0=\fiverm
  \textfont1=\teni \scriptfont1=\seveni \scriptscriptfont1=\fivei
  \textfont2=\tensy \scriptfont2=\sevensy \scriptscriptfont2=\fivesy
  \textfont3=\tenex \scriptfont3=\tenex \scriptscriptfont3=\tenex
  \def\it{\fam\itfam\tenit}\textfont\itfam=\tenit
  \def\sl{\fam\slfam\tensl}\textfont\slfam=\tensl
  \def\bf{\fam\bffam\tenbf}\textfont\bffam=\tenbf \scriptfont\bffam=\sevenbf
  \scriptscriptfont\bffam=\fivebf
  \normalbaselineskip=12pt
  \let\sc=\eightrm
  \let\big=\tenbig
  \setbox\strutbox=\hbox{\vrule height8.5pt depth3.5pt width\z@}%
  \normalbaselines\rm}

\def\twelvepoint{\def\rm{\fam0\twelverm}%
  \textfont0=\twelverm \scriptfont0=\ninerm \scriptscriptfont0=\sevenrm
  \textfont1=\twelvei \scriptfont1=\ninei \scriptscriptfont1=\seveni
  \textfont2=\twelvesy \scriptfont2=\ninesy \scriptscriptfont2=\sevensy
  \textfont3=\tenex \scriptfont3=\tenex \scriptscriptfont3=\tenex
  \def\it{\fam\itfam\twelveit}\textfont\itfam=\twelveit
  \def\sl{\fam\slfam\twelvesl}\textfont\slfam=\twelvesl
  \def\bf{\fam\bffam\twelvebf}\textfont\bffam=\twelvebf%
  \scriptfont\bffam=\ninebf
  \scriptscriptfont\bffam=\sevenbf
  \normalbaselineskip=12pt
  \let\sc=\eightrm
  \let\big=\tenbig
  \setbox\strutbox=\hbox{\vrule height8.5pt depth3.5pt width\z@}%
  \normalbaselines\rm}

\def\fourteenpoint{\def\rm{\fam0\fourteenrm}%
  \textfont0=\fourteenrm \scriptfont0=\tenrm \scriptscriptfont0=\sevenrm
  \textfont1=\fourteeni \scriptfont1=\teni \scriptscriptfont1=\seveni
  \textfont2=\fourteensy \scriptfont2=\tensy \scriptscriptfont2=\sevensy
  \textfont3=\tenex \scriptfont3=\tenex \scriptscriptfont3=\tenex
  \def\it{\fam\itfam\fourteenit}\textfont\itfam=\fourteenit
  \def\sl{\fam\slfam\fourteensl}\textfont\slfam=\fourteensl
  \def\bf{\fam\bffam\fourteenbf}\textfont\bffam=\fourteenbf%
  \scriptfont\bffam=\tenbf
  \scriptscriptfont\bffam=\sevenbf
  \normalbaselineskip=17pt
  \let\sc=\elevenrm
  \let\big=\tenbig
  \setbox\strutbox=\hbox{\vrule height8.5pt depth3.5pt width\z@}%
  \normalbaselines\rm}

\def\seventeenpoint{\def\rm{\fam0\seventeenrm}%
  \textfont0=\seventeenrm \scriptfont0=\fourteenrm \scriptscriptfont0=\tenrm
  \textfont1=\seventeeni \scriptfont1=\fourteeni \scriptscriptfont1=\teni
  \textfont2=\seventeensy \scriptfont2=\fourteensy \scriptscriptfont2=\tensy
  \textfont3=\tenex \scriptfont3=\tenex \scriptscriptfont3=\tenex
  \def\it{\fam\itfam\seventeenit}\textfont\itfam=\seventeenit
  \def\sl{\fam\slfam\seventeensl}\textfont\slfam=\seventeensl
  \def\bf{\fam\bffam\seventeenbf}\textfont\bffam=\seventeenbf%
  \scriptfont\bffam=\fourteenbf
  \scriptscriptfont\bffam=\twelvebf
  \normalbaselineskip=21pt
  \let\sc=\fourteenrm
  \let\big=\tenbig
  \setbox\strutbox=\hbox{\vrule height 12pt depth 6pt width\z@}%
  \normalbaselines\rm}

\def\ninepoint{\def\rm{\fam0\ninerm}%
  \textfont0=\ninerm \scriptfont0=\sixrm \scriptscriptfont0=\fiverm
  \textfont1=\ninei \scriptfont1=\sixi \scriptscriptfont1=\fivei
  \textfont2=\ninesy \scriptfont2=\sixsy \scriptscriptfont2=\fivesy
  \textfont3=\tenex \scriptfont3=\tenex \scriptscriptfont3=\tenex
  \def\it{\fam\itfam\nineit}\textfont\itfam=\nineit
  \def\sl{\fam\slfam\ninesl}\textfont\slfam=\ninesl
  \def\bf{\fam\bffam\ninebf}\textfont\bffam=\ninebf \scriptfont\bffam=\sixbf
  \scriptscriptfont\bffam=\fivebf
  \normalbaselineskip=11pt
  \let\sc=\sevenrm
  \let\big=\ninebig
  \setbox\strutbox=\hbox{\vrule height8pt depth3pt width\z@}%
  \normalbaselines\rm}

\def\eightpoint{\def\rm{\fam0\eightrm}%
  \textfont0=\eightrm \scriptfont0=\sixrm \scriptscriptfont0=\fiverm%
  \textfont1=\eighti \scriptfont1=\sixi \scriptscriptfont1=\fivei%
  \textfont2=\eightsy \scriptfont2=\sixsy \scriptscriptfont2=\fivesy%
  \textfont3=\tenex \scriptfont3=\tenex \scriptscriptfont3=\tenex%
  \def\it{\fam\itfam\eightit}\textfont\itfam=\eightit%
  \def\sl{\fam\slfam\eightsl}\textfont\slfam=\eightsl%
  \def\bf{\fam\bffam\eightbf}\textfont\bffam=\eightbf \scriptfont\bffam=\sixbf%
  \scriptscriptfont\bffam=\fivebf%
  \normalbaselineskip=9pt%
  \let\sc=\sixrm%
  \let\big=\eightbig%
  \setbox\strutbox=\hbox{\vrule height7pt depth2pt width\z@}%
  \normalbaselines\rm}

\def\tenbig#1{{\hbox{$\left#1\vbox to8.5pt{}\right.\n@space$}}}
\def\ninebig#1{{\hbox{$\textfont0=\tenrm\textfont2=\tensy
  \left#1\vbox to7.25pt{}\right.\n@space$}}}
\def\eightbig#1{{\hbox{$\textfont0=\ninerm\textfont2=\ninesy
  \left#1\vbox to6.5pt{}\right.\n@space$}}}

\def\footnote#1{\edef\@sf{\spacefactor\the\spacefactor}#1\@sf
      \insert\footins\bgroup\eightpoint
      \interlinepenalty100 \let\par=\endgraf
        \leftskip=\z@skip \rightskip=\z@skip
        \splittopskip=10pt plus 1pt minus 1pt \floatingpenalty=20000
        \smallskip\item{#1}\bgroup\strut\aftergroup\@foot\let\next}
\skip\footins=12pt plus 2pt minus 4pt 
\dimen\footins=30pc 

\newinsert\margin
\dimen\margin=\maxdimen
\def\titlefont{\seventeenpoint}
\loadtruetwelvepoint 
\loadtrueseventeenpoint

\def\eatOne#1{}
\def\ifundef#1{\expandafter\ifx%
\csname\expandafter\eatOne\string#1\endcsname\relax}
\def\notTrue{\iffalse}\def\isTrue{\iftrue}
\def\ifdef#1{{\ifundef#1%
\aftergroup\notTrue\else\aftergroup\isTrue\fi}}
\def\use#1{\ifundef#1\linemessage{Warning: \string#1 is undefined.}%
{\tt \string#1}\else#1\fi}


\global\newcount\refno \global\refno=1
\newwrite\rfile
\newlinechar=`\^^J
\def\@ref#1#2{\the\refno\n@ref#1{#2}}
\def\n@ref#1#2{\xdef#1{\the\refno}%
\ifnum\refno=1\immediate\openout\rfile=\jobname.refs\fi%
\immediate\write\rfile{\noexpand\item{[\noexpand#1]\ }#2.}%
\global\advance\refno by1}
\def\nref{\n@ref} 
\def\ref{\@ref}   
\def\lref#1#2{\the\refno\xdef#1{\the\refno}%
\ifnum\refno=1\immediate\openout\rfile=\jobname.refs\fi%
\immediate\write\rfile{\noexpand\item{[\noexpand#1]\ }#2\semi}%
\global\advance\refno by1}
\def\cref#1{\immediate\write\rfile{#1\semi}}

\def\preref#1#2{\gdef#1{\@ref#1{#2}}}

\def\semi{;\hfil\noexpand\break}

\def\listrefs{\vfill\eject\immediate\closeout\rfile
\centerline{{\bf References}}\smallskip\frenchspacing%
\input \jobname.refs\vfill\eject\nonfrenchspacing}

\def\inputAuxIfPresent#1{\immediate\openin1=#1
\ifeof1\message{No file \auxfileName; I'll create one.
}\else\closein1\relax\input\auxfileName\fi%
}
\def\NPB{Nucl.\ Phys.\ B}
\def\PRL{Phys.\ Rev.\ Lett.\ }

\def\ZPC{Z.\ Phys.\ C}

\newif\ifWritingAuxFile
\newwrite\auxfile
\def\SetUpAuxFile{%
\xdef\auxfileName{\jobname.aux}%
\inputAuxIfPresent{\auxfileName}%
\WritingAuxFiletrue%
\immediate\openout\auxfile=\auxfileName}

\def\L{\left(}\def\R{\right)}

\def\LB{\left[}\def\RB{\right]}


\catcode`\@=\active
\catcode`@=12  
\catcode`\"=\active


\def\Tr{\mathop{\rm Tr}\nolimits}

\def\mod{\mathop{\rm mod}\nolimits}

\def\pol{\varepsilon}

\def\c{\,\cdot\,}
\def\ksl{\slashed{k}}
\def\Ksl{\slashed{K}}

\def\L{\left(}\def\R{\right)}

\def\spa#1.#2{\left\langle#1\,#2\right\rangle}
\def\spb#1.#2{\left[#1\,#2\right]}
\def\lor#1.#2{\left(#1\,#2\right)}
\def\sand#1.#2.#3{%
\left\langle\smash{#1}{\vphantom1}^{-}\right|{#2}%
\left|\smash{#3}{\vphantom1}^{-}\right\rangle}
\def\sandp#1.#2.#3{%
\left\langle\smash{#1}{\vphantom1}^{-}\right|{#2}%
\left|\smash{#3}{\vphantom1}^{+}\right\rangle}
\def\sandpp#1.#2.#3{%
\left\langle\smash{#1}{\vphantom1}^{+}\right|{#2}%
\left|\smash{#3}{\vphantom1}^{+}\right\rangle}
\catcode`@=11  
\def\meqalign#1{\,\vcenter{\openup1\jot\m@th
   \ialign{\strut\hfil$\displaystyle{##}$ && $\displaystyle{{}##}$\hfil
             \crcr#1\crcr}}\,}
\catcode`@=12  

\newread\epsffilein    
\newif\ifepsffileok    
\newif\ifepsfbbfound   
\newif\ifepsfverbose   
\newdimen\epsfxsize    
\newdimen\epsfysize    
\newdimen\epsftsize    
\newdimen\epsfrsize    
\newdimen\epsftmp      
\newdimen\pspoints     
\pspoints=1bp          
\epsfxsize=0pt         
\epsfysize=0pt         
\def\epsfbox#1{\global\def\epsfllx{72}\global\def\epsflly{72}%
   \global\def\epsfurx{540}\global\def\epsfury{720}%
   \def\lbracket{[}\def\testit{#1}\ifx\testit\lbracket
   \let\next=\epsfgetlitbb\else\let\next=\epsfnormal\fi\next{#1}}%
\def\epsfgetlitbb#1#2 #3 #4 #5]#6{\epsfgrab #2 #3 #4 #5 .\\%
   \epsfsetgraph{#6}}%
\def\epsfnormal#1{\epsfgetbb{#1}\epsfsetgraph{#1}}%
\def\epsfgetbb#1{%
%
%
\openin\epsffilein=#1
\ifeof\epsffilein\errmessage{I couldn't open #1, will ignore it}\else
%
%
   {\epsffileoktrue \chardef\other=12
    \def\do##1{\catcode`##1=\other}\dospecials \catcode`\ =10
    \loop
       \read\epsffilein to \epsffileline
       \ifeof\epsffilein\epsffileokfalse\else
%
%
          \expandafter\epsfaux\epsffileline:. \\%
       \fi
   \ifepsffileok\repeat
   \ifepsfbbfound\else
    \ifepsfverbose\message{No bounding box comment in #1; using defaults}\fi\fi
   }\closein\epsffilein\fi}%
%
%
\def\epsfclipstring{}
\def\epsfsetgraph#1{%
   \epsfrsize=\epsfury\pspoints
   \advance\epsfrsize by-\epsflly\pspoints
   \epsftsize=\epsfurx\pspoints
   \advance\epsftsize by-\epsfllx\pspoints
%
%
   \epsfxsize\epsfsize\epsftsize\epsfrsize
   \ifnum\epsfxsize=0 \ifnum\epsfysize=0
      \epsfxsize=\epsftsize \epsfysize=\epsfrsize
      \epsfrsize=0pt
%
%
     \else\epsftmp=\epsftsize \divide\epsftmp\epsfrsize
       \epsfxsize=\epsfysize \multiply\epsfxsize\epsftmp
       \multiply\epsftmp\epsfrsize \advance\epsftsize-\epsftmp
       \epsftmp=\epsfysize
       \loop \advance\epsftsize\epsftsize \divide\epsftmp 2
       \ifnum\epsftmp>0
          \ifnum\epsftsize<\epsfrsize\else
             \advance\epsftsize-\epsfrsize \advance\epsfxsize\epsftmp \fi
       \repeat
       \epsfrsize=0pt
     \fi
   \else \ifnum\epsfysize=0
     \epsftmp=\epsfrsize \divide\epsftmp\epsftsize
     \epsfysize=\epsfxsize \multiply\epsfysize\epsftmp
     \multiply\epsftmp\epsftsize \advance\epsfrsize-\epsftmp
     \epsftmp=\epsfxsize
     \loop \advance\epsfrsize\epsfrsize \divide\epsftmp 2
     \ifnum\epsftmp>0
        \ifnum\epsfrsize<\epsftsize\else
           \advance\epsfrsize-\epsftsize \advance\epsfysize\epsftmp \fi
     \repeat
     \epsfrsize=0pt
    \else
     \epsfrsize=\epsfysize
    \fi
   \fi
%
%
   \ifepsfverbose\message{#1: width=\the\epsfxsize, height=\the\epsfysize}\fi
   \epsftmp=10\epsfxsize \divide\epsftmp\pspoints
   \vbox to\epsfysize{\vfil\hbox to\epsfxsize{%
      \ifnum\epsfrsize=0\relax
        \includegraphics{#1}%
      \else
        \epsfrsize=10\epsfysize \divide\epsfrsize\pspoints
        \includegraphics{#1}%
      \fi
      \hfil}}%
\global\epsfxsize=0pt\global\epsfysize=0pt}%
%
%
{\catcode`\%=12 \global\let\epsfpercent=
%
%
\long\def\epsfaux#1#2:#3\\{\ifx#1\epsfpercent
   \def\testit{#2}\ifx\testit\epsfbblit
      \epsfgrab #3 . . . \\%
      \epsffileokfalse
      \global\epsfbbfoundtrue
   \fi\else\ifx#1\par\else\epsffileokfalse\fi\fi}%
%
%
\def\epsfempty{}%
\def\epsfgrab #1 #2 #3 #4 #5\\{%
\global\def\epsfllx{#1}\ifx\epsfllx\epsfempty
      \epsfgrab #2 #3 #4 #5 .\\\else
   \global\def\epsflly{#2}%
   \global\def\epsfurx{#3}\global\def\epsfury{#4}\fi}%
%
%
\def\epsfsize#1#2{\epsfxsize}
%
%


\SetUpAuxFile
\loadfourteenpoint
\hfuzz 60 pt
\FiguresInTexttrue

\def\Split{{\rm Split}}
\def\eps{\epsilon}
\def\e{\epsilon}
\def\c{\cdot}
\def\Li{\mathop{\hbox{\rm Li}}\nolimits}
\def\tree{{\rm tree}}

\def\fact{{\rm fact}}
\def\nonfact{{\rm non-fact}}

\def\cg{c_\Gamma^{\vphantom{M^X_X}}}
\def\rg{r_\Gamma^{\vphantom{M^X_X}}}
\def\rS{r_S^{\vphantom{M^X_X}}}
\def\tauG{\tau_\Gamma^{\vphantom{{M^X_X\over Z^X_X}}}}

\def\Fact{{\cal F}}
\def\Ord{{\cal O}}
\def\Soft{{\cal S}}
\def\Collinear{{\cal C}}
\def\half{{\textstyle{1\over2}}}
\def\hf{{\textstyle{1\over2}}}

\def\Psl{\not{\hbox{\kern-2.1pt $P$}}}
\def\Ksl{\not{\hbox{\kern-2.1pt $K$}}}

\def\discOne{d_1}
\def\discTwo{d_2}
\def\discThree{b}

\def\Fs#1#2{F^{{#1}}_{n:#2}}
\def\Fone{\Fs{\rm 1m}}
\def\Feasy{\Fs{{\rm 2m}\,e}}
\def\Fhard{\Fs{{\rm 2m}\,h}}
\def\Fthree{\Fs{\rm 3m}}
\def\Ffour{\Fs{\rm 4m}}
\def\Wsix#1{W_6^{(#1)}}


\nopagenumbers

\noindent
hep-ph/9503236 \hfill UCLA/TEP/95-6


\Title{Factorization in One-Loop Gauge Theory}

\vskip 3 truecm
\centerline{Zvi Bern and Gordon Chalmers\footnote{$^\star$}{%
Address after September 1, 1995: Institute for Theoretical Physics,
SUNY, Stony Brook, NY 11794.}}
\vskip .2 cm
\baselineskip=13pt
\centerline{\it Department of Physics}
\centerline{\it UCLA}
\centerline{\it Los Angeles, CA 90024}
\vglue 0.3cm


\vglue 2 cm
\centerline{\tenrm ABSTRACT}
\vglue 0.3cm
{\rightskip=3pc
\leftskip=3pc
\tenrm\baselineskip=12pt
\noindent
Factorization properties of one-loop gauge theory amplitudes have been
used as checks on explicitly computed amplitudes and in the
construction of ans\"atze for higher-point ones. In massless theories,
such as QCD at high energies, infrared divergences complicate
factorization.  Here we prove that factorization in such theories is
described by a set of universal functions.  In particular, a proof of
the universality of one-loop splitting functions as the momenta of two
particles become collinear is presented.  Factorization in
multi-particle channels is also given.  The discontinuity functions
that appear in the splitting functions may also be used to obtain
infrared divergent box integral functions from finite ones.}
\baselineskip 15 pt

\vglue 0.3cm

\vfill
\eject

\footline={\hss\tenrm\folio\hss}



\preref\NLOTwoJets{%
S.D.\ Ellis, Z. Kunszt and D.E.\ Soper,
Phys.\ Rev.\ D40:2188 (1989);
Phys.\ Rev.\ Lett.\ 64:2121 (1990);
Phys.\ Rev.\ Lett.\ 69:1496 (1992)\semi
F. Aversa, M. Greco, P. Chiappetta and J.P.\
Guillet, Phys.\ Rev.\ Lett.\ 65:401 (1990)\semi
F. Aversa, L. Gonzales, M. Greco, P. Chiappetta
and J.P.\ Guillet, Z. Phys.\ C49:459 (1991)}

\preref\FiveGluon{%
Z. Bern, L. Dixon and D.A. Kosower, Phys.\ Rev. Lett.\
70:2677 (1993)}

\preref\Kunsztqqqqg{%
Z. Kunszt, A. Signer and Z. Tr\'ocs\'anyi, Phys.\ Lett.\ B336:529 (1994),
hep-ph/9405386}

\preref\TreeColor{%
J.E.\ Paton and Chan Hong-Mo, Nucl.\ Phys. B10:519 (1969)\semi
F.A. Berends and W.T. Giele, Nucl.\ Phys.\ B294:700 (1987)\semi
M.\ Mangano, S. Parke and Z.\ Xu, Nucl.\ Phys.\ B298:653 (1988)\semi
M.\ Mangano, Nucl.\ Phys.\ B309:461 (1988)}

\preref\ManganoReview{%
M. Mangano and S.J. Parke, Phys.\ Rep.\ 200:301 (1991)}

\preref\GG{W.T.\ Giele and E.W.N.\ Glover,
Phys.\ Rev.\ D46:1980 (1992)\semi
W.T.\ Giele, E.W.N.\ Glover and D. A. Kosower,
Nucl.\ Phys.\ B403:633 (1993)}

\preref\KunsztSoper{Z. Kunszt and D. Soper, Phys.\ Rev.\ D46:192 (1992)}

\preref\SpinorHelicity{%
F.A.\ Berends, R.\ Kleiss, P.\ De Causmaecker, R.\ Gastmans and T.\ T.\ Wu,
        Phys.\ Lett.\ 103B:124 (1981)\semi
P.\ De Causmaeker, R.\ Gastmans,  W.\ Troost and  T.T.\ Wu,
Nucl. Phys. B206:53 (1982)\semi
R.\ Kleiss and W.J.\ Stirling,
   Nucl.\ Phys.\ B262:235 (1985)\semi
   J.F.\ Gunion and Z.\ Kunszt, Phys.\ Lett.\ 161B:333 (1985)\semi
Z. Xu, D.-H.\ Zhang and L. Chang, Nucl.\ Phys.\ B291:392 (1987)}

\preref\Color{%
Z. Bern and D.A.\ Kosower, Nucl.\ Phys.\ B362:389 (1991)}

\preref\Long{Z. Bern and D.A.\ Kosower \NPB 379:451 (1992)}

\preref\StringBased{
Z. Bern and D.A.\ Kosower, \PRL 66:1669 (1991)\semi
Z. Bern, Phys.\ Lett.\  296B:85 (1992)\semi
K. Roland, Phys.\ Lett.\ 289B:148 (1992)\semi
M.J.\ Strassler,  Nucl.\ Phys.\ B385:145 (1992)\semi
C.S. Lam, Nucl.\ Phys.\ B397:143 (1993); Phys.\ Rev.\ D48:873 (1993)\semi
Z. Bern, D.C. Dunbar and T. Shimada, Phys.\ Lett.\ 312B:277 (1993),
hep-th/9307001\semi
G. Cristofano, R. Marotta and K. Roland, Nucl.\ Phys.\ B392:345 (1993)\semi
M.G.\ Schmidt and C. Schubert, Phys.\ Lett.\ 318B:438 (1993);
Phys.\ Lett.\ B331:69 (1994)\semi
D. Fliegner, M.G.\ Schmidt and C. Schubert, Z.\ Phys.\ C64:111 (1994),
hep-ph/9401221\semi
D.C.\ Dunbar and P.S. Norridge, Nucl.\ Phys.\ B433:181 (1995),
hep-th/9408014\semi
P. Di Vecchia, A. Lerda, L. Magnea and R. Marotta, preprint hep-th/9502156}

\preref\Tasi{
Z. Bern, hep-ph/9304249, in {\it Proceedings of Theoretical
Advanced Study Institute in High Energy Physics (TASI 92)},
eds.\ J. Harvey and J. Polchinski (World Scientific, 1993)\semi
Z.\ Bern and A.\ Morgan, Phys.\ Rev.\ D49:6155 (1994), hep-ph/9312218}

\preref\Mapping{Z. Bern and D.C.\ Dunbar,  Nucl.\ Phys.\ B379:562 (1992)}

\preref\GSB{M.B.\ Green, J.H.\ Schwarz and L.\ Brink,
 Nucl.\ Phys.\ B198:474 (1982)}

\preref\Cutting{L.D.\ Landau, Nucl.\ Phys.\ 13:181 (1959)\semi
 S. Mandelstam, Phys.\ Rev.\ 112:1344 (1958), 115:1741 (1959)\semi
 R.E.\ Cutkosky, J.\ Math.\ Phys.\ 1:429 (1960)}

\preref\AP{G. Altarelli and G. Parisi, Nucl.\ Phys.\ B126:298, (1977)\semi
G.\ Curci, W.\ Furmanski and R.\ Petronzio, Nucl.\ Phys.\ B175:27
(1980)}

\preref\Susy{%
M.T.\ Grisaru, H.N.\ Pendleton and P.\ van Nieuwenhuizen,
Phys. Rev. {D15}:996 (1977)\semi
M.T. Grisaru and H.N. Pendleton, Nucl.\ Phys.\ B124:81 (1977)\semi
S.J. Parke and T. Taylor, Phys.\ Lett.\ 157B:81 (1985)\semi
Z. Kunszt, Nucl.\ Phys.\ B271:333 (1986)}

\preref\KunsztFourPoint{%
Z. Kunszt, A. Signer and Z. Tr\'ocs\'anyi, Nucl.\ Phys.\ B411:397 (1994)}

\preref\SusyFour{Z. Bern, L. Dixon, D.C. Dunbar and D.A. Kosower,
Nucl.\ Phys.\ B425:217 (1994), hep-ph/9403226}

\preref\SusyOne{Z. Bern, L. Dixon, D.C. Dunbar and D.A. Kosower,
Nucl.\ Phys.\ B435:59 (1995), hep-ph/9409265}

\preref\BDKconf{Z. Bern, L. Dixon and D.A. Kosower, hep-th/9311026,
in {\it Proceedings of Strings 1993}, eds. M.B. Halpern, A. Sevrin
and G. Rivlis (World Scientific, 1994), hep-th/9311026}

\preref\AllPlus{Z. Bern, G. Chalmers, L. Dixon and D.A. Kosower,
Phys.\ Rev.\ Lett.\ 72:2134 (1994), hep-ph/9312333}

\preref\Mahlon{G.D.\ Mahlon, Phys.\ Rev.\ D49:2197 (1994);
               Phys.\ Rev.\ D49:4438 (1994)}

\preref\Siegel{W. Siegel, Phys.\ Lett.\ 84B:193 (1979)\semi
D.M.\ Capper, D.R.T.\ Jones and P. van Nieuwenhuizen, Nucl.\ Phys.\
B167:479 (1980)}

\preref\CollinsBook{J.C.\ Collins, {\it Renormalization}
(Cambridge University Press, 1984)}

\preref\Recursive{F.A.\ Berends and W.T.\ Giele, \NPB 306:759 (1988)\semi
D.A.\ Kosower, \NPB335:23 (1990)}

\preref\Background{
L.F.\ Abbott, Nucl.\ Phys.\ B185:189 (1981)\semi
L.F. Abbott, M.T. Grisaru and R.K. Schaefer,
Nucl.\ Phys.\ B229:372 (1983)}

\preref\GN{J.L.\ Gervais and A. Neveu, Nucl.\ Phys.\ B46:381 (1972)}

\preref\IntegralsShort{Z. Bern, L. Dixon and D.A. Kosower,
Phys.\ Lett.\ 302B:299 (1993); erratum {\it ibid.} 318B:649 (1993)}

\preref\IntegralsLong{Z. Bern, L. Dixon and D.A. Kosower,
\NPB 412:751 (1994), hep-ph/9306240}

\preref\ParkeTaylor{S.J.\ Parke and T.R.\ Taylor, \PRL 56:2459
(1986)}

\preref\Fermion{Z. Bern, L. Dixon and D.A. Kosower,
Nucl.\ Phys.\ B437:259 (1995), hep-ph/9409393}

\preref\Lewin{L.\ Lewin, {\it Dilogarithms and Associated Functions\/}
(Macdonald, 1958)}

\preref\TreeCollinear{F.A. Berends and W.T. Giele, Nucl.\ Phys.\
B313:595 (1989)}

\preref\GordonConf{G. Chalmers, hep-ph/9405393,
in {\it Proceedings of the XXII ITEP International Winter School
of Physics} (Gordon and Breach, 1995)}

\preref\Morgan{E.W.N.\ Glover and A. Morgan, Z. Phys.\ C60:175 (1993)}

\preref\KunsztSingular{%
Z. Kunszt, A. Signer and Z. Tr\'ocs\'anyi, Nucl.\ Phys.\
B420:550 (1994)}

\preref\FourMassBox{A. Denner, U. Nierste and R. Scharf,
  \NPB{367:637 (1991)}\semi
N.I.\ Usyukina and A.I.\ Davydychev, Phys.\ Lett.\ {298B:363 (1993)};
Phys.\ Lett.\ {305B:136 (1993)}}

\preref\DimensionalRegularization{G. 't\ Hooft and M. Veltman, Nucl.\
Phys.\ B44:189 (1972)}

\preref\PV{G.\ Passarino and M.\ Veltman, Nucl.\ Phys.\ {B160:151} (1979)}

\preref\Reduction{L.M.\ Brown and R.P.\ Feynman, Phys.\ Rev.\ 85:231
(1952)\semi
L.M.\ Brown, Nuovo Cimento {21:3878} (1961)\semi
G. 't Hooft and M. Veltman, \NPB{153:365 (1979)}\semi R.G. Stuart,
Comp.\ Phys.\ Comm.\ 48:367 (1988)\semi R.G. Stuart and A. Gongora,
Comp.\ Phys.\ Comm.\ 56:337 (1990)}

\preref\VNV{
W. van Neerven and J.A.M. Vermaseren, Phys.\ Lett.\ 137B:241 (1984)\semi
G.J. van Oldenborgh and J.A.M. Vermaseren, \ZPC{46:425 (1990)}}

\preref\OtherInt{
B. Petersson,  J. Math. Phys., 6:1955 (1965)\semi
G. Kallen and J.S. Toll, J. Math.\ Phys., 6:299 (1965)\semi
D.B. Melrose, Il Nuovo Cimento 40A:181 (1965)}

\preref\MutaBook{T. Muta,
  {\it Foundations of Quantum Chromodynamics: an introduction to
   perturbative methods in gauge theories}, (World
   Scientific, 1987)}

\preref\GordonNotes{G.\ Chalmers, (unpublished)}


\section{Introduction}
\tagsection\IntroSection

In recent years a variety of one-loop gauge theory amplitudes with
five or more external legs, including all virtual corrections to
three-jet production at hadron colliders
[\use\FiveGluon,\use\Kunsztqqqqg,\use\Fermion], have been computed.  A
number of formal developments in computational techniques, including
spinor helicity methods [\use\SpinorHelicity], string-based methods
[\use\Long,\use\StringBased], supersymmetry relations [\use\Susy],
recursive techniques [\use\Recursive,\use\Mahlon], and unitarity
methods [\use\SusyFour,\use\SusyOne] have allowed the computation of a
large variety of new amplitudes.

Factorization properties
[\use\TreeCollinear,\use\FiveGluon,\use\AllPlus,\use\SusyFour,\use\Fermion]
provide strong consistency checks on newly computed amplitudes.
Furthermore, they provide a means for obtaining higher-point
amplitudes from lower-point ones.  To do this one constructs a
function that has the correct factorization properties in all
channels, which is then an ansatz for the amplitude.  An explicit
construction of $n$-gluon amplitudes with all identical helicities
derived from this technique has been given in
refs.~[\use\AllPlus,\use\BDKconf] and verified via recursive
techniques~[\use\Mahlon].

The limiting form of one-loop massless gauge theory amplitudes as the
momenta of two legs become collinear has been given, although not
proven, in terms of splitting functions in previous papers
[\use\AllPlus,\use\GordonConf,\SusyFour,\use\Fermion].  In particular,
a tabulation of the splitting functions in QCD has been given in
refs.~[\use\SusyFour,\use\Fermion] from the collinear limits of
calculated five-point amplitudes.  Ample evidence of the universality
of the splitting functions is provided by the large number of
explicitly known one-loop helicity amplitudes in massless gauge theory
[\use\Long,\use\KunsztFourPoint,\use\FiveGluon,\use\Kunsztqqqqg,\use\BDKconf,%
\use\AllPlus,\use\Mahlon,\use\SusyFour,\use\SusyOne].  For
multi-particle factorization less information is available; except for
the recently computed set of one-loop six-point amplitudes in $N=4$
supersymmetric gauge theory [\use\SusyOne], no examples have been
constructed with non-trivial factorization properties.  (Mahlon
[\use\Mahlon] has also constructed a set of helicity amplitudes
containing multi-particle poles, but with simple factorization
properties due to a lack of infrared singularities.)  Other
properties at loop level which are known are for the unpolarized
differential cross-sections, and give the well known Altarelli-Parisi
splitting functions [\use\AP].  The Altarelli-Parisi splitting
functions are for cross-sections which include both real emission and
virtual diagrams; the splitting functions that we discuss in this
paper are for the virtual (i.e. loop) diagrams alone.

In this paper we provide a general proof of the universal behavior of
massless amplitudes, independent of the number of legs, when
factorized on a collinear or multi-particle pole.  Amplitudes
which lack infrared divergences factorize straightforwardly as one
would expect.  For amplitudes with infrared singularities the
situation is more subtle: massless gauge theory amplitudes generally
do {\it not} factorize in any simple sense since there are
contributions which cannot be interpreted directly in terms of
lower-point amplitudes.  Nevertheless, we will prove here that the
non-factorization may be described in terms of a small set of {\it
factorization functions}, given in this paper, whose coefficients are
fixed by the known infrared divergences.  We also give a practical
procedure for calculating the splitting functions of any one-loop
amplitude from two- and three-point diagrams.

In previous papers [\use\AllPlus,\use\GordonConf] proofs of
the universality of the (two-particle) one-loop splitting functions
for the case of $n$ external gluons with a virtual scalar loop were
outlined.  (This was sufficient for the particular helicity amplitudes
considered in these papers, because of supersymmetry identities
[\use\Susy] which relate the contributions of virtual gluons and fermions
to those of scalars.)  The proof was based on an analysis of
Feynman diagrams in the limit that the momenta of two external legs
become collinear.  This type of proof also works for a fermion loop
[\use\GordonNotes], but a generalization to the case of virtual gluons
and multi-particle factorization is non-trivial due to complications
in quantifying non-factorizing contributions.

One expects, however, that the universality of the splitting functions
should follow from general field theory considerations and not from
the details of particular diagrams.  Our proof will make this explicit
and is valid for any particle content and external states.  The key
behind the proof is that we can link all non-factorizing contributions
to infrared divergences present in the one-loop amplitudes.  We will
show that only a small number of functions may enter and that their
coefficients are fixed by their infrared divergences.  Since the
divergences have a known form
[\use\KunsztSoper,\use\GG,\use\KunsztSingular], all functions entering
into factorization are fixed and universal; this is the main result of
this paper.

A spin-off from our analysis is that the same functions that
contribute to the loop splitting functions may also be used to
efficiently obtain all infrared divergent box integrals by taking
limits of known infrared finite massive box integrals
[\use\FourMassBox].  To illustrate this, we reproduce the box
integrals necessary for calculating amplitudes in massless gauge
theory given in	ref.~[\use\IntegralsLong].

First, in Section~\use\ReviewSection\ we review some properties of
gauge theory amplitudes that we will find useful in our subsequent
discussion.  In Section~\use\CollinearSection\ we present our results
for the one-loop splitting functions; an explicit sample calculation
is presented for obtaining the splitting functions from three-point
diagrams.  In section~\use\MultiSection\ we present analogous results
for multi-particle factorization and a non-trivial example.
The actual proof of these results are presented in
Section~\use\ProofSection\ and the appendices.  In
section~\use\DiscontinuitySection\ we show that the same functions
which appear in the non-factorizing contributions are useful for
obtaining infrared divergent box integral functions from infrared
finite ones.  We also provide a variety of appendices on integrals and
their properties which are useful in the proof.

\section{Review of Previous Results}
\tagsection\ReviewSection

We now briefly review some previous results and conventions
that we will use in this paper: the color ordering of non-abelian
scattering amplitudes, the spinor helicity method, tree-level
two-particle collinear limits and the structure of infrared and
ultraviolet singularities.  We review (and slightly modify) the
integral reduction method for calculating any one-loop integral in
terms of box, triangle and bubble functions in
appendix~\use\ReductionAppendix.

\subsection{Color-Ordered Amplitudes}

Tree-level $SU(N_c)$ gauge theory amplitudes can be written in terms
of independent color-ordered partial amplitudes multiplied by an
associated color trace [\use\TreeColor].  This is extensively
discussed in the review article of Mangano and Parke
[\use\ManganoReview], whose normalizations and conventions we follow.
(In particular, we normalize fundamental representation color matrices
as $\Tr[T^a T^b] = \delta^{ab}$.)  One of the key features of the
partial amplitudes is that the external legs have a fixed ordering.

At one-loop, although the analogous decomposition [\use\Color] is a
bit more complicated, gauge theory amplitudes may be conveniently
written in terms of gauge-invariant `primitive' amplitudes
[\use\Fermion] which also have fixed ordering of external legs.  The
$n$-gluon primitive amplitudes correspond to the leading-color partial
amplitudes, $A^{\rm loop}_{n;1}$.  Furthermore, the complete amplitude
may be expressed in terms of appropriate permutation sums over
primitive amplitudes multiplied by ordered color factors.  Therefore,
all results obtained in this paper for primitive amplitudes may be
converted to results for full amplitudes. For amplitudes with only external
gluons, or two quarks and the rest gluons, explicit expressions giving
the full amplitudes in terms of these primitive amplitudes may be
found in refs.~[\use\SusyFour,\use\Fermion].  For this paper, the main
property needed is that a decomposition of one-loop $SU(N_c)$ gauge
theory amplitudes exists in terms of a set of color decomposed
amplitudes where the ordering of external legs is fixed; this is
convenient, although not necessary for our discussion.

\subsection{Spinor Helicity}

In explicit calculations it is usually convenient to use a
spinor helicity basis [\use\SpinorHelicity], where
all quantities are rewritten in terms of
Weyl spinors $\vert k^{\pm} \rangle$.  In the formulation
of Xu, Zhang and Chang
the polarization vectors are expressed as
$$
\pol^{(+)}_\mu (k;q) =
     {\sand{q}.{\gamma_\mu}.k
      \over \sqrt2 \spa{q}.k}\, ,\hskip 1cm
\pol^{(-)}_\mu (k,q) =
     {\sandpp{q}.{\gamma_\mu}.k
      \over \sqrt{2} \spb{k}.q} \, ,
\eqn\HelicityDef
$$
where $q$ is an arbitrary null `reference momentum' which drops out of
the final gauge-invariant amplitudes.  The reader is referred to the
article of Xu, Zhang and Chang for further details. For the purposes
of this presentation we note that
$$
\langle k_i^{-} \vert k_j^{+} \rangle \equiv \langle ij \rangle =
\sqrt{2k_i \c k_j} \exp(i\phi), \quad\quad\quad
\langle k_i^{+} \vert k_j^{-} \rangle \equiv [ij] =
-\sqrt{2k_i \c k_j} \exp(-i\phi) \, ,
\anoneqn
$$
where $\phi$ is a phase.  These spinor products vanish in the limit
$k_i \c k_j\rightarrow 0$, are anti-symmetric and satisfy
$$
\spa{i}.j \spb{j}.i = 2 k_i \cdot k_j \, .
\anoneqn
$$

\subsection{Tree-Level Two-particle Collinear Factorization}

Consider first an $n$-point tree-level partial amplitude
$A_n^{\rm tree}(1,2,\ldots,n)$ with a fixed ordering of external legs and
an arbitrary helicity configuration
$(\lambda_1,\lambda_2,\ldots,\lambda_n)$, where each $\lambda_i =
\pm$.  As the momenta of two {\it neighboring\/} legs $a$ and $b$
become collinear the leading behavior of the amplitudes is given by
[\use\ParkeTaylor,\use\TreeCollinear]
$$
A_{n}^{\rm tree}\ \mathop{\longrightarrow}^{a \parallel b}\
\sum_{\lambda=\pm}
 \Split^{\rm tree}_{-\lambda}(a^{\lambda_a},b^{\lambda_b})\,
      A_{n-1}^{\rm tree}(\ldots K^\lambda \ldots)\, ,
\eqn\treesplit
$$
where the non-vanishing splitting functions (or amplitudes) diverge as
$1/\sqrt{s_{ab}}$ in the collinear limit $s_{ab}=(k_a+k_b)^2
\rightarrow0$.  In this equation and all subsequent ones we extract
all coupling constants from the color-ordered amplitudes.
The collinear limit is defined by $k_a = z\,K$ and
$k_b = (1-z)\,K$, where the null vector
$K$ is the sum of the collinear momenta;
$\lambda$ is the helicity of the intermediate state with momentum
$K$.  The tree splitting functions $\Split^{\rm tree}_{-\lambda}
(a^{\lambda_a},b^{\lambda_b})$ may be found in ref.~[\use\ManganoReview].
The $g \rightarrow gg$ splitting functions are
$$
\eqalign{
&\Split^{\tree}_{-}(a^{-},b^{-}) = 0 \,,  \hskip 3.4 cm
\Split^{\tree}_{-}(a^{+},b^{+})={1\over \sqrt{z(1-z)}
\langle ab \rangle} \; ,  \cr
& \Split^{\tree}_{-}(a^{+},b^{-})=-{z^{2}\over
\sqrt{z(1-z)} [ab]} \, ,
\hskip 1. cm
\Split^{\tree}_{-}(a^{-},b^{+})=-{(1-z)^{2}\over
\sqrt{z(1-z)} [ab]} \, ,  \cr }
\anoneqn
$$
which explicitly exhibit the $1/\sqrt{s_{ab}}$ pole.  All remaining
$g \rightarrow gg$ splitting functions may be obtained by parity.

\subsection{Infrared and Ultraviolet Singularities in One-Loop Amplitudes}

In massless gauge theory one encounters three types of singularities
when evaluating virtual corrections with fixed particle number.
Besides the usual ultraviolet singularities one encounters both soft
and collinear infrared ones. (By `soft' we refer to the $\eps^{-2}$
terms in dimensionally regulated amplitudes.)  These types of
singularities must cancel in final physical cross-sections containing
both real and virtual corrections, but are present in the individual
parts.  The singularities in an $n$-point massless QCD amplitude with
a fixed ordering of legs are of the form
[\use\KunsztSoper,\use\GG,\use\KunsztSingular]
$$
A_{n}^{\rm loop} \Bigr|_{\rm singular} =
 \cg A^\tree_n  \biggl[ - {1\over\eps^2} \sum_{j = 1}^n \Soft_j^{[n]}
    \L{\mu^2 \over - s_{j, j+1} } \R^\eps + \Collinear^{[n]} {1\over\eps}
\biggr] \, ,
\eqn\singular
$$
where

\vskip -.6 cm
$$
c_\Gamma\ =\ {(4 \pi)^\eps \over 16 \pi^2 }
{\Gamma(1+\eps)\Gamma^2(1-\eps)\over\Gamma(1-2\eps)}\, ,
\eqn\Prefactor
$$
with $\eps = (4-D)/2$ the dimensional regularization parameter and
$\mu$ the renormalization scale.  The parameters $\Soft_j^{[n]}$ are
the coefficients of the soft singularities and $\Collinear^{[n]}$ is
the sum of the coefficients of collinear and ultraviolet
divergences, which depend on the particle content of the amplitude.
Although we will write the soft singularities in the form appearing in
eq.~(\use\singular), the functional form of the amplitudes are taken,
as usual, to be expanded in $\eps$.  This expansion is implied in
eq.~(\use\singular) and in all subsequent expressions.

The main property that we will make use of is that in massless QCD the
coefficients $\Soft_j^{[n]}$ and $\Collinear^{[n]}$ are known.  For
primitive amplitudes which have been stripped of all color factors and
have a fixed ordering of legs, these coefficients are particularly
simple.  For such amplitudes, $\Soft_j^{[n]}$ is $0$ or $1$ depending
on the particle type of the internal loop line connecting legs $j$ and
$j+1$ in the `parent diagram', which is the one where all external
legs are directly attached to the loop by three-vertices.  If the
propagator between legs $j$ and $j+1$ is a gluon then $\Soft_j^{[n]}
=1$, while if it is a fermion or scalar then $\Soft_j^{[n]} =0$.
(Explicit examples are found in ref.~[\use\Fermion].)

The collinear infrared singularities for $n$-point amplitudes are
$$
\Collinear^{[n]}_{\rm IR} =
- \sum_{a=1}^n {\gamma (a)} \, ,
\eqn\IRCollinear
$$
where the sum is over all $n$ legs and depends on whether leg $a$ is
a gluon $g$ or a fermion $q$,
$$
\eqalign{
\gamma(g) & = {11 \over 6} - {1\over 3} {n_f \over N_c}
- {1\over 6} {n_s \over N_c} , \cr
\gamma(q) &= {3 \over 4} \Bigl( 1 - {1\over N_c^2} \Bigr) \, , \cr}
\eqn\collinear
$$
where $n_{\! f}$ is the number of fermions, $n_s$ the number of
scalars (which is zero in QCD) and $N_c$ the number of colors.  Our
conventions follow the ones in
refs.~[\use\FiveGluon,\use\SusyFour,\use\Fermion]; in particular we
have extracted an overall factor of $N_c$ from leading color
amplitudes so that contributions from fundamental representation loops
carry a factor of $1/N_c$.  The normalization for each complex scalar
is non-standard and represents a total of four states in the $(N_c +
\overline N_c)$ representation (instead of the usual two; the
rationale for this choice is to maintain supersymmetry identities with
the four states of fundamental representation Dirac fermions).  There
are also ultraviolet singularities
$$
\Collinear^{[n]}_{\rm UV} =
(n-2) \Bigl({11\over 6} - {1\over 3} {n_{\! f} \over N_c}
-{1\over 6} {n_s \over N_c} \Bigr) \, .
\eqn\ultraviolet
$$
The coefficient $\Collinear^{[n]}$ appearing in eq.~(\use\singular) is
the sum
$$
\Collinear^{[n]} =
\Collinear^{[n]}_{\rm UV} + \Collinear^{[n]}_{\rm IR} \, ,
\eqn\CollDiv
$$
of the infrared and ultraviolet contributions in
eqs.~(\use\IRCollinear) and (\use\ultraviolet).

\vskip -.4 truecm
\section{Collinear Limits}
\tagsection\CollinearSection

\vskip -.3 truecm
\subsection{General Results}

In this section we present our results for the factorization of one-loop
amplitudes as the momenta of two color-adjacent external legs become
collinear; in subsequent sections we present the proof.  The behavior
of one-loop amplitudes in the collinear limit found from explicit
calculations of amplitudes with five or more legs, is
[\use\FiveGluon,\use\SusyFour,\use\Fermion]
$$
\hskip -.1 cm \eqalign{
A_{n}^{\rm loop}\ \mathop{\longrightarrow}^{a \parallel b}\
\sum_{\lambda=\pm}  \biggl\{
  \Split^{\rm tree}_{-\lambda}(a^{\lambda_a},b^{\lambda_b})\,
&
      A_{n-1}^{\rm loop}(\ldots K^\lambda\ldots)
  +\Split_{-\lambda}^{\rm loop} (a^{\lambda_a},b^{\lambda_b})\,
      A_{n-1}^{\rm tree}(\ldots K^\lambda\ldots) \biggr\} \, ,
\cr}
\eqn\loopsplit
$$
where the $A_n^{\rm loop}$ and $A_n^{\rm tree}$ are color-decomposed
one-loop and tree amplitudes with a fixed ordering of legs and $a$ and
$b$ are consecutive in the ordering.  This is a natural generalization
of the tree-level factorization and is exactly what one would expect
in massive theories with no infrared divergences, since in this case
individual diagrams respect the factorization depicted in
\fig\CollinearFactFigure, where legs $i$ and $i+1$ correspond to
$a$ and $b$.

\vskip -.9 cm
\LoadFigure\CollinearFactFigure{\baselineskip 13 pt
\noindent\narrower\ninerm The `naive' factorization of one-loop
amplitudes in the limit as two external momenta become collinear,
where $K = k_i + k_{i+1}$.
The shaded disc represents the sum over tree diagrams
and the annulus the sum over one-loop diagrams.}
{\epsfysize 1.4 truein}{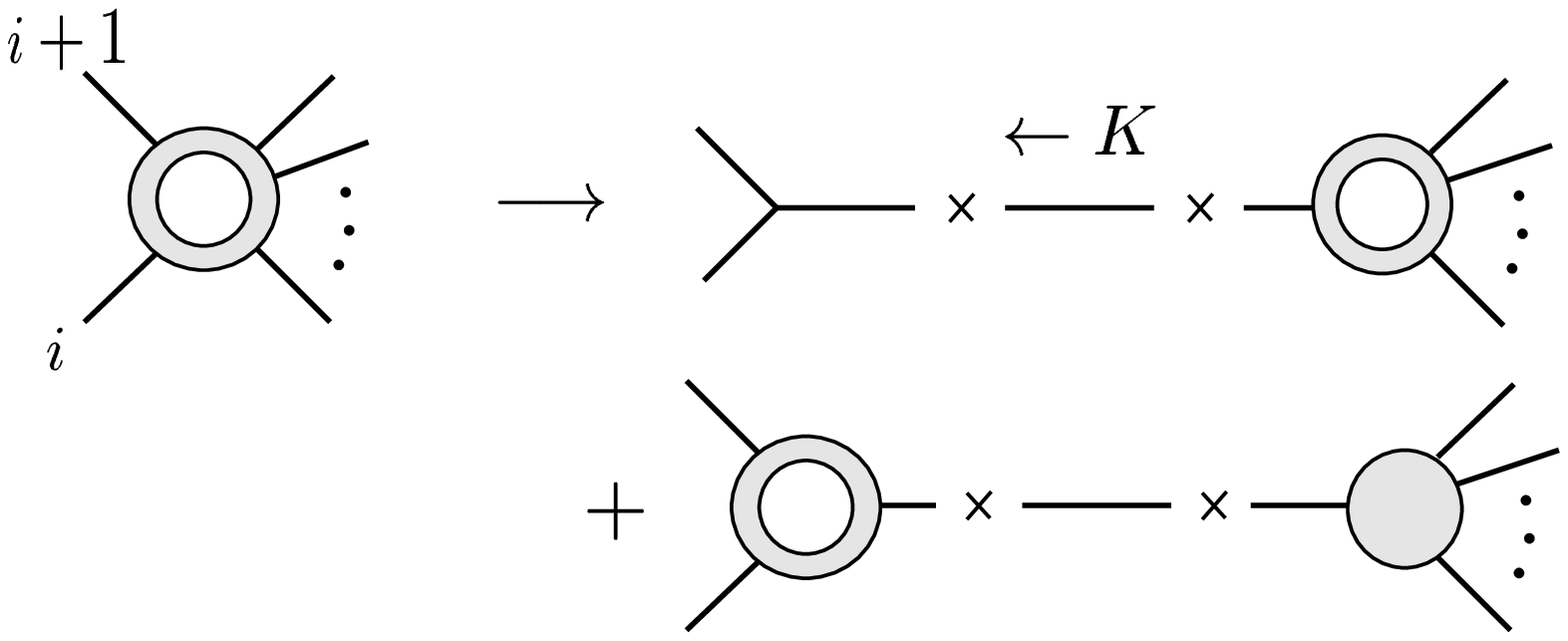}{}

For infrared divergent theories the situation is much more subtle
since individual diagrams may not have a smooth behavior as the
intermediate momentum becomes massless.  We shall, however, prove that
the splitting functions $\Split_{-\lambda}^{\rm loop}$ are independent
of the number of external legs $n$, and that they may be obtained by
calculating three-point diagrams followed by adding in a particular
set of functions fixed by the singularities in $\eps$ through
eq.~(\use\singular).

The following analysis is performed on bare amplitudes (before
renormalization).  For massless amplitudes it is slightly more
convenient to leave all quantities as bare, but at any point in the
discussion one may convert to renormalized quantities by performing
the appropriate ultraviolet subtraction in all quantities; this will
be done for the final splitting functions.  We will also use the
dimensional reduction [\use\Siegel] or equivalently the
four-dimensional helicity (FDH) regularization schemes
[\use\Long]. These schemes are convenient since they do not introduce
any $\eps$-dependence coming from the contraction of tensors which
count the number of states circulating in the loop. It is
straightforward to convert the splitting functions to conventional
[\use\CollinsBook] or 't~Hooft-Veltman schemes
[\use\DimensionalRegularization]; through $\Ord(\eps^0)$ the splitting
functions shift by constants multiplied by tree splitting functions,
as given in refs.~[\use\SusyFour,\use\Fermion]. A discussion of scheme
conversions has been given in ref.~[\use\KunsztFourPoint].

In general, infrared divergent amplitudes will not factorize naively.
As a simple example of non-factorization, consider an infrared
singularity of the form
$$
-\cg {1\over \eps^2} (-s_{23})^{-\eps} A_n^{\tree} = -\cg \Bigl[
{1\over\eps^2} - {1\over \eps} \ln(- s_{23}) + \cdots \Bigr] A_n^{\tree} \,,
\eqn\SoftExample
$$
in the collinear limit $k_1 = z K$ and $k_2 = (1-z)K$.  This
singularity contains the logarithmic term
$$
{1\over \eps} \ln(- s_{23}) \rightarrow {1\over \eps}
\ln\bigl( - (1-z) s_{K3}\bigr)\, ,
\anoneqn
$$
with $k_i^2 = 0$,
which introduces a $\ln(1-z)/\eps$ not belonging with either
the `naively factorized' diagrams on the left- or right-hand-side of
the tree-pole in fig.~\use\CollinearFactFigure\ (for $i=1$).
It therefore cannot be interpreted as a factorizing
contribution.

Our results for the limits of amplitudes as legs $i$ and $i+1$
become collinear ($k_i \rightarrow z K$ and
$k_{i+1}\rightarrow (1-z) K$) are as follows.  As the two legs become
collinear, loop integrals may not have smooth limits and develop
infrared divergences.  The possible discontinuities, which describe
the off-shell to on-shell transition, are described by a set of
universal `discontinuity functions'.  Two discontinuity functions
which play an explicit role in computing the loop splitting functions
are
$$
\eqalign{
\discThree(s_{i, i+1}) &=
{1\over\eps(1-2\eps)} \L{\mu^2 \over -s_{i,i+1}}\R^\eps \, ,  \cr
\discOne(s_{i, i+1}) &= {1\over\eps^2} \L{\mu^2 \over -s_{i,i+1}}\R^\eps
\,.\cr}
\eqn\Discontinuities
$$
(There are additional discontinuity functions to be discussed in
section~\use\ProofSection, but they are not needed for our discussion
of the factorizing diagrams.)

Now consider the computation of loop splitting functions which are composed
of factorizing and non-factorizing pieces,
$$
\Split^{\rm loop} = \Split^{\rm fact} + \Split^{\rm non-fact} \,.
\eqn\totalsplit
$$

The first step in obtaining the factorizing contributions is to
compute the diagrams depicted in \fig\MasterSplitLoopFigure.  Observe
that diagrams with bubbles on external lines are not included.  This
is due to the dimensional regularization prescription that massless
on-shell bubble diagrams vanish [\use\MutaBook], which is interpreted
as a complete cancellation of infrared and ultraviolet divergences.
(Below we also describe the diagrams necessary for calculating the
splitting functions in a massive theory.) The diagrams in
fig.~\use\MasterSplitLoopFigure\ yield the general form
$$
{\cal D} = B_2 {1\over \eps^2} \L{\mu^2 \over - s_{i, i+1}}\R^\eps
+ B_1 {1\over \eps} + B_0  \, ,
\eqn\FactSubtract
$$
where $B_1$ and $B_2$ are rational functions depending on the particle
content and on the type of external legs.  Depending on the type of
off-shell leg, represented by the dotted line in
fig.~\use\MasterSplitLoopFigure, ${\cal D}$ may have uncontracted
spinor or vector indices.

\vskip -1.0 cm
\LoadFigure\MasterSplitLoopFigure{\baselineskip 13 pt
\noindent\narrower\ninerm The diagrams in a massless
theory (ignoring tadpoles) that need to be calculated to
obtain the factorizing contribution to the loop splitting function.
The dotted line represents the off-shell
leg on which the collinear factorization is performed.}
{\epsfysize
1.4truein}{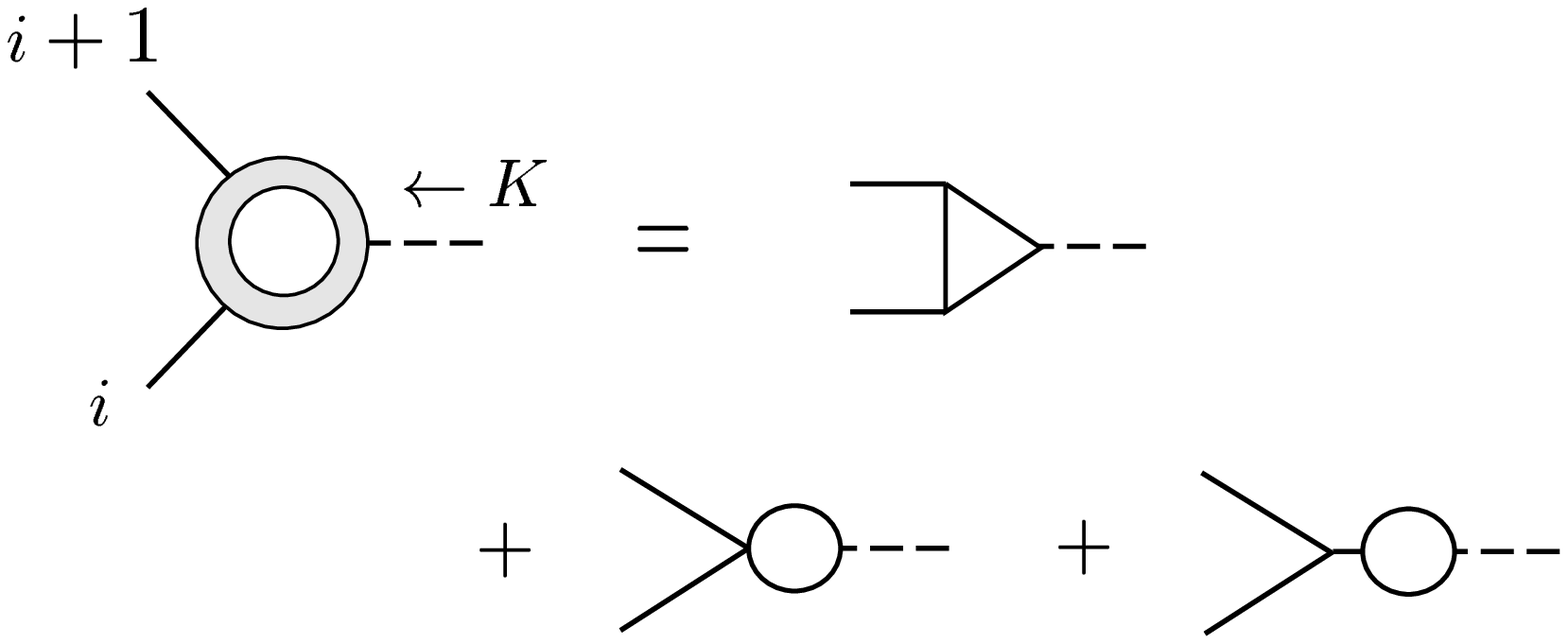}{}

It is convenient to collect all singular terms, including those in
eq.~(\use\FactSubtract), into the non-factorizing category. We achieve
this by subtracting all singularities in $\eps$ from
eq.~(\use\FactSubtract) using the discontinuity functions
(\use\Discontinuities), and then adding them back into the
non-factorizing set discussed below. This collects all non-smooth
behavior into the non-factorizing contributions.  (In general, the
coefficients of the singularities, $B_1$ and $B_2$, depend on gauge
choices and the processes under consideration.)  Subtracting out the
divergences from the factorizing diagrams yields
$$
{\cal D\,}' = B_2 \LB
 {1\over \eps^2}\L{\mu^2 \over - s_{i, i+1}}\R^\eps -
 \discOne(s_{i, i+1}) \RB
+ B_1 \LB {1\over \eps} - \discThree(s_{i,i+1}) \RB + B_0 \,,
\eqn\factorizing
$$
which we label as the `factorizing' contribution to the splitting
functions.  By construction this is completely free of singularities
in $\eps$.  The result ${\cal D\,}'$ may then be contracted against
the $(n-1)$-point tree diagrams, as in the last diagram in
fig.~\use\CollinearFactFigure, and converted to an expression in terms
of spinor helicity.  The conversion is performed by inserting a
complete set of helicity states on the intermediate factorized leg and
then taking the collinear limit; this yields
$$
\sum_{\lambda = \pm} \Split^{\rm fact}_{-\lambda} (i, i+1)
A_{n-1}^{\rm tree}
(\ldots K^\lambda \ldots) \, .
\eqn\factsplit
$$
In this way we obtain the factorizing contribution to the loop splitting
functions. In the next sub-section we present an explicit example.

The second type of contribution is the `non-factorizing' one, which
comes from non-smooth behavior in any of the diagrams.  This includes
the non-smooth contributions which were subtracted from the diagrams
in fig.~\use\MasterSplitLoopFigure\ and any other diagrams which
generate kinematic poles from the loop integrals.  These contributions are
proportional to discontinuity functions (\use\Discontinuities) and a
limited set of integral functions containing poles in $s_{i,i+1}$ to
be discussed in section~\use\ProofSection.

We find that the non-factorizing contributions to the loop splitting
functions are proportional to the tree-level splitting functions, so
we have
$$
 \Split^{\nonfact}_{-\lambda}(a^{\lambda_{a}},b^{\lambda_{b}})
 \ =\ \cg
 \times \Split^{\rm \tree}_{-\lambda}(a^{\lambda_a},b^{\lambda_{b}})
 \times \rS(a,b) \,,
\eqn\genrsdef
$$
where $\rS$ contains no spinor products or helicity dependence, and $a$
and $b$ are legs $i$ and $i+1$.  The contributions to $\rS$ are given
in Table~1 in terms of the singularities of the amplitudes
(\use\singular).  The proportionality of $\Split^{\nonfact}$ to
$\Split^{\rm \tree}$ follows from the appearance of singularities in
all non-factorizing contributions and from the proportionality of the
singularities in loop amplitudes to tree amplitudes.

In Table~1 the coefficients for the singularities $\Soft^{[n]}_{i-1}$,
$\Soft^{[n]}_i$, $\Soft^{[n]}_{i+1}$, and
$\Collinear^{[n]}$ are for the $n$-point amplitude, and the
coefficient $\Collinear^{[n-1]}$ is for the $(n-1)$-point loop
amplitude described by the second set of diagrams on the
right-hand-side of fig.~\use\CollinearFactFigure.  The additive
contribution to $\rS$ is given by the coefficient in the first column
multiplied by the corresponding terms in the third column.  Note that
$\rS$ depends only on the particle types of legs $i$, $i+1$ (and the
fused leg) through the dependence on $\Soft_j^{[n]}$ and
$\Collinear^{[n]} -
\Collinear^{[n-1]}$.

\vskip .5 cm
\hskip -.5 truecm
\def\disp{\displaystyle \vphantom{\L{A^X_X\over B^X_X}\R} }
\hbox{
\def\tend{\cr \noalign{\hrule}}

\vbox{\offinterlineskip
{
\hrule
\halign{
        &\vrule#
        &\quad\hfil\strut#\hfil\quad\vrule
        &\quad\hfil\strut#\hfil\quad\vrule
        & \quad\hfil\strut # \hfil 
        \cr
height13pt  & {\bf Coefficient }& {\bf Singularity }
    &{\bf Non-Factorizing Contribution to $\rS$}  &\tend
height17pt & $\disp \Soft^{[n]}_{i-1}$ & $\disp -{1\over \e^2}
 \Bigl( {\mu^2 \over -s_{i-1, i}} \Bigr)^{\e}$
& $\disp
 {1\over \eps^2} \Bigl({\mu^2 \over  -s_{i, i+1}} \Bigr)^{\eps}
-{1\over \eps^2} \Bigl({\mu^2 \over  -zs_{i, i+1}} \Bigr)^{\eps}
    - \Li_2 (1-z) $ &\tend
height17pt  & $\disp \Soft^{[n]}_{i+1}$ &  $\disp - {1\over \e^2}
 \Bigl( {\mu^2 \over -s_{i+1, i+2}} \Bigr)^{\e}$
& $\disp
 {1\over \eps^2} \Bigl({\mu^2 \over  -s_{i, i+1}} \Bigr)^{\eps}
-{1\over \eps^2} \Bigl({\mu^2 \over -(1-z)s_{i, i+1}} \Bigr)^{\eps}
    - \Li_2(z) $  &\tend
height17pt & $\disp \Soft^{[n]}_i $
     & $\disp - {1\over \e^2} \Bigl( {\mu^2 \over -s_{i, i+1}} \Bigr)^{\e}$
     & $\disp - {1\over \e^2}
             \Bigl( {\mu^2 \over -s_{i, i+1}} \Bigr)^{\e}$ & \tend
height17pt & $\disp \Collinear^{[n]} - \Collinear^{[n-1]} $
     & $\disp  {1\over \e}$
     & $\disp {1\over \e (1-2\eps)}
             \Bigl( {\mu^2 \over -s_{i, i+1}} \Bigr)^{\e}$ & \tend
}
}
}
}
\vskip .2 cm
\nobreak
{\baselineskip 12 pt
\narrower\smallskip\noindent\ninerm
{\ninebf Table 1:}  The `non-factorizing' contributions to the
loop factorization functions in the $s_{i,i+1}$ channel.
The coefficients in the first column are the coefficients
of the contributions to $\rS$ given in the third column.
\smallskip}

The first two entries appearing in the third column are the collinear
limits of discontinuity functions summed with specific box functions
$$
\eqalign{
& \discOne(s_{i,i+1}) + \mu^{2\eps} F_{n:i+2}^{\rm 1 m}
\Bigr|_{k_i\parallel k_{i+1}} \,, \cr
& \discOne(s_{i,i+1}) + \mu^{2\eps} F_{n:i+3}^{\rm 1 m}
\Bigr|_{k_i\parallel k_{i+1}} \,, \cr}
\eqn\FullSplit
$$
where $F_{n:i+2}^{\rm 1 m}$ and $F_{n:i+3}^{\rm 1 m}$ are the
single-external-mass (or off-shell) box functions given through
$\Ord(\eps^0)$ in eqs.~(\use\NotationChangeBoxes{e}) and
(\use\Fboxes{e}).  (The form in eq.~(\use\FullSplit) is valid to
any order in $\eps$.)

The terms of higher order in $\eps$ can become important when performing
phase-space integrals for $n+1$ parton contributions to $n$-jet final
states at next-to-next-to-leading-order (NNLO) when using, for example, the
formalism of ref.~[\use\GG].  The splitting functions may be used to
compute analytically the phase-space integrals in the collinear
regions; in these regions the phase space integrals contain powers of
$\eps^{-1}$ which can cancel against higher order terms in the
splitting functions to leave finite results.  The higher-order in
$\eps$ terms in the splitting functions may be obtained (within the
dimensional reduction [\use\Siegel] or FDH [\use\Long] schemes) by
simply keeping as many orders as desired from both the diagrams in
eq.~(\use\factorizing), and from Table~1, using the form in
eq.~(\use\FullSplit) for the first two entries. The other functions
appearing in the table are given in a form valid to all orders in
$\eps$.  In performing the phase space integral it is convenient not
to convert to a helicity basis, but to leave all expressions in terms
of the formal polarization vectors.  (If one were to convert to a
helicity form one would have to account for the fact that the momentum
over which one performs the phase space integral is actually in
$4-2\eps$ dimensions and not in four dimensions; this type of subtlety
has, however, already been addressed by Mahlon [\use\Mahlon] in the
context of recursion relations for amplitudes.)

In summary, the total contribution to the loop splitting function
(\use\totalsplit) is given by the sum of the factorizing and
non-factorizing contributions.  The factorizing contributions
(\use\factsplit) are independent of the number of external legs since
they always come from the same three-point diagrams depicted in
fig.~\use\MasterSplitLoopFigure.  The non-factorizing contributions
(\use\genrsdef) are also universal because their coefficients are
fixed by the singular terms (\use\singular).  Thus, for any number of
external legs the loop splitting functions appearing in
eq.~(\use\loopsplit) depend only on the `local' properties within the
diagrams: the helicity and particle type of the two collinear legs and
the virtual matter content.  For the color-ordered primitive
amplitudes of ref.~[\use\Fermion] the splitting functions also depend
on the routing of the fermion through the diagrams.

\subsection{Collinear Limit Example}

As an example, consider a one-loop $n$-gluon partial amplitude in a
theory with $n_{\! f}$ massless fermions, $n_s$ massless
complex scalars, and $N_c$
colors in the limit that legs 1 and 2 become
collinear: $k_1 = zK$ and $k_2 = (1-z) K$.  First we compute the
factorizing contributions by evaluating the two- and three-point loop
diagrams of the type in fig.~\use\MasterSplitLoopFigure\ (for $i=1$);
the ones with fermion loops are depicted in
\fig\MasslessSplitLoopFigure. The gluon and scalar loop diagrams are
similar except that one must include diagrams with four-point contact
vertices.  In performing the calculation, the intermediate leg (the
dashed line) with momentum $(-k_1-k_2)$ should be left off-shell in
the initial part of the calculation.  These diagrams are conveniently
evaluated [\use\Mapping] using color-ordered Feynman background field
gauge [\use\Background] for the loop and Gervais-Neveu gauge
[\use\GN] for the tree parts of the diagrams.  Through all orders
in $\eps$ we have
$$
{\cal D}^\mu_{\rm Background} =
{i \over \sqrt{2}}
{\tauG\over 3}
 \Bigl( 1+ {n_s\over N_c} - {n_{\! f}\over N_c} \Bigr)
\Bigl[ \pol_1 \c \pol_2 -
{\pol_1 \c k_2\, \pol_2 \c k_1 \over k_1 \c k_2}
\Bigr](k_1-k_2)^{\mu} \, ,
\eqn\BackgroundResult
$$
where
$$
\tauG \equiv {6\over(4\pi)^{2-\eps}}
 {\Gamma(1+\eps) \Gamma^2(1-\eps) \over \Gamma(4-2\eps)}
\L { \mu^2 \over -s_{12}} \R^\eps
 = {1\over 16 \pi^2} + \Ord(\eps) \, .
\anoneqn
$$
In the computation we have used the dimensional reduction
scheme, while in the conventional [\use\CollinsBook] or 't Hooft-Veltman
[\use\DimensionalRegularization] schemes there would be an extra
overall factor of $(1-\eps)$ in the gluon loop contributions, which
has no effect through $\Ord(\eps^0)$.  The fermions and
scalars are taken to be in the fundamental representation.  (For the adjoint
representation the $n_s$ and $n_f$ terms would not have a factor of
$1/N_c$.)  A background field Ward identity between the two and
three-point function cancels the divergences between the separate
diagrams leaving the finite result in eq.~(\use\BackgroundResult).

One may of course use other gauges, although in general these yield
more complicated expressions.  For example, with color-ordered
Feynman gauge the gluon loop diagrams are
$$
\eqalign{
{\cal D}_{\rm Feynman}^\mu = {i\over \sqrt{2}} &
\biggl\{
\cg \, \discOne(s_{12})
\Bigl[-{3 \over 4} \pol_1\c \pol_2 \, (k_1 -k_2)^\mu
   +  \pol_1^\mu \, \pol_2\cdot k_1
   -  \pol_2^\mu \, \pol_1\cdot k_2  \Bigr] \cr
&
+
{1\over 2}\, \cg\,  \discThree(s_{12})
\Bigl[ - \pol_1\c\pol_2 (k_1 -  k_2)^\mu
              + {1\over 4} \pol_1\c\pol_2 (k_1 +  k_2)^\mu
                 + \pol_2 \cdot k_1 \pol_1^\mu
                 -\pol_1 \cdot k_2 \pol_2^\mu \Bigr] \cr
&
+ { \tauG\over 3}
    \Bigl[ \pol_1\cdot \pol_2 -
           {\pol_1\cdot k_2 \,\pol_2\cdot k_1 \over k_1 \cdot k_2} \Bigr]
     \Bigl( (k_1 - k_2)^\mu + {1\over8} (3 - 2\eps)(k_1 +  k_2)^\mu \Bigr)
\biggr\}
\,, \cr }
\eqn\FeynmanResult
$$
which has been written in a form to expose the discontinuity functions
(\use\Discontinuities) contained within the diagrams.  The result for this
gauge (\use\FeynmanResult) looks rather different than the one for
background field gauge (\use\BackgroundResult) and is a reflection of
the gauge dependence of the diagrams due to the off-shell
intermediate leg.

\vskip -.8 cm
\LoadFigure\MasslessSplitLoopFigure{\baselineskip 13 pt
\noindent\narrower\ninerm The diagrams for the
factorizing massless fermion loop contributions to the
$g \rightarrow gg$ loop splitting functions.}
{\epsfysize 1.1truein}{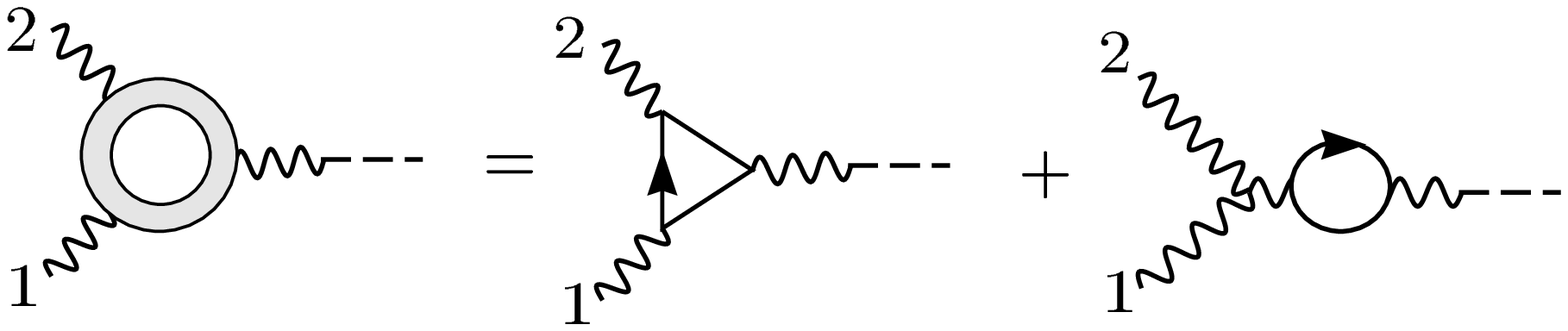}{}

As discussed above, the first step in obtaining the contributions to
the loop splitting function after having calculated the factorizing
diagrams in fig.~\use\MasterSplitLoopFigure\ (or
fig.~\use\MasslessSplitLoopFigure\ for fermions) is to push all
singularities in $\eps$ into the non-factorizing category using the
discontinuity functions (\use\Discontinuities).  After moving the
singularities from the Feynman gauge result (\use\FeynmanResult) into
the non-factorizing contributions, it agrees with the background field
gauge result (\use\BackgroundResult), up to terms proportional to
$(k_1 + k_2)^\mu$.  Since the free index $\mu$ contracts against a
tree-level conserved current, these terms are irrelevant. Thus, in
either gauge, we obtain the `factorizing' contribution as
$$
\hskip -.4 cm
A_n^{\rm fact} \ \mathop{\longrightarrow}^{1\parallel 2}\
 \Bigl( 1+ {n_s\over N_c} - {n_f\over N_c} \Bigr)
{\tauG \over 6} \; \left[ \pol_1 \c \pol_2 -
{\pol_1 \c k_2 \pol_2 \c k_1 \over k_1 \c k_2}
\right] {(k_1-k_2)^{\mu} \over \sqrt{2} k_1\c k_2}
\eta_{\mu\nu}  A^{\tree}_{n-1}(K,3,\ldots,n)^{\nu}\,,
\eqn\FiniteSplit
$$
where
$$
A^{\tree}_{n-1}(K,3,\ldots,n)^{\nu} = {\partial \over \partial \pol_\nu(K)}
A^{\tree}_{n-1}(K,3,\ldots,n) \, .
\anoneqn
$$
and $\eta_{\mu\nu}$ is the Minkowski metric.  Although we obtain the
same result in either gauge, the natural choice for the calculation is
background field gauge, since one directly obtains the physical
part from eq.~(\use\BackgroundResult).  (The form in eq.~(\use\FiniteSplit)
is valid to higher order in $\eps$ and thus would be the appropriate
expression to use for NNLO phase-space integrals.)

The expression (\use\FiniteSplit) may be converted to spinor helicity
notation by inserting a complete set of helicity states on the
intermediate leg [\use\Recursive] using an off-shell generalization of
the helicity formalism. For on-shell momentum $K$ from
eq.~(\use\HelicityDef) we have
$$
\eqalign{
\pol^+_\mu(-K, q)\, \pol^-_\nu(K,q) &  \equiv
{\sand q.{\gamma_\mu \Ksl \gamma_\nu }.q  \over
2 \sand q.\Ksl.q } \, .\cr}
\eqn\OffShellHel
$$
The right-hand-side of this equation is well defined for off-shell
$K$, so we may take it as the definition for a product of off-shell
polarization vectors; we need only define the product of polarizations
since this is the only combination that appears when inserting a
complete set of helicity states.  It is not difficult to verify with
eq.~(\use\OffShellHel) the identity
$$
\eta_{\mu\nu}  =
- \pol^+_\mu(-K, q)\, \pol^-_\nu(K,q) -
                \pol^-_\mu(-K, q)\, \pol^+_\nu(K,q)
+ {K_\mu q_\nu + q_\mu K_\nu \over K\cdot q} \, ,
\eqn\EtaInsert
$$
which then allows us to replace the $\eta_{\mu\nu}$ with
off-shell polarization vectors plus a piece which will vanish by
current conservation.  For $K=k_1 + k_2$ we have
$$
(k_2 - k_1)\cdot \pol^+(-K, q)\, \pol^-_\nu(K,q)
=  {\spa{q}.{k_2} \spb{k_2}.{k_1} \sand{k_1}.{\gamma_\nu}.{q}
  - \spa{q}.{k_1} \spb{k_1}.{k_2} \sand{k_2}.{\gamma_\nu}.{q}
    \over 2 \sand q.\Ksl.q }  \,.
\anoneqn
$$
We parametrize the collinear limit by $k_1
\rightarrow z K$ and $k_2 \rightarrow (1-z)K$ with $K^2 \rightarrow 0$,
to obtain
$$
(k_2-k_1) \c \pol^+(-K,q)\, \pol^-_\nu(K,q)
   \rightarrow -\sqrt{2z(1-z)} \spb{k_1}.{k_2}
   \pol^-_\nu(K,q) \, ,
\anoneqn
$$
where the polarization vector on the right-hand side is on-shell.
Thus after inserting a complete set of helicity states using
eq.~(\use\EtaInsert) (through $\Ord(\eps^0)$) we obtain from
eq.~(\use\FiniteSplit)
$$
A_n^{\rm fact} \ \mathop{\longrightarrow}^{1\parallel 2}\ \Bigl(1+{n_s\over
N_c}- {n_f\over N_c}\Bigr)
\sum_{\lambda=\pm} \Split^{\rm fact}_{-\lambda}(1,2)
A^{\tree}_{n-1} (K^{\lambda},3,\ldots,n) \, ,
\eqn\SplitFact
$$
where
$$
\eqalign{
& \Split^{\rm fact}_{\pm}(1^{+},2^{-})=
\Split^{\rm fact}_{\pm}(1^{-},2^{+})=0\; ,  \cr
& \Split^{\rm fact}_{+}(1^{+},2^{+})= -
{1\over 48\pi^2}
\sqrt{z(1-z)} {[12] \over \langle 12 \rangle^2} \; , \cr
& \Split^{\rm fact}_{-}(1^{+},2^{+})=
{1\over 48\pi^2} \sqrt{z(1-z)} {1\over \langle 12 \rangle}  \,,  \cr}
\anoneqn
$$
are the factorizing contributions to the loop splitting functions.

To obtain the non-factorizing contribution to the splitting function
we note that for $n$ external gluons $\Soft_i^{[n]} = 1$ for all $i$ and
$\Collinear^{[n]} = \Collinear^{[n-1]}$, so that from
eq.~(\use\singular) the singular terms in $\eps$ are
$$
A_n^{\rm loop} \Bigr|_{\rm singular} =
- \cg A^\tree_n  \biggl[{1\over\eps^2} \sum_{j = 1}^n
    \L{\mu^2 \over - s_{j, j+1} } \R^\eps
+ {2\over \eps}
\Bigl({11\over 6} - {1\over 3} {n_{\! f} \over N_c}
- {1\over 6} {n_s \over N_c}  \Bigr) \biggr] \,.
\anoneqn
$$
{}From Table~1, the non-factorizing contribution to $\rS$ for this
example is the sum of the first three entries in the last column, given by
$$
\rS(1,2) = {1\over \eps^2} \Bigl({\mu^2 \over  -s_{12}} \Bigr)^{\eps}
-{1\over \eps^2} \Bigl({\mu^2 \over  -zs_{1 2}} \Bigr)^{\eps}
-{1\over \eps^2} \Bigl({\mu^2 \over -(1-z)s_{1 2}} \Bigr)^{\eps}
      - \Li_2 (1-z)  - \Li_2(z) + \Ord(\eps)\,.
\anoneqn
$$
After simplifying $\rS$ using the dilogarithm identity
$$
\Li_2 (1-z) + \Li_2(z) =  -\ln(z)\ln(1-z) + {\pi^2 \over 6} \,,
\anoneqn
$$
we obtain from eq.~(\use\genrsdef) the non-factorizing contributions
to the $g \rightarrow gg$ splitting functions,
$$
\eqalign{
\Split_{-\lambda}^{\nonfact}(1, 2)
& =  \cg \Split_{-\lambda}^\tree(1,2) \biggl[
- {1 \over \eps^2} \L {\mu^2 \over z(1-z) (-s_{12})}\R^\e + 2 \ln (z) \ln(1-z)
  - {\pi^2 \over 6} \biggr]  \,. \cr}
\eqn\NonFactSplit
$$

Combining the results (\use\SplitFact) and (\use\NonFactSplit) yields
the total contribution
$$
\Split^{\rm loop}(1,2) = \Bigl(1 + {n_s \over N_c} - {n_{\! f}\over N_c} \Bigr)
\Split^\fact(1,2) + \Split^\nonfact(1,2) \, ,
\anoneqn
$$
in agreement with the results in ref.~[\use\SusyFour] obtained
from taking the explicit collinear limits of five-gluon
amplitudes~[\use\FiveGluon].

Finally by performing the ultra-violet subtraction using (\use\ultraviolet)
we obtain the corresponding
splitting function for renormalized amplitudes as
$$
\eqalign{
\Split_{\rm ren}^{\rm loop} (1,2) & =
\Bigl(1 + {n_s \over N_c} - {n_{\! f}\over N_c} \Bigr)
\Split^\fact(1,2) + \Split^\nonfact(1,2) \cr
& \hskip 2 cm
- \cg \, {1\over \eps} \Bigl({11\over 6} - {1\over 3} {n_{\! f} \over N_c}
-{1\over 6} {n_s \over N_c} \Bigr) \Split^{\rm tree}(1,2) \, , \cr}
\eqn\MasslessSplitRen
$$
where all remaining powers of $\eps^{-1}$ are infrared singularities.

The loop splitting functions for amplitudes with external fermions and
gluons may similarly be obtained; these have already been extracted
from four- and five-parton calculations and tabulated in
refs.~[\use\SusyFour,\use\Fermion].

\subsection{Collinear Factorization with Massive Fermions
            and Scalars}

For amplitudes with no infrared singularities the splitting functions
are entirely determined by the naively factorizing contributions.
Consider, for example, the collinear limit of gluon amplitudes in a
theory with massive fermion or scalar loops. Since the mass cuts off
any singular behavior as legs go on-shell, all loop integrals are
smooth in the collinear limit and the amplitude naively factorizes;
the collinear poles only come from external gluon propagators. Thus
massive fermion contributions to the splitting functions may be
directly determined by calculating only the diagrams in
\fig\MassiveSplitLoopFigure.  (For a scalar loop there are additional diagrams
containing the four-point interaction.)

Following the usual renormalization procedure, half of each bubble on
the external line is associated with external wavefunction
renormalization.  Note that the set of diagrams that contribute to the
massive loops is different from that of massless loops (given in
fig.~\use\MasslessSplitLoopFigure).  The difference between the two
sets is due to the dimensional regularization prescription that
massless bubbles on external legs vanish; in
fig.~\use\MasslessSplitLoopFigure\ the two diagrams with bubbles on
external lines vanish and are therefore not included. For the massive
case in fig.~\use\MassiveSplitLoopFigure\ half of the bubble on the
`internal' line on which the factorization is being performed belongs
with the $(n-1)$-point amplitude and not with the loop splitting
function.

\vskip -.8 cm
\LoadFigure\MassiveSplitLoopFigure{\baselineskip 13 pt
\noindent\narrower\ninerm The diagrams representing
contributions of massive fermions to the $g\rightarrow gg$
loop splitting function; the wavy lines represent gluons, and the
dashed lines represent the fact that only half of each bubble is included.}
{\epsfysize 1.2truein}{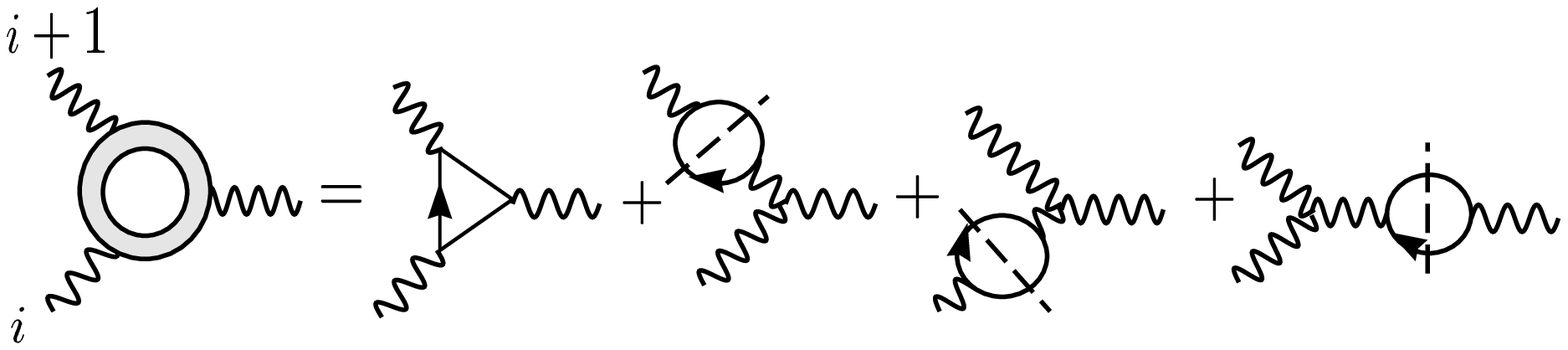}{}

The calculation of the diagrams in fig.~\use\MassiveSplitLoopFigure\
is straightforward and gives the entire contribution to the loop
splitting functions.  This gives the contribution of massive fermion
and scalar loops to the $g \rightarrow gg$ splitting functions as%
\footnote{$^*$}{As in
refs.~[\use\FiveGluon,\use\Fermion], each scalar here contains four
states (to match the four states of Dirac fermions)
so that $n_s$ must be divided by two for comparisons to
conventional normalizations of scalars.}
$$
\Split^{\rm loop}_{-\lambda} (a, b) =
-{1\over 3} {1 \over (4\pi)^{2 -\eps}} \Gamma(\eps)
 \biggl[{n_{\! f} \over N_c} \Bigl( {\mu^2 \over m_{\! f}^2} \Bigr)^\eps
+ {1\over 2} {n_s \over N_c} \Bigl( {\mu^2 \over m_s^2} \Bigr)^\eps \biggr]
\Split^{\rm tree}_{-\lambda} (a, b) \, ,
\eqn\MassiveSplit
$$
where $n_{\! f}$ and $n_s$ are the number of fermions and scalars
and $m_{\! f}$ and $m_s$ the corresponding masses.

The proportionality of these contributions to the tree splitting
functions may be understood from the structure of the effective
action (for $k_1\c k_2 \ll m^2$).
The contribution to the effective Lagrangian from a massive
fermion or scalar is
$$
{\cal L}_{\rm eff} = -{Z_A\over 4}  F^a_{\alpha\beta}
F^{a\alpha\beta} + \Ord\Bigl({1\over m^2} \Bigr) \, ,
\anoneqn
$$
where $Z_A$ is the wavefunction renormalization.  The higher order
terms are suppressed in the collinear limit because they contain
$1/m^2$ instead of $1/k_1\c k_2$, leaving only a renormalization
of the tree splitting functions.  In the massless case this
argument breaks down and there is no reason to expect the loop
splitting functions to be proportional to the tree ones.

Subtracting off ultra-violet singularities (\use\ultraviolet) from
eq.~(\use\MassiveSplit) gives
the splitting functions from massive fermions or scalars
for renormalized gluon amplitudes as
$$
\Split^{\rm renorm}_{-\lambda} (a, b) =
-{1\over 3} {1 \over (4\pi)^2}
 \biggr[ {n_{\! f} \over N_c} \ln\Bigl( {\mu^2 \over m_{\! f}^2} \Bigr)
+ {1\over 2} {n_s \over N_c} \ln\Bigl( {\mu^2 \over m_s^2 } \Bigr)\biggr]
\Split^{\rm tree}_{-\lambda} (a, b)  + \Ord(\eps) \, .
\anoneqn
$$
The gluon loop contributions in this theory may be obtained from
eq.~(\use\MasslessSplitRen) by taking the number of massless scalar
and fermions to vanish, $n_{\! f} = n_s = 0$.

\vskip -.4 truecm
\section{Multi-particle Factorization}
\tagsection\MultiSection

Consider now multi-particle factorization which we will show has
analogous behavior to the two-particle collinear limits.
As in the latter case,
this non-smooth behavior may be linked to the infrared singularities
appearing in massless amplitudes.

\subsection{General Considerations}

We will prove that the factorization properties
for $(k_i + k_{i+1} + \cdots + k_{i+r-1})^2 = t_i^{[r]}
\equiv K^2 \rightarrow 0$ (with $r>2$) are described by the universal
formula,
$$
\eqalign{
A_{n}^{\rm loop}\
\mathop{\longrightarrow}^{K^2 \rightarrow 0}\
\sum_{\lambda=\pm}  \biggl[
&
      A_{r+1}^{\rm loop}(k_i, \cdots k_{i+r-1}, K^\lambda) \, {1 \over K^2} \,
      A_{n-r+1}^{\rm tree}(K^{-\lambda}, k_{i+r}, \cdots, k_{i-1})
\cr
+
& \null
     A_{r+1}^{\rm tree}(k_i, \cdots k_{i+r-1}, K^\lambda) \, {1\over K^2} \,
      A_{n-r+1}^{\rm loop}(K^{-\lambda}, k_{i+r}, \cdots, k_{i-1})
\cr
+
& \null
     A_{r+1}^{\rm tree}(k_i, \cdots k_{i+r-1}, K^\lambda) \, {1\over K^2} \,
      A_{n-r+1}^{\rm tree}(K^{-\lambda}, k_{i+r}, \cdots, k_{i-1}) \,
      \cg\,  \Fact_n(K^2;k_1, \ldots, k_n) \biggr] \,,
\cr}
\eqn\loopfact
$$
where the one-loop {\it factorization function} $\Fact_n$ is
independent of helicities.  This formula is similar to the one for an
amplitude which factorizes naively, as depicted in
\fig\MultiFactFigure, except that $\Fact_n$ may contain kinematic
invariants with momenta from both sides of the pole in $K^2$; for
example $\ln(-t_{i-1}^{[2]}) = \ln(-s_{i-1, i})$ is one such
logarithm. (In fig.~\use\MultiFactFigure\ we have made the bubble on
the intermediate leg explicit; in fig.~\use\CollinearFactFigure, for
collinear limits, the bubble diagram was implicitly included in the
last diagram on the right-hand-side, as depicted in
fig.~\use\MasterSplitLoopFigure.)
 As for the splitting functions the factorization
function is composed of factorizing and non-factorizing components,
$$
\Fact_n = \Fact_n^{\rm fact} + \Fact_n^{\rm non-fact}\,.
\eqn\FactAndNonFact
$$
For convenience we have extracted an overall factor
of $\cg$ from the factorization function.

\vskip -1. cm
\LoadFigure\MultiFactFigure
{\baselineskip 13 pt
\noindent\narrower\ninerm The diagrams with an explicit pole
in $K^2 = t_i^{[r]}$ coming from a tree propagator.}
{\epsfysize 1.5truein}{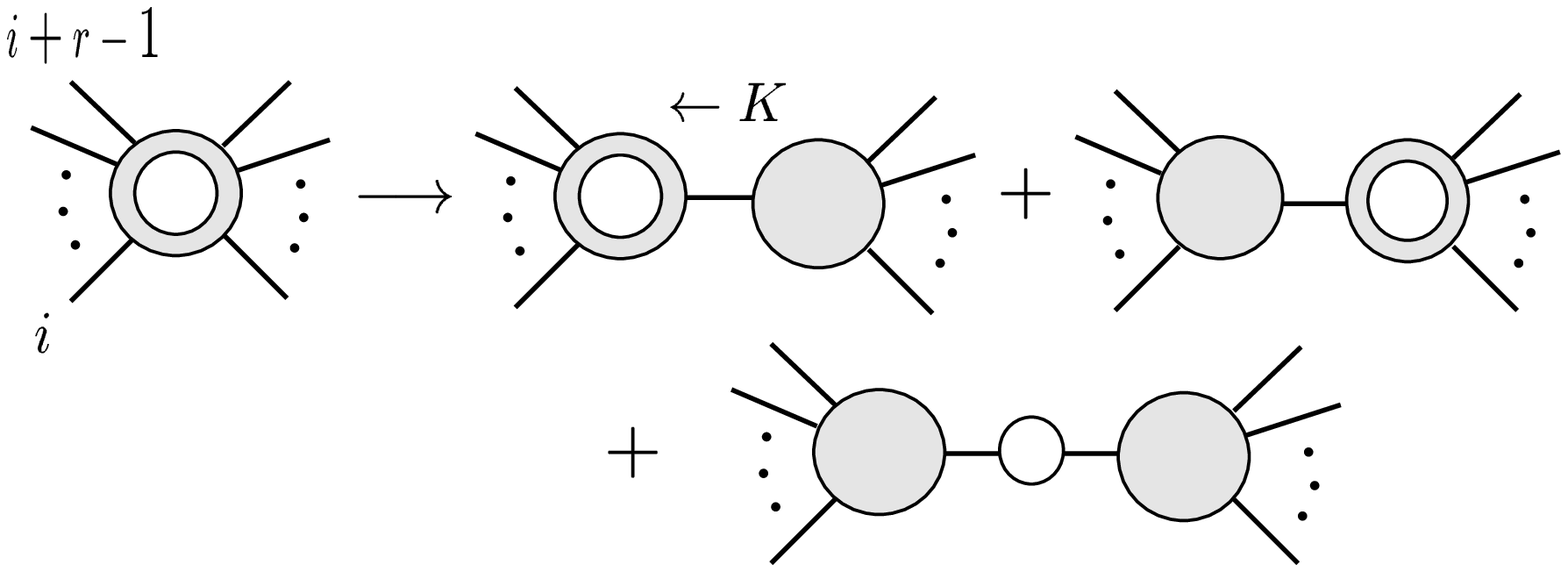}{}

One of the results that we shall prove is that only a few integral
functions may contribute to a pole in any multi-particle channel.  Furthermore,
the singular terms tightly constrain the coefficients of the allowed
integral functions.  The factorizing contributions in a multi-particle
channel are composed only of integral functions depicted in
\fig\FactFigure; all external momenta from one of the two sides of the
multi-particle pole are part of one leg of these integral functions.
If $t_i^{[r]}$ constitutes a kinematic invariant of an entire leg of an
integral function there will be non-factorization from the discontinuous
limit.  There are also two integral functions which may enter into
non-factorization: the box functions
$I_{4:r-1;i+1}$ and $I_{4:n-r-1;i+r+1}$ in
\fig\NonFactBoxesFigure.  (The values of these box functions are
given in appendix~\use\IntegralsAppendix.)  These two box functions
have the property that they contain explicit poles in $K^2$ yet they
would not appear in any of the naively factorized diagrams in
fig.~\MultiFactFigure.

\vskip -1. cm
\LoadFigure\FactFigure
{\baselineskip 13 pt
\noindent\narrower\ninerm The integral functions which contribute
to diagrams with a tree pole in the  $K^2 = t_i^{[r]}$ channel.}
{\epsfysize 1.1truein}{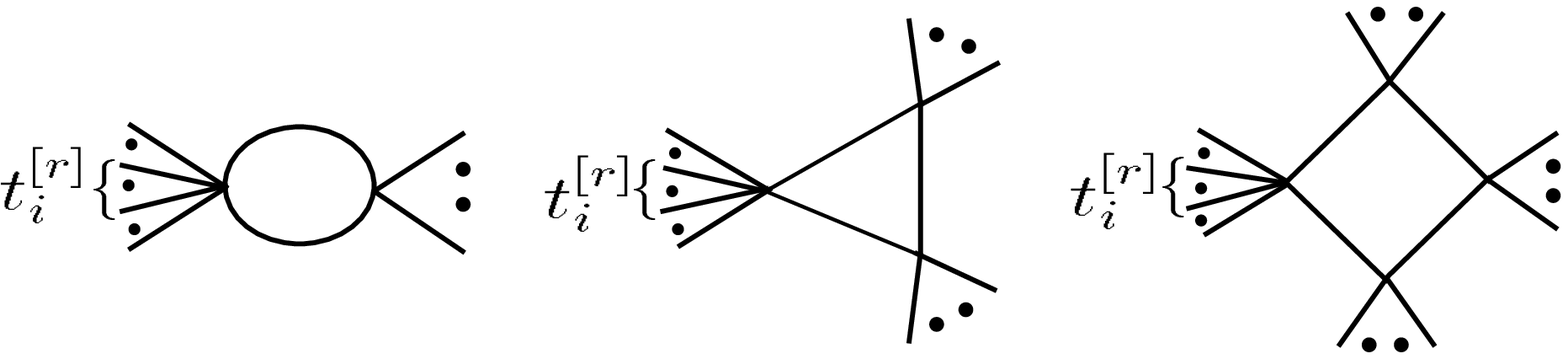}{}

\vskip -1. cm

\LoadFigure\NonFactBoxesFigure
{\baselineskip 13 pt
\noindent\narrower\ninerm The two boxes functions which contribute
to non-factorization in the channel $t_i^{[r]}$, indicated by the dashed
line.}
{\epsfysize 1.3truein}{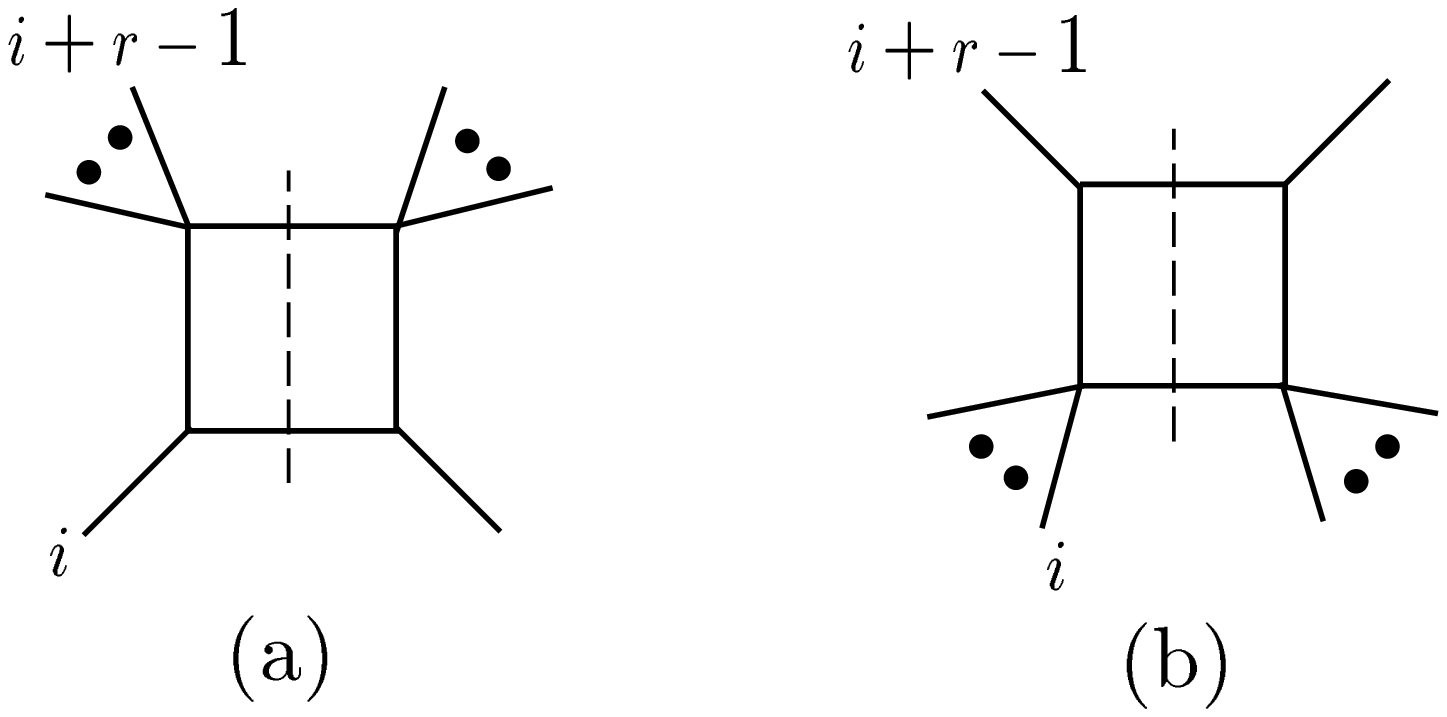}{}

The procedure for obtaining the factorization function is similar
to the one for obtaining the loop splitting functions.  First
compute the factorizing contribution given by the bubble loop
in the third class of diagrams on the right-hand-side of figure~5.
This contribution is of the form
$$
C_1 {1\over \eps} + C_0 \, .
\eqn\FactBubble
$$
In this case there is no $\eps^{-2}$ singularity since bubble diagrams
have at worst $\eps^{-1}$ ultraviolet
singularities (in a Feynman-like gauge).
Note that the bubble loop is gauge dependent.  As was the case
for two-particle
factorization we must subtract an appropriate discontinuity function
to obtain
$$
C_1 \LB {1\over \eps} - \discThree(t_i^{[r]}) \RB + C_0  \,,
\eqn\FactBubbleFinal
$$
which is finite.  After inserting a complete set of helicity states we
obtain the factorizing contribution $\Fact_n^{\rm fact}$ from
eq.~(\use\FactBubbleFinal).  (In some cases, such as factorizing on a
fermion line, this contribution vanishes.)

The second type of contribution to the factorization function,
$\Fact_n^{\rm non-fact}$, are the
non-factorizing ones given in Table~2.  The coefficients of the
contributing terms in the third column are given in the first column.  As
for two-particle collinear limits the non-factorizing terms are fixed
by the singular terms (\use\singular).  The box
functions $F^{{\rm 2m} \, h}$ appearing in the third column of Table~2
are defined in eqs.~(\use\Fboxes) and (\use\NotationChangeBoxes) and
correspond to the kinematic configurations depicted in
fig.~\use\NonFactBoxesFigure.

\vskip .5 cm
\hskip -1.5 truecm
\def\disp{\displaystyle \vphantom{\L{A^X_X\over B^X_X}\R} }
\hbox{
\def\tend{\cr \noalign{\hrule}}

\vbox{\offinterlineskip
{
\hrule
\halign{
        &\vrule#
        &\quad\hfil\strut#\hfil\quad\vrule
        &\quad\hfil\strut#\hfil\quad\vrule
        & \quad\hfil\strut # \hfil 
        \cr
height13pt & {\bf Coefficient} &{\bf Singularity}
    &{\bf Factorization Function Contribution}  &\tend
height17pt & $\disp \Soft^{[n]}_{i-1}$ & $\disp -{1\over \e^2}
 \Bigl( {\mu^2 \over -s_{i-1, i}} \Bigr)^{\e}$
& $\disp 2 (\mu^2)^\eps \Bigl( \Fhard{r-1;i+1}
      + {1 \over \eps^2} (-t_i^{[r]})^{-\eps} \Bigr)$
 & \tend
height17pt  &$\disp \Soft^{[n]}_{i+r-1}$ &$\disp -{1\over \e^2}
 \Bigl( {\mu^2 \over -s_{i+r-1, i+r}} \Bigr)^{\e}$
& $\disp 2(\mu^2)^\eps \Bigl( \Fhard{n-r-1;i+r+1}
      + {1 \over \eps^2} (-t_i^{[r]})^{-\eps} \Bigr)$
&\tend
height17pt & $\disp \Collinear^{[n]} - \Collinear^{[n-1]}$
& $\disp {1\over \e}$
& $ \disp  (\mu^2)^\eps {1 \over \eps (1-2 \eps)}(-t_i^{[r]})^{-\eps} $
 & \tend
}
}
}
}
\vskip .2 cm
\nobreak
{\baselineskip 13 pt
\narrower\smallskip\noindent\ninerm
{\ninebf Table 2:}  The `non-factorizing' contributions to the
factorization functions in the channel $t_i^{[r]} \rightarrow 0$.
The three coefficients in the first column are the coefficients
of the contributions to the factorization function given in the third
column.
\smallskip}

The total contribution to the factorization function $\Fact_n$ is
the sum of terms in Table~3 along with ones in
eq.~(\use\FactBubbleFinal).

\subsection{Multi-Particle Factorization Example}

We now present an example to illustrate the use of multi-particle
factorization to provide checks on explicitly calculated results.
This is analogous to the types of checks that have been
performed using collinear limits both for tree [\use\ParkeTaylor,
\use\TreeCollinear] and loop [\use\AllPlus,\use\SusyFour,\use\Fermion]
amplitudes.  Multi-particle factorization can also be helpful for
constructing ans\"atze for higher-point amplitudes.

Consider the one-loop $N=4$ supersymmetric six-gluon amplitude
$A_{6;1}^{N=4}(1^+,2^+,3^+,4^-,5^-,6^-)$, which has been computed in
ref.~[\use\SusyOne]. In the $t_1^{[3]} =(k_1+k_2+k_3)^2$ channel, the
right-hand-side of eq.~(\use\loopfact) vanishes since $A^{\rm tree}
(1^+, 2^+, 3^+, K^\pm) = 0$.  Thus the amplitude does not contain a
pole in $t_1^{[3]}$.  The $t_2^{[3]}$ channel, however, will contain a
pole, since tree and loop amplitudes appearing on the right-hand-side
of eq.~(\use\loopfact) do not vanish.  Since we are dealing with $N=4$
super-Yang-Mills, only box functions may enter [\use\SusyFour] in the
integral reduction. Therefore, the only integral functions which may contribute
to the coefficient of the $t_2^{[3]}$ pole are given by the four box
functions $ F_{6:2}^{1{\rm m}}\, , F_{6:5}^{1{\rm m}} \,,
F_{6:2;3}^{2{\rm m}\, h}\,, F_{6:2;6}^{2{\rm m}\, h}\,, $ which are
depicted in \fig\BoxesSixFigure\ and defined in eqs.~(\use\Fboxes) and
(\use\NotationChangeBoxes).  For the first two integral functions the
pole comes from a tree propagator, while for the second two integral
functions the loop integral itself generates the pole. (The $F$
functions have the kinematic poles scaled out, but the original
integrals defined in eq.~(\use\NPointLoopIntegral) and related to the
$F$ in eq.~(\use\RescaledBoxes) contain them.)

\vskip -.8 cm
\LoadFigure\BoxesSixFigure
{\baselineskip 13 pt
\noindent\narrower\ninerm The four box functions which may appear as
coefficients of a $t_2^{[3]}$ pole in a six-point amplitude; the dashed lines
represent the $t_2^{[3]}$ channel.}
{\epsfysize 2.6truein}{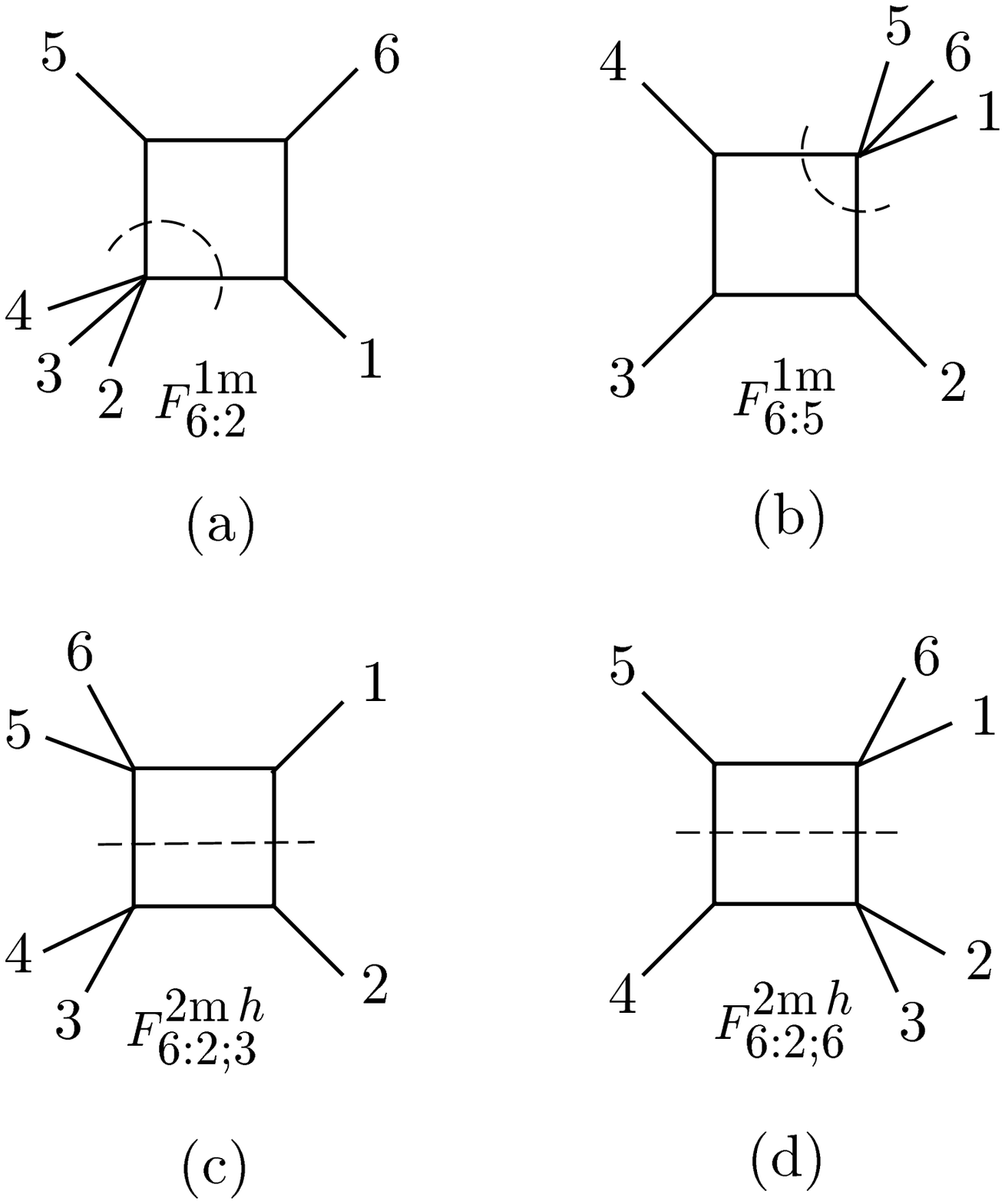}{}

For the $N=4$ supermultiplet with one gluon, four Weyl fermions
and three complex scalars (with two states each) all ultraviolet and
collinear singularities cancel leaving only the soft ones,
$$
\eqalign{
A_{6;1}^{N=4}(1^+,2^+,3^+,4^-,5^-,6^-) \Bigr|_{\rm singular}
    &  = - \cg A_{6}^{\rm tree}(1^+,2^+,3^+,4^-,5^-,6^-)
    {1\over\e^2} \sum_{j=1}^6
         \left( { \mu^2 \over -s_{j,j+1} } \right)^\e \, . \cr }
\eqn\SusyFourSingular
$$
The tree amplitude in this equation contains a pole in the
$t_2^{[3]}$ channel given by
$$
A_{6}^{\rm tree}(1^+,2^+,3^+,4^-,5^-,6^-) =
A_4^{\rm tree} (2^+, 3^+, 4^-, K^-)
 \, {1\over t_2^{[3]}} \,  A_4^{\rm tree}(K^+, 5^-, 6^-, 1^+)
+ \hbox{non-pole} \, .
\anoneqn
$$
The four box functions which appear as coefficients of the $t_2^{[3]}$
pole in the amplitude $A_{6;1}^{N=4}$ have infrared singular behavior,
$$
\eqalign{
F_{6:2}^{1{\rm m}} \Bigr|_{\rm singular} & =  -{1\over \eps^2}
\Bigl[ (-s_{61})^{-\eps} + (-s_{56})^{-\eps}
- (-t_2^{[3]})^{-\eps} \Bigr]\,, \cr
F_{6:5}^{1{\rm m}} \Bigr|_{\rm singular} & =  -{1\over \eps^2}
\Bigl[ (-s_{34})^{-\eps} + (-s_{23})^{-\eps}
- (-t_2^{[3]})^{-\eps} \Bigr]\,, \cr
F_{6:2;3}^{2{\rm m}h} \Bigr|_{\rm singular} & =  -{1\over 2\eps^2}
\Bigl[(-s_{12})^{-\eps} + 2(-t_2^{[3]})^{-\eps} - (-s_{56})^{-\eps}
 - (-s_{34})^{-\eps} \Bigr]\,,\cr
F_{6:2;6}^{2{\rm m}h} \Bigr|_{\rm singular} & =  -{1\over 2\eps^2}
\Bigl[(-s_{45})^{-\eps} + 2(-t_2^{[3]})^{-\eps} - (-s_{23})^{-\eps}
 - (-s_{61})^{-\eps} \Bigr] \, ,\cr
}
\eqn\BoxSingular
$$
and we have redundant constraints for determining their coefficients from
eq.~(\use\SusyFourSingular). This fixes the relative
coefficients of the four box functions to be unity so that
$$
\eqalign{
A_{6;1}^{N=4}(1^+,2^+,3^+,4^-,5^-,6^-)  & =
\cg \mu^{2\e} A_4^{\rm tree} (2^+, 3^+, 4^-, K^-)
 \, {1\over t_2^{[3]}} \,  A_4^{\rm tree}(K^+, 5^-, 6^-, 1^+) \cr
& \hskip 2 cm \times
2 \Bigl( F_{6:2}^{1{\rm m}} + F_{6:5}^{1{\rm m}} + F_{6:2;3}^{2{\rm m}\,h} +
F_{6:2;6}^{2{\rm m}\, h} \Bigr) + \hbox{non-pole} \,, \cr}
\eqn\MultiFactExampleA
$$
where $K = k_2 + k_3 + k_4$ and the tree amplitudes with the other
helicity configuration of the intermediate line vanish.  Thus,
in the full amplitude, the $t_2^{[3]}$ pole must multiply the four box
functions in fig.~\use\BoxesSixFigure\ with a relative coefficient of
unity.

{}From Table~2 we may also read off the factorization function in the
$t_2^{[3]}$ channel to be
$$
\eqalign{
\Fact_6^{N=4} & (t_2^{[3]}; k_1, \ldots, k_6) \cr
&= 2 (\mu^2)^\eps \Bigl(
F_{6:2;3}^{2{\rm m} h} + F_{6:2;6}^{2{\rm m}h} + {2 \over \eps^2}
  (-t_2^{[3]})^{-\e} \Bigr) \cr
&=\
2 (\mu^2)^\eps \biggl[
 -{1\over\e^2} \Bigl[ (- s_{12})^{-\e}
              - (-s_{56})^{-\e} - (-s_{34})^{-\e} \Bigr]
   - {1\over2\e^2}
    { (-s_{56})^{-\e}(-s_{34})^{-\e}
     \over (- s_{12})^{-\e} }
   + {1\over 2} \ln^2\left({s_{12}\over t_2^{[3]} }\right)\cr
& \hskip .5 cm
   - {1\over 2} \ln^2 \left( {s_{34} \over t_2^{[3]} }\right)
   - {1\over 2} \ln^2 \left( {s_{56} \over t_2^{[3]} }\right)
   - {\pi^2 \over 3}\biggr]
+ \{k_2 \leftrightarrow k_5, k_3 \leftrightarrow k_6,
          k_4 \leftrightarrow k_1 \} \, .
\cr}
\anoneqn
$$
Note that the integral functions do not undergo any particularly
large simplification as $t_2^{[3]} \rightarrow 0$, in contrast to the
simplification appearing in the splitting functions.

Finally, after explicitly taking the limit $t_2^{[3]} \rightarrow 0$
we may rewrite eq.~(\use\MultiFactExampleA) as
$$
\eqalign{
A_{6;1}^{N=4}(1^+,2^+,3^+,4^-,5^-,6^-)  & \longrightarrow
\cg \mu^{2\e} A_4^{\rm tree} (2^+, 3^+, 4^-, K^-)
 \, {1\over t_2^{[3]}} \,  A_4^{\rm tree}(K^+, 5^-, 6^-, 1^+)  \cr
& \hskip .5 cm \times
2 \Bigl( F^{\rm 0m}(k_2, k_3, k_4, K) + F^{\rm 0m}(K, k_5, k_6, k_1)
         + \Fact_6^{N=4} \Bigr) \,, \cr}
\eqn\MultiLimitForm
$$
where the $F^{\rm 0m}$ zero-external-mass box function is given in
eq.~(\use\Fboxes{f}).  The three terms in eq.~(\use\MultiLimitForm)
correspond to the ones in
eq.~(\use\loopfact), after re-expressing the functions in terms of the
four-point [\use\GSB,\use\SusyFour] loop amplitudes.

We may obtain the pole in the $t_3^{[3]}$ channel by symmetry
under reflection of the external states, so that the same results hold but
with relabeled indices
$$
A_{6;1}^{N=4}(1^+,2^+,3^+,4^-,5^-,6^-) =
\cg \mu^{2\e} A_4^{\rm tree} {1\over t_3^{[3]}} A_4^{\rm tree} \times
2 \Bigl( F_{6:3}^{1{\rm m}} + F_{6:6}^{1{\rm m}} + F_{6:2;1}^{2{\rm m}\,h} +
F_{6:2;4}^{2{\rm m}\, h} \Bigr) + \hbox{non-pole} \,.
\eqn\MultiFactExampleB
$$

Now compare the results in eqs.~(\use\MultiFactExampleA) and
(\use\MultiFactExampleB) against the explicitly computed result of
ref.~[\use\SusyOne]
$$
A_{6;1}^{N=4}(1^+,2^+,3^+,4^-,5^-,6^-)\ =\
 \cg\ \LB B_1\,\Wsix1+B_2\,\Wsix2+B_3\,\Wsix3\RB,
\eqn\pppmmmloop
$$
where
$$
\eqalign{
  B_1 &=
    i \, {(\spb1.2\spa2.4+\spb1.3\spa3.4)
   \, (\spb3.1\spa1.6+\spb3.2\spa2.6) \, (t_1^{[3]})^3
  \over \spa1.2\spa2.3\spb4.5\spb5.6\
  (t_1^{[3]}t_3^{[3]}-s_{12}s_{45}) \,
    (t_1^{[3]}t_2^{[3]}-s_{23}s_{56}) } \, ,\cr
  B_2\ &=\ \left({ \langle1^+| (\ksl_2+\ksl_3) |4^+\rangle
         \over t_2^{[3]} } \right)^4  \ B_1 \vert_{j\to j+1}
       + \left({ \spb2.3\spa5.6 \over t_2^{[3]} } \right)^4
            \ B_1^\dagger \vert_{j\to j+1}\ , \cr
  B_3\ &=\ \left({ \langle3^+| (\ksl_1+\ksl_2) |6^+\rangle
         \over t_3^{[3]} } \right)^4  \ B_1 \vert_{j\to j-1}
       + \left({ \spb1.2\spa4.5 \over t_3^{[3]} } \right)^4
            \ B_1^\dagger \vert_{j\to j-1}\, ,  \cr}
\eqn\pppmmmdef
$$
and
$$
\eqalign{
  \Wsix{i}\ &\equiv\ F^{1 \rm m}_{6:i} + F^{1\rm m}_{6:i+3}
                 + F^{2 {\rm m}\, h}_{6:2;i+1}
                 + F^{2 {\rm m}\, h}_{6:2;i+4} \, . \cr}
\eqn\Wdef
$$
The $B_i^\dagger$ means complex conjugating spinor products in $B_i$,
$\spa{k}.{j}\leftrightarrow\spb{j}.k$ without complex conjugating
factors of $i$, and the subscript $j\to j+1$ means applying a
cyclic permutation of the six momenta $k_i$,
$\{1,2,3,4,5,6\}\rightarrow \{2,3,4,5,6,1\}$.  The $B_i$ satisfy the
condition
$$
B_1 + B_2 + B_3 = 2 A_6^{\rm tree}(1^+, 2^+, 3^+, 4^-, 5^-, 6^-) \,.
\anoneqn
$$
The amplitude (\use\pppmmmloop) clearly satisfies the multi-particle
pole conditions described above: it does not contain a pole in the
$t_1^{[3]}$ channel and the coefficients of the $t_2^{[3]}$ and
$t_3^{[3]}$ poles match the ones in eqs.~(\use\MultiFactExampleA) and
(\use\MultiFactExampleB).  Thus the amplitude (\use\pppmmmdef) is
consistent with the constraints imposed by multi-particle
factorization, providing a further stringent check on its correctness.

The same type of analysis may be performed for larger numbers of
external legs.  However, the singularities in $\eps$ will not uniquely
specify the coefficients of the boxes since the number of boxes
proliferates.  Nevertheless, all functions entering into the
factorization function $\Fact_n$ are uniquely specified from the
singularities in $\eps$.

\vskip -.3 truecm
\section{Proof of Universal Factorization}
\tagsection\ProofSection

In this section we prove the factorization properties of amplitudes
described in the two previous sections.  Our proof is based on
identifying all potential poles in a generic amplitude that
arise either from tree propagators or within loop momentum integrals. We
use the integral reduction procedure
[\use\Reduction,\use\PV,\use\OtherInt,\use\VNV,\use\IntegralsShort,%
\use\IntegralsLong], reviewed (and slightly modified)
in appendix~\use\ReductionAppendix, to make the poles from loop
integrals explicit.  One might expect
that all non-factorizing contributions should be related to the infrared
divergences; our method of proof makes this connection explicit.

To summarize appendix~\use\ReductionAppendix, any one-loop Feynman
diagram may be expressed as a linear combination of integral functions
with four or fewer legs multiplied by rational `reduction
coefficients'.  In the standard integral reduction procedure one
obtains coefficients of integral functions which can contain
dependence on $\eps$.  Since we are linking all non-factorizing
contributions to infrared singularities, any $\eps$-dependence would
have to be tracked because it could induce finite shifts in the
splitting and factorization functions when multiplying divergent
integrals.  Rather than dealing with $\eps$-dependent reduction
coefficients, we find it simpler to keep all $\eps$-dependence in the
basis of integral functions. Thus, in the appendix we modify the usual
reduction procedure in order to eliminate $\eps$-dependence in the
reduction coefficients; through $\Ord(\eps^0)$
this leads to the basis of integral functions:
(a) $D=4-2\eps$ scalar bubbles, triangles and boxes, (b) $D=6-2\eps$
scalar bubbles and triangles, and (c) $D=8-2\eps$ scalar boxes.  In
the dimensional reduction or FDH schemes, where the number of states
is always fixed at their four-dimensional value, there are no other
sources of $\eps$-dependence.  Thus, all amplitudes may be expressed
as linear combinations of these integral functions with rational
coefficients containing no $\eps$-dependence.

If the reduction coefficients contain poles or if the integral
functions contain discontinuities or poles, the amplitudes may not
factorize in a simple way. After having performed the integral
reduction procedure on all Feynman diagrams contributing to an
amplitude, poles in kinematic variables arise from three sources:%
\footnote{$^\dagger$}
{We are assuming that the spinor helicity reference momenta are chosen
so as to not introduce any additional poles in the channel of
interest.}

\vskip .2 cm

\item{1)} diagrams which have a pole coming from an explicit tree propagator,

\item{2)}  poles in the integral reduction coefficients, and

\item{3)} explicit poles appearing within the box and triangle
integral functions in the basis.

\vskip .2 cm

\noindent
The naively factorizing contributions are of the first type but
all three can lead to non-factorization.  We now systematically
collect the non-factorizing pieces to the loop splitting
function associated with each of the three types of kinematic poles.
Subsequently, we will show that they are
uniquely fixed by the singularities in $\eps$ via
eq.~(\use\singular); this is because no non-zero linear combination can be
constructed which is free of $\eps^{-1}$.

\subsection{Kinematic Poles from Tree Propagators}

Consider the Feynman diagrams containing a tree propagator which has a
kinematic pole in the variable $t_i^{[r]}$, as depicted in
fig.~\use\MultiFactFigure. In the limit that the kinematic variable
vanishes one expects the amplitude to factorize on this pole since the
diagrams break into a product of lower-point tree and loop amplitudes.
However, if additional infrared divergences develop in the loop
integrals as the kinematic invariant vanishes there will be
discontinuities in the integral functions.  In the off-shell case
($t_i^{[r]} \not = 0$) there may be, for example, terms such as
$\ln(-t_i^{[r]})$ which correspond to poles in $\eps$ in the on-shell
case ($t_i^{[r]} = 0$); in this case the off-shell to on-shell
transition is discontinuous, and there is a `discontinuity function'
modifying the tree pole.

To find the non-factorizing contributions in the loop integrals of the
diagrams in fig.~\use\MultiFactFigure (including the case of collinear
limits), we categorize all discontinuities in the integral functions
(coming from the reduction) as the kinematic invariant $K^2$ vanishes.
(We denote momenta which are sums of external momenta $k_i$ by an
upper case letter; the lower case $k_i$ we reserve for on-shell
external momenta.)  Out of all possible integral functions appearing
after the reduction, discussed in appendix~\use\ReductionAppendix,
only those where one leg has momentum $K$ have a discontinuous
off-shell ($K^2 \not = 0$) to on-shell ($K^2 = 0$) transition.  We now
step through the possible discontinuities coming from
these integrals, which are collected in
appendix~\use\IntegralsAppendix .

Consider first discontinuity functions arising from the $D=4-2\eps$ bubble
function.  Since bubbles are a function of a single
momentum, the only ones which can have a non-smooth limit as
$K^2\rightarrow 0$ are those with momentum $K$ flowing through them.
Bubble integrals with on-shell (massless) legs vanish by a standard
dimensional regularization prescription [\use\MutaBook], so the
off-shell bubble (\use\BubbleInt) is itself a discontinuity function. Thus
we have the  discontinuity function given by
$$
\eqalign{
b(K^2) = {1\over r_\Gamma} I_2[1](K^2) & = {1 \over \eps (1-2\eps)}
(-K^2)^{-\eps} \,, \cr}
\eqn\BubbleDisc
$$
where $\rg$ is the ratio of Gamma-functions in eq.~(\use\RGdef).

Consider now the discontinuity functions arising from the $D=4-2\eps$
triangle
integrals, depicted in \fig\TriDiscExFigure.   Starting from the
three-external-mass integral, depicted in fig.~\use\TriDiscExFigure{a}
and explicitly defined in eqs.~(\use\RescaledTri{a}) and (\use\Triangles{a}),
and taking $K^2_3 \rightarrow 0$ one obtains
$$
T_3^{3\rm m}(K^2_1, K^2_2, K^2_3) \longrightarrow
T_3^{2\rm m}(K^2_1, K^2_2) - \discTwo(K^2_3; K^2_1, K^2_2)
+ \discTwo(K^2_3; K^2_2, K^2_1) \, ,
\eqn\DiscTwoDef
$$
where $T_3^{2\rm m}$ is the two-external-mass triangle depicted in
fig.~\use\TriDiscExFigure{b} and explicitly given in
eqs.~(\use\RescaledTri{b}) and (\use\Triangles{b}). Equation~
(\use\DiscTwoDef) defines the discontinuity function
$\discTwo$ to any finite order in $\eps$.
Through $\Ord(\eps^0)$, it is
$$
\discTwo(K^2_3; K^2_1, K^2_2) \equiv
{1\over 2\eps^2} (-K^2_3)^{-\eps} - {1\over 2\eps^2}
{(-K^2_3)^{-\eps} (-K^2_1)^{-\eps} \over (-K^2_2)^{-\eps}}
-\Li_2\Bigl(1- {K^2_1\over K^2_2} \Bigr) \, ,
\eqn\DiscTwo
$$
and satisfies the property
$$
\discTwo(K^2_3; K^2_1, K^2_2) =
- \discTwo(K^2_3; K^2_2, K^2_1) \, ,
\eqn\DiscProperty
$$
as may be checked using the dilogarithm identity
$$
\Li_2(1-x) + \Li_2(1- x^{-1}) = - {1\over 2} \ln^2(x) \, .
\anoneqn
$$

\vskip -1. cm
\LoadFigure\TriDiscExFigure
{\baselineskip 13 pt
\noindent\narrower\ninerm The triangle functions generate two types
of discontinuity function as one of the external legs becomes
on-shell (or massless).}
{\epsfysize 1.2 truein}{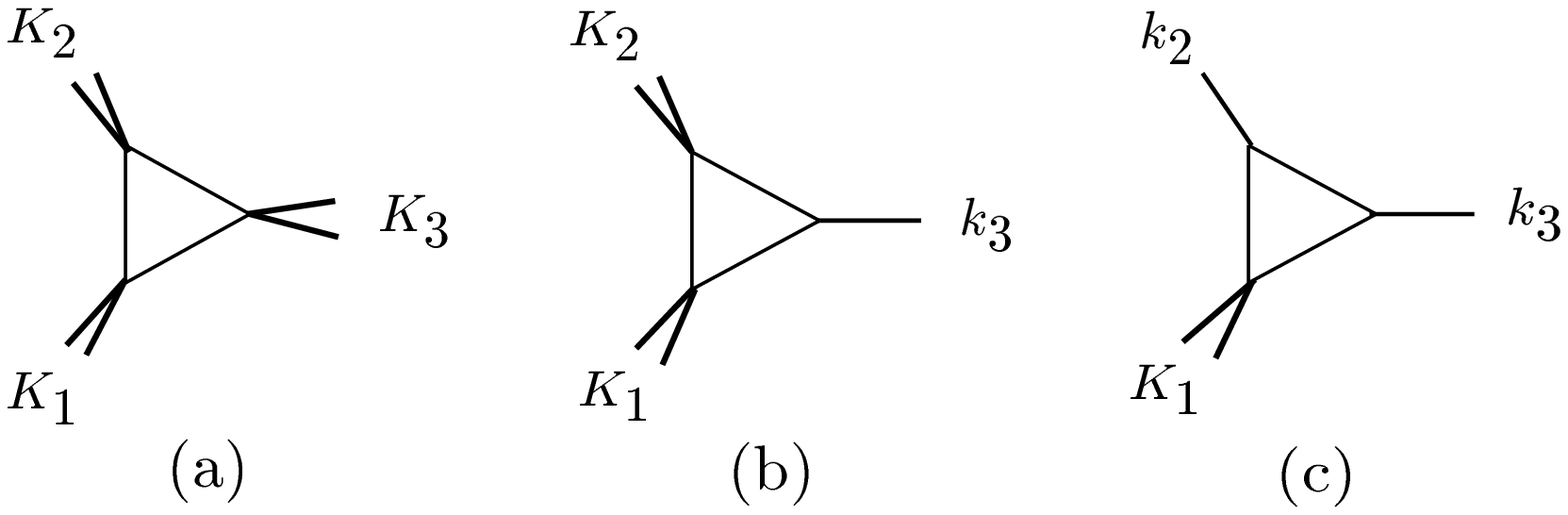}{}

Next, by taking $K_2^2 \rightarrow 0$ the two-external-mass scalar
triangle in fig.~\use\TriDiscExFigure{b} becomes
$$
\eqalign{
T_{3}^{2 \rm m}(K^2_1,K^2_2) & \longrightarrow
\Bigl[ {1 \over \eps^2} (-K^2_1)^{-\eps} - {1\over\eps^2} (-K^2_2)^{-\eps}
\Bigr] \cr
& = T_{3}^{1 \rm m}(K_1^2) - \discOne(K^2_2) \, , \cr }
\anoneqn
$$
where the discontinuity function is
$$
\discOne(K^2) \equiv {1\over \eps^2} (-K^2)^{-\eps} \,.
\eqn\DiscOne
$$
Finally, by taking $K_1^2 \rightarrow 0$ for the one-external-mass
triangle depicted in fig.~\use\TriDiscExFigure{c} we obtain
$$
T_3^{1 \rm m} (K_1^2) =  \discOne(K_1^2) \, .
\anoneqn
$$
Note that the zero-external-mass triangle vanishes.

Now consider discontinuities in the $D=4-2\eps$ box.  These are
readily obtainable from the general relationship between box and
triangle functions [\use\IntegralsShort]
$$
 I_4[1]\ =\ {1\over2} \Biggl[
     \sum_{i=1}^4 c_i\ I_{3}^{(i)}[1]
    \ +\ (-1+2\eps)\, c_0\ I_4^{D=6-2\eps}[1] \Biggr]\ ,
\eqn\FourToThree
$$
where
$$
c_i = \sum_{j=1}^4 S^{-1}_{ij} \, , \hskip 2 cm
c_0 = \sum_{i=1}^4 c_i\ =\ \sum_{i,j=1}^4 S^{-1}_{ij} \,,
\hskip 2 cm   S_{ij} = -{1\over 2} (p_{i-1} - p_{j-1})^2 \, .
\eqn\cdef
$$
Here $p_i = \sum_{j=1}^{i} K_j$, where the $K_j$ are the external momenta
of the box functions and $p_0 = 0$.

The six-dimensional box function $I_4^{D=6-2\eps}[1]$ in
eq.~(\use\FourToThree) (expanded to a
finite order around $\eps =0$) has a smooth transition between
external massive and massless kinematics since it does not contain
infrared divergences and the loop integral converges uniformly.  Thus
from eq.~(\FourToThree), the only discontinuity functions that may
appear in box functions are linear combinations of ones appearing for
triangle functions (\use\DiscOne) and (\use\DiscTwo) (and explicitly
listed in Table~5 in section~6) and there is no need for a separate
analysis.

Finally, consider the higher-dimensional integrals. These higher
dimension boxes, triangles and bubbles asise from the reduction of
tensor integrals, following the procedure discussed in
appendix~\use\ReductionAppendix. (If one were keeping terms of
$\Ord(\eps)$ and beyond, then the basis is further enlarged to include
higher-dimensional integrals with $n\ge 5$ legs.)  In general, these
integrals are smooth as an external leg makes an off-
to on-shell transition since no infrared divergences develop.
Therefore they will not contribute a discontinuity function. There
are, however, two exceptions which need a closer examination:
$$
\eqalign{
I_2^{D=6-2\eps}[1](K^2)&= -{\rg \over 2\eps(1-2\eps)(3-2\eps)}
(-K^2)^{1-\eps} \, , \cr
I_3^{{\rm 1m}, D= 6-2\eps}[1](K^2) & =
                   {\rg\over 2\eps (1-\eps) (1-2\eps)}\,
(-K^2)^{-\eps} \, .\cr}
\eqn\BubbleTrouble
$$
These integral functions are proportional to $I_2$ in
eq.~(\use\BubbleDisc), but with $\eps$-dependent coefficients.  The
possibility of the integrals (\use\BubbleTrouble) contributing to
non-factorization would be problematic since their difference is
simply a constant, which would not be linked to a singularity in
$\eps$.  In appendix~\use\HigherDimAppendix\ we show that the integrals
in eq.~(\use\BubbleTrouble) do not contribute for $K^2 \rightarrow 0$,
leaving only $D=4-2\eps$ integrals as source of discontinuity
functions; there are no additional discontinuity functions from any of
the higher-dimensional integrals.

Converting to the $n$-point kinematic variables in the $K^2 =
t_i^{[r]} \rightarrow 0$ channel, the discontinuity functions which
may enter from the loop diagrams in fig.~\use\MultiFactFigure\ are thus
$$
\eqalign{
&\discOne(t_i^{[r]})  = {1\over \eps^2} (-t_i^{[r]})^{-\eps} \,, \cr
&\discTwo(t_i^{[r]}; t_{i+r}^{[r']}, t_i^{[r+r']})
  = {1\over 2\eps^2} (-t_i^{[r]})^{-\eps}
-{1\over 2\eps^2} { (-t_i^{[r]})^{-\eps}
 (-t_{i+r}^{[r']})^{-\eps} \over (-t_i^{[r+r']})^{-\eps} }
-\Li_2 \Bigl(1- {t_{i+r}^{[r']}\over t_i^{[r+r']}} \Bigr) \,, \cr
&\discThree(t_i^{[r]}) = {1\over \eps (1- 2\eps)} (-t_i^{[r]})^{-\eps} \,
,\cr }
\eqn\Discont
$$
where $ 2\le r'\le n - r -2$.  These discontinuity functions describe
the additional contributions one would find from diagrams containing
a tree propagator pole in $t_i^{[r]}$.
For example, as $K^2 = t_i^{[r]} \rightarrow 0$, the
first set of diagrams on the right-hand-side
of fig.~\use\MultiFactFigure\ behave as
$$
\sum_{\lambda=\pm}  A^{\rm loop}_{r+1} (\ldots, K^\lambda)\, {1\over K^2} \,
A^{\rm tree}_{n-r+1} (K^{-\lambda}, \ldots) +
{1\over K^2} \sum \hbox{discontinuities} \,,
\anoneqn
$$
where the discontinuities are the ones in eq.~(\use\Discont)
multiplying rational coefficients (determined in
section~\use\ProofSection{.4}) where the pole in $K^2$
comes from the intermediate tree propagator.  That is, the diagrams
factorize as one might naively expect, except that there are possible
additional discontinuity contributions multiplying the explicit
tree pole.

The origin of the discontinuities in eq.~(\use\Discont) may be traced
back to the implicit expansion of amplitudes in $\eps$; if one were to
keep the full functional behavior in $\eps$ there would be no
discontinuity for $\eps <0$ since they can be analytically continued
to zero.  Note that the discontinuity functions contain non-singular
(in $\eps$) contributions, including the dilogarithm; these finite
contributions are linked to the infrared singular ones because they
are part of a single function.

\subsection{Kinematic Poles from Integral Reduction Coefficients}

As discussed in appendix~\use\PoleAppendix\ poles may appear in the
reduction coefficients as two external momenta become collinear (or as
an external momentum becomes soft) and not in multi-particle channels
for general kinematics (except for the tensor bubble
(\use\BubbleIntegral) which generates a multi-particle pole that is
canceled by the $D=6-2\eps$ bubble).  For convenience we choose the
$s_{12}$ channel.  Only loop diagrams of the type in
\fig\IntPoleFigure\ may generate potential $s_{12}$ poles
in the reduction coefficients.  (For gluon amplitudes with scalar
loops these types of diagrams were explicitly analyzed in
refs.~[\use\AllPlus,\use\GordonConf].)  We will present a general
argument and analysis that all such poles are proportional to the
discontinuity functions (\use\Discont).

\vskip -1. cm
\LoadFigure\IntPoleFigure
{\baselineskip 13 pt
\noindent\narrower\ninerm The class of loop integral which may generate
a pole in $s_{12}$.}
{\epsfysize .9 truein}{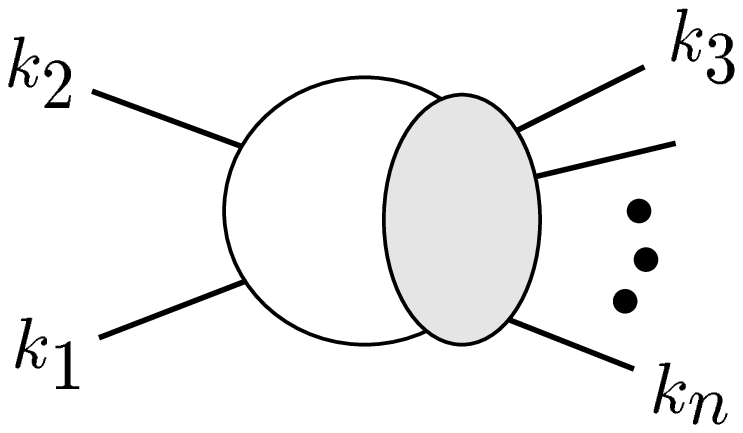}{}

For the case of loop integrals which have a uniform non-vanishing mass
in the loop (see eq.~\use\tensorloopint), it is a simple matter to
argue directly from the loop momentum representation that there are no
massless poles.  The mass acts as an infrared cutoff which prevents
the integral from diverging as a kinematic variable vanishes.
Alternatively, in the integral reduction framework, the
reduction coefficients explicitly given in
appendix~\use\ReductionAppendix\ do appear to contain the
poles collected in appendix~\use\PoleAppendix; for the case of a
uniform mass in the loop these poles are necessarily spurious (i.e.,
the residue vanishes) since massive loop integrals cannot contain massless
poles.  (The spurious nature of these poles
was first noted by Brown [\use\Reduction].)  The specific details of
how the residues of poles in the reduction coefficients all vanish is
in general quite intricate and may involve Abel's [\use\Lewin]
dilogarithm identity; nevertheless they must vanish.

As shown in appendix~\use\PoleAppendix, the reduction
coefficients for the case of a uniform mass in the loop contain
exactly the same apparent poles as for the massless case.  In the
massive loop case the residue of all massless poles cancel so one might
expect the residues to also vanish for a massless loop.   Indeed
by taking the massive loop integral
residues (which vanish) and taking $m\rightarrow 0$
one obviously still obtains zero since the limit is smooth; all potential
non-smoothness would be of the form $\ln(m)$, but such terms (as well as
all others in the residue) vanish.

We must, however, be a bit more careful for the massless case
since we must take the limits in the reverse ordering: first we set $m
=0$ and then we extract the residue of the kinematic pole.  After
setting the internal mass to zero, the extraction of the residue by
taking the appropriate kinematic variable to vanish may no longer be
smooth since new infrared divergences may develop.  The new
singularities can prevent the necessary identities for the vanishing
of the residues in the massive case from holding.  Thus we need to
collect the new singularities and all associated discontinuities from
the integral functions.  This situation is completely identical to the
case where the pole arises from a tree propagator.  The analysis is
completely identical and the non-smooth behavior of any integral
function when a kinematic variable vanishes is described by the
discontinuity functions (\use\Discont).  Thus any (non-spurious) poles
in the reduction coefficients must give a contribution to splitting or
factorization functions proportional to discontinuity functions.

We have performed a number of explicit checks to verify the above
general argument.  Consider for example the
scalar pentagon integrals with the kinematic configuration depicted in
\fig\PentagonPoleFigure, where momenta $k_1$ and $k_2$ are on-shell and the
remaining $K_i$ are either on-shell or off-shell.  As discussed in
appendix~\use\PoleAppendix , for this kinematic configuration the
reduction coefficients contain poles in $s_{12}$.

\vskip .9 cm
\LoadFigure\PentagonPoleFigure
{\baselineskip 13 pt
\noindent\narrower\ninerm
The original $n$-point integral may be reduced into a sum of
pentagons; the type of pentagon depicted here is the only one which
can generate poles in $s_{12}$.}  {\epsfysize 1.1
truein}{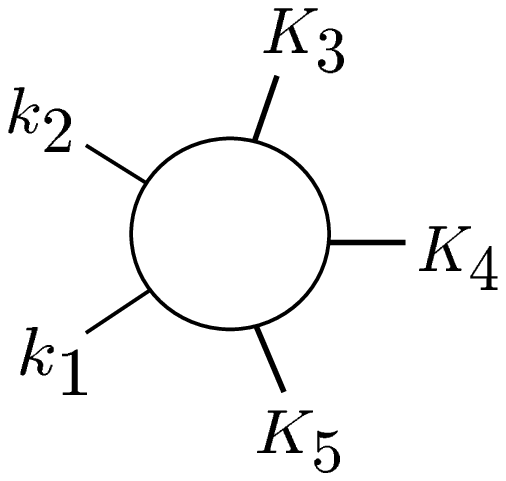}{}

{}From the discussion in appendix~\use\ReductionAppendix,
the pentagon integrals are
$$
I_5 = \half \mathop{\sum}_{i=1}^5 c_i I_{4}^{(i)} +
 \eps c_0 \, I_5^{D=6-2\eps} \,,
\eqn\pentagonreduce
$$
where
$$
c_i = \mathop{\sum}_{j=1}^5 S_{ij}^{-1} \, , \hskip 2 cm
c_0 = \sum_{i=1}^5 c_i \, , \hskip 2 cm
S_{ij} = - \half (p_{i-1} - p_{j-1})^2 \,.
\anoneqn
$$
For on-shell $k_1$ and $k_2$ it is straightforward to check that as
$s_{12} \rightarrow 0$
$$
S_{ij}^{-1} \longrightarrow  (-1)^{i+j+1} {1\over s_{12}}
{\sqrt{\det S^{(i)}} \; \sqrt{\det S^{(j)}} \over
       \sqrt{\det S^{(1)}}\;  \sqrt{\det S^{(3)} }} + \Ord(1)\, ,
\anoneqn
$$
where
$$
\eqalign{
&
4\sqrt{\det S^{(1)}} = s_{23} s_{34} - s_{51} K_3^2 \, , \hskip 2 cm
4\sqrt{\det S^{(2)}} = s_{34} s_{45}  - K_3^2 K_5^2\, ,\cr
&
4\sqrt{\det S^{(3)}} = s_{45} s_{51} - s_{23} K_5^2 \, , \hskip 2 cm
4\sqrt{\det S^{(4)}} = s_{51} s_{12} \, , \cr
& \hskip 4  cm
4\sqrt{\det S^{(5)}} = s_{12} s_{23} \,. \cr}
\anoneqn
$$
The $S^{(i)}_{mn}$ are the matrices for the daughter box diagram
obtained by removing the internal propagator between leg $(i-1)$ and
leg $i$.  (This follows the same labeling convention as for the box
diagrams themselves.)  Since $\sqrt{\det S^{(4)}}$ and $\sqrt{\det
S^{(5)}}$ are both proportional to $s_{12}$, only the three
coefficients $c_1, c_2$ and $c_3$ contain a pole in $s_{12}$ and are
relevant.

In all cases, the coefficients $c_4$ and $c_5$ do
not contain any singularities.  Rather, the integrals $I^{(4)}_4$
and $I^{(5)}_4$ themselves contain a divergence.  Since we are
only investigating the singular behavior of the reduction
coefficients, we only consider the partial sum over
$c_1$, $c_2$, and $c_3$
$$
\eqalign{
& c_1 I^{(1)}_4 + c_2 I^{(2)}_4 + c_3 I^{(3)}_4
 = - {r_\Gamma \over 2}
{1\over s_{12}} {1\over  \sqrt{\det S^{(1)}}\;  \sqrt{\det S^{(3)} }}
\Bigl( \sum_{j=1}^5 (-1)^j \sqrt{\det S^{(j)}} \Bigr)
\Bigl( F_4^{(1)} - F_4^{(2)} + F_4^{(3)} \Bigr) \,, \cr}
\anoneqn
$$
where we used eq.~(\use\GeneralFdefn) to scale out the overall
kinematic denominator from each integral function.

As explained above, for any configuration of $K_3$, $K_4$ and $K_5$,
being massive ($K_i^2\not=0$) or massless ($K_i^2 = 0$) we have
$$
\eqalign{
\mathop{\rm lim}_{k_1\parallel k_2} \;
(F_4^{(1)} - F_4^{(2)} + F_4^{(3)})
= \alpha \,\discOne(s_{12}) + \beta\, \discTwo(s_{12}; K_3^2, s_{45}})
+ \gamma\, \discTwo(s_{12}; K_5^2, s_{34})  \,,
\eqn\DiscIdentity
$$
where the $\alpha, \beta$ and $\gamma$ are constants.
The combination of box functions
$F_4^{(1)}$,  $F_4^{(2)}$, and $F_4^{(3)}$ is displayed in
\fig\IdentityFigure.  We have
explicitly verified through $\Ord(\eps^0)$ that this holds for all
possible kinematic configurations.  For each situation
the real numbers $\alpha$, $\beta$, and $\gamma$ were found by
explicitly performing the limit for all types of box integrals
(i.e. for the different cases of whether or not $K_3^2 = 0$, $K_4^2 =
0$, or $K_5^2 = 0$).  In all cases the discontinuity functions come from
$F_4^{(2)}$, which is the only one of the three box
functions which acquires new infrared singularities as $s_{12}
\rightarrow 0$.  (See table~5 in section~6 for a list of all box
discontinuities.) The identity (\use\DiscIdentity) is quite
non-trivial when actually performing the limit on the integrals in
appendix~\use\IntegralsAppendix\ since many dilogarithm identities
must be used in the process.  Nevertheless, in a theory with a mass
such identities necessarily exist to cancel the pole in the reduction
coefficients; in the massless limit these identities break down in the
infrared, leaving behind discontinuity functions containing
singularities in $\eps$.

\vskip -.8 cm
\LoadFigure\IdentityFigure
{\baselineskip 13 pt
\noindent\narrower\ninerm An example of an non-trivial identity
in the collinear limit $k_1 \parallel k_2$ is that this sum of box
functions reduces to discontinuity functions; $k_1$ and $k_2$ are
on-shell while $K_3$, $K_4$ and $K_5$ are either on- or off-shell.}
{\epsfysize 1.3truein}{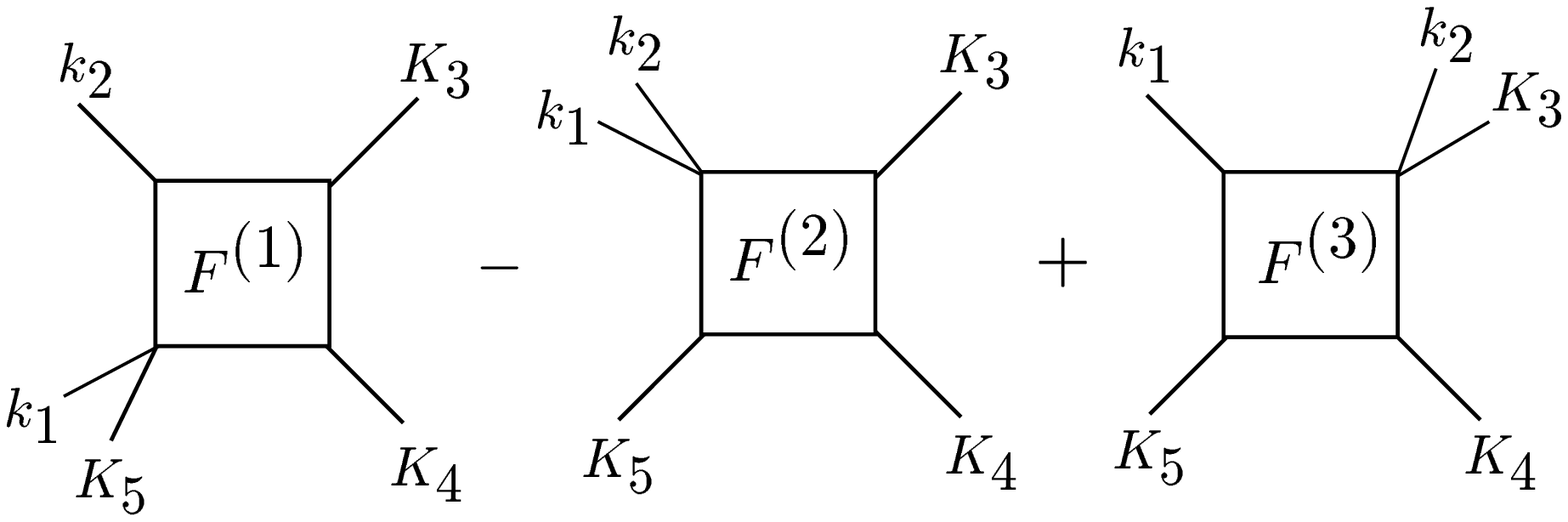}{}

As a second check, the general five-point integral with one power of
loop momenta is [\use\IntegralsShort,\use\IntegralsLong]
$$
\eqalign{
I_{5}[l^{\mu}] & =
\sum_{i=i}^4 p_i^{\mu} \sum_{j=1}^{5} S_{i+1,j}^{-1} I^{(j)}_4
 \mathop{\longrightarrow}^{k_1 \parallel k_2}\
 {\rg\over 2} \sum_{i=2}^5 p_i^{\mu}
(-1)^{i+1} {\sqrt{\det S^{(i+1)}}  \over
       s_{12} \sqrt{\det S^{(1)}}\;  \sqrt{\det S^{(3)} }}
     \sum_{j=1}^{3} (-1)^{j}
           F^{(j)}_4 \, , \cr}
\eqn\onepower
$$
so once again identity (\use\DiscIdentity) guarantees that any pole in
$s_{12}$ is proportional to a linear combination of discontinuity
functions. One can continue in this way to show that any poles in
$s_{12}$ are proportional to linear combinations of discontinuity
functions even for higher powers of loop momentum in the numerator.
We have also verified that the poles in the reduction coefficients of
scalar hexagons are also proportional to discontinuity functions.

The explicit reduction results above are complete for an $N=4$
supersymmetric theory since the only integrals encountered
[\use\SusyFour] that can produce a collinear pole in reduction coefficients
(using the reduction basis discussed in appendix~\use\PoleAppendix) are
the ones discussed above.  A proof of the lack of collinear poles in
integrals for the case of external gluons with scalar and fermion
loops has previously been constructed by a direct diagrammatic
analysis [\use\AllPlus,\use\GordonConf].  Since gluon loops may be
interpreted in terms of a linear combination of scalar, fermion and
$N=4$ loops [\use\FiveGluon,\use\Tasi], this provides a complete check
of our general argument for $n$-gluon amplitudes.

In summary, all contributions with a kinematic pole in the reduction
coefficients are proportional to
$$
\sum \; \hbox{discontinuities} \, ,
\anoneqn
$$
where the discontinuities are again the ones in eq.~(\use\Discont).
The general argument we have presented above holds for any type of
massless pole. Besides collinear poles, the reduction coefficients may
contain poles as momenta become soft or ones not found in amplitudes,
such as in non-adjacent kinematic variables (e.g. $s_{13}$); in view
of the above general argument all such poles must either be spurious
or proportional to discontinuity functions.

\subsection{Kinematic Poles from Integral Functions}

The third source of kinematic poles are in the explicit forms of the
integral functions contained in the reduction basis.  In
eqs.~(\use\RescaledTri) and (\use\RescaledBoxes) of
appendix~\use\IntegralsAppendix\ overall dimensions have been
explicitly pulled out from the $D=4-2\eps$ scalar integral functions;
the only ones that have kinematic poles in two- or multi-particle
channel are $I_3^{1 {\rm m}}$, $I_4^{1 {\rm m}}$ and $I_4^{2 {\rm m}
h}$.  In a two-particle channel $s_{i, i+1}$ the integral functions
$$
\eqalign{
& I_{4:i+2}^{1 \rm m} \equiv
- {2 \rg \over s_{i, i+1} s_{i-1, i}} F^{1 \rm m}_{n:i+2} \, ,
\hskip 2 cm
 I_{4:i+3}^{1 \rm m} \equiv
- {2 \rg \over s_{i, i+1} s_{i+1, i+2}} F_{n:i+3}^{1 \rm m} \, , \cr
& I_{4:r;i+2}^{2 {\rm m} h} \equiv
- {2 \rg \over s_{i, i+1} t_{i+1}^{[r+1]}} F_{n:r;i+2}^{2{\rm m} h} \, ,
\hskip 1.6 cm
 I_{3:i}^{1 \rm m} \equiv  - {\rg \over s_{i, i+1}} T_{n:i}^{1 \rm m}
= - {\rg \over s_{i,i+1}} \discOne(s_{i, i+1})\, , \cr}
\eqn\TwoPoleFunc
$$
contain poles, which are depicted in \fig\TwoPoleIntegralsFigure .
These integrals arise from the reduction of the type of diagrams in
fig.~\use\IntPoleFigure.
All other scalar integrals in the basis do not diverge as $s_{i, i+1}
\rightarrow 0$.  We also note that the triangle function $T_{n:i}^{1
\rm m}$ is equal to the discontinuity function $\discOne(s_{i,i+1})$
defined in eq.~(\use\DiscOne).

\vskip -.8 cm
\LoadFigure\TwoPoleIntegralsFigure
{\baselineskip 13 pt
\noindent\narrower\ninerm The integral functions which have a pole in
$s_{i, i+1}$.}
{\epsfysize 2.6truein}{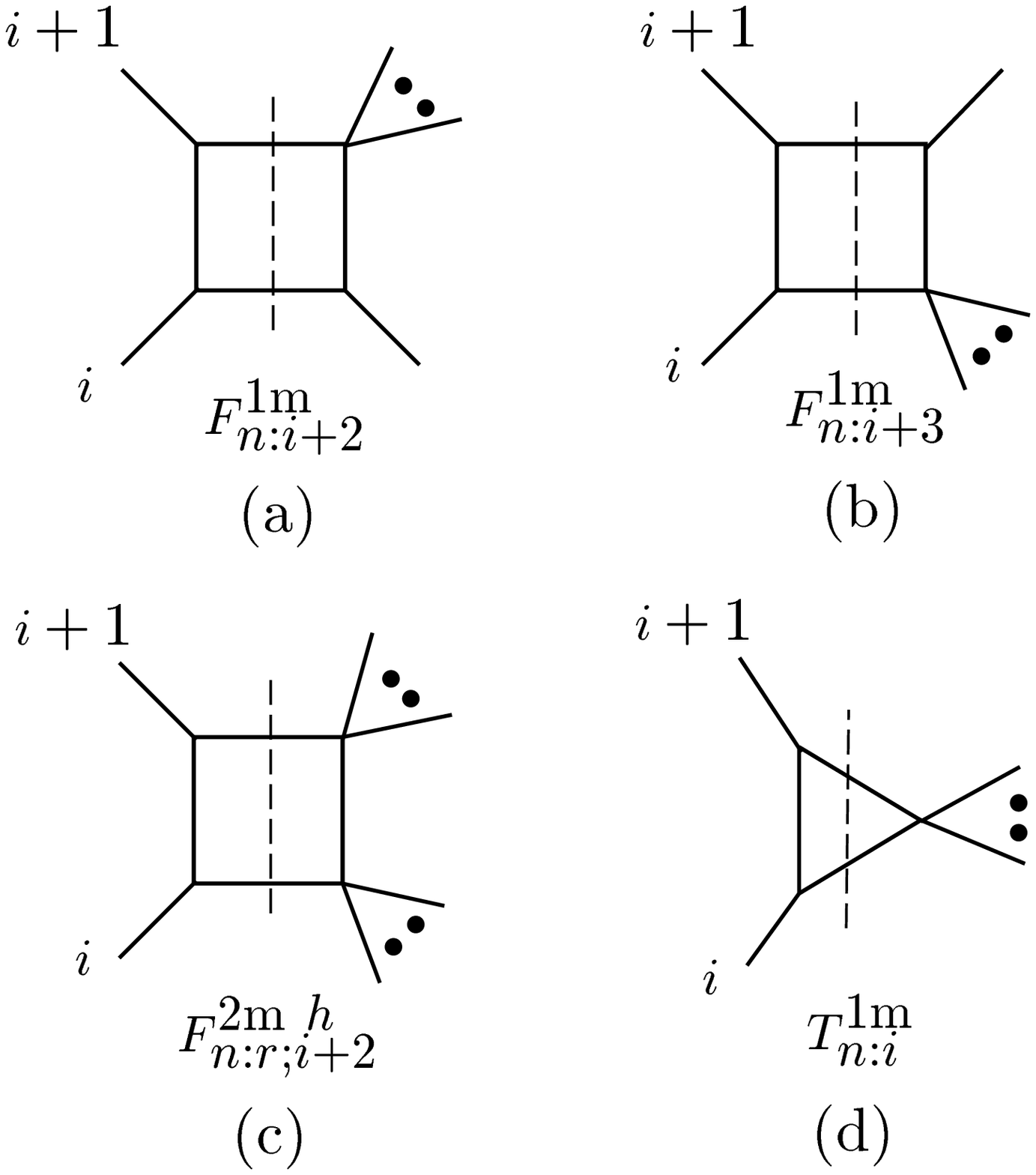}{}

The case of multi-particle poles is similar,
and only the two integrals (depicted in fig.~\use\NonFactBoxesFigure)
$$
I_{4:r-1;i+1}^{2 {\rm m} h} = {-2\, \rg \over t_{i-1}^{[2]} t_{i}^{[r]} }
F_{n:r-1;i+1}^{2 {\rm m} h} \, , \hskip 1.3 cm
I_{4:n-r-1;i+r+1}^{2 {\rm m} h} = {-2\,\rg \over t_{i+r-1}^{[2]} t_{i}^{[r]}}
F_{n:n-r-1;i+r+1}^{2 {\rm m} h} \; ,
\anoneqn
$$
contain poles in the channel $t_i^{[r]}$ ($r>2$).

Finally there are the higher-dimension integrals.  Such integrals,
however, cannot contain (non-spurious) poles.  In general the nearest
neighbor $t_i^{[r]}$ poles in the integrals come from isolated regions
in the loop momentum integral.  In six- or higher-dimensions the
measure $d^D l$ prevents the occurrence of a pole from the dominant
region of the loop momentum integral, where propagators become
singular.  This type of analysis is similar to the previously
performed analysis of the suppression of poles in $n$-gluon diagrams
containing internal scalars [\use\AllPlus,\use\GordonConf].

\subsection{Linking all contributions to singularities}

Having found all possible non-factorizing forms containing poles in
the channel of interest, we now go on to fix their coefficients by
requiring that the singular terms are consistent with the known
behavior in eq.~(\use\singular).

First consider two-particle collinear limits.  The loop diagrams in
fig.~\use\MasterSplitLoopFigure, containing the factorizing
contributions to the loop splitting functions, may contain
singularities in $\eps$ which depend on the details (and gauge
choices) of the process under discussion. A procedure which simplifies
our discussion is to collect all singularities in the non-factorizing
set. We therefore always subtract and add the discontinuity functions
$\discThree(s_{i,i+1})$ and $\discOne(s_{i,i+1})$ given in
eq.~(\use\Discont) with coefficients adjusted to move the
singularities to the non-factorizing contributions.  (See
section~\use\CollinearSection.2 for an explicit example of this
procedure.)

In Table~3 we collect all non-factorizing contributions to
$\Split^{\rm loop}$ in eq.~(\use\loopsplit), along with the singular
terms which we use to fix the coefficients.  These singular terms
satisfy the property that there is no non-zero linear combination
which is free of poles in $\eps$.  Since all singular terms appearing
in one-loop amplitudes are necessarily proportional to tree
amplitudes, the final non-factorizing part of the loop
splitting functions must be proportional to tree-level splitting
functions, as given in eq.~(\use\genrsdef).

To fix the coefficients we systematically step through Table~3.
Starting with the first row we have the
potential contribution to $\Split^{\rm loop}$,
$$
F^{\rm 1m}_{n:i+2} \ \mathop{\longrightarrow}^{k_i \parallel k_{i+1}}\
 -{1\over \eps^2} \Bigl( -zs_{i,i+1} \Bigr)^{-\eps} - \Li_2 ( 1-z )
+ \Ord(\eps) \,,
\eqn\FContribution
$$
which is the only non-factorizing function containing a $\ln(z)/\eps$.
We may fix the coefficient of this term by comparing it to the
coefficient of the same term in the original $n$-point amplitude,
on the left-hand-side of eq.~(\use\loopsplit).  In the collinear limit
where $k_i = z K$ and $k_{i+1} = (1-z) K$, from eq.~(\use\singular),
we have
$$
-\cg \Soft_{i-1}^{[n]} {1\over \eps^2} \L {\mu^2 \over - s_{i-1,i}} \R^\eps
A_n^{\rm tree}
\ \mathop{\longrightarrow}^{k_i \parallel k_{i+1}}\
-\cg \,\Soft_{i-1}^{[n]} \LB
{1\over \eps^2}  + {1\over \eps} \ln\Bigl({\mu^2 \over - s_{i-1, K} } \Bigr)
- {1\over \eps}\, \ln(z) \RB \sum_{\lambda = \pm}
\Split^{\rm tree} A_{n-1}^{\rm tree}
+ \Ord(\eps^0) \, ,
\eqn\NPointContribution
$$
where $s_{i-1,K} = (k_{i-1} + K)^2$. This is the only singular
term in the original $n$-point amplitude containing a
$\ln(z)/\eps$ in the collinear limit.

\vskip .5 cm
\centerline{
\def\tend{\cr \noalign{\hrule}}

\vbox{\offinterlineskip
\hrule
\halign{
        \vrule#
        &\quad\hfil#\hfil\quad\vrule
       &\quad\hfil\strut#\hfil\quad\vrule
        &\quad\hfil\strut#\hfil\quad\vrule
        \cr
height13pt & {\bf Potential Contribution}  &{\bf Selected Singular Part}
 &{\bf Collinear Limit}
 \tend
height17pt
& $F^{\rm 1m}_{n:i+2}$
& $\disp{1\over\eps}\ln(-s_{i-1,i})$
& $\disp{1\over\eps}\ln(-z s_{i-1, K})$ \tend
height17pt & $F^{\rm 1m}_{n:i+3}$
& $\disp{1\over\eps}\ln(- s_{i+1,i+2} )$
& $\disp{1\over\eps}\ln(- (1-z) s_{K,i+2} )$ \tend
height17pt & $F^{2 {\rm m} h}_{n:r;i+2}$
& $\disp{1\over \eps}\ln(-t_{i+1}^{[r+1]})$
& $\disp{1\over \eps}\ln(-(1-z) t_K^{[r+1]} - z t_{i+2}^{[r]})$ \tend
height17pt
& $\discTwo(s_{i,i+1}; t_{i+2}^{[r]}, t_i^{[r+2]})$
& $\disp -{1\over 2 \eps}\ln(-t_{i+2}^{[r]}/t_i^{[r+2]})$
& $\disp -{1\over 2 \eps}\ln(-t_{i+2}^{[r]}/t_{K}^{[r+1]})$ \tend
height17pt &
$\discOne(s_{i, i+1})$
& $\disp {1\over\eps}\ln(-s_{i,i+1})$
& $\disp {1\over\eps}\ln(-s_{i,i+1})$ \tend
height17pt &
$\discThree(s_{i, i+1})$
& $\disp{1\over\eps}$
& $\disp{1\over\eps}$ \tend
\cr
}
}
}
\vskip .13in
\nobreak
{\baselineskip 13 pt
\narrower\smallskip\noindent\ninerm
{\ninebf Table 3:} Particular singular terms in the second column
used to fix the coefficients of
potential non-factorizing contributions to $\Split^{\rm loop}$.
\smallskip}

\vskip .1 cm

Matching the coefficients of the $\ln(z)/\eps$ in the potential contribution
(\use\FContribution) to the one in the $n$-point
amplitude (\use\NPointContribution) gives the splitting function
contribution
$$
\cg\, \Soft_{i-1}^{[n]}\, \Split^{\rm tree} \Bigl(
\discOne(s_{i, i+1})
+ \mu^{2\eps} F^{\rm 1m}_{n:i+2} \Bigr|_{k_i\parallel k_{i+1}} \Bigr) \,,
\eqn\FContributionB
$$
which is thus fixed by the coefficient $\Soft^{[n]}_{i-1}$ in the
original $n$-point amplitude.  Since $F^{\rm 1m}_{n:i+2}$ also
contains a $\ln(-s_{i,i+1})/\eps$ it is convenient to subtract out
this this singularity using the discontinuity function
$\discOne(s_{i,i+1})$, so that we only adjust the coefficient of one
singularity at a time.  In the collinear limit the singular term
(\use\NPointContribution) in the amplitude also contains a
$\ln(-s_{i-1, K})/\eps$ which exactly matches a corresponding
singularity in the loop amplitude $A_{n-1}^{\rm loop}$ on the
right-hand-side of eq.~(\use\loopsplit).  Pulling out the pre-factor
$\cg \Split^{\rm tree}$, as in eq.~(\use\genrsdef), gives the
first term in eq.~(\use\FullSplit).  The collinear limit $k_i
\parallel k_{i+1}$ of eq.~(\use\FContributionB) gives the first entry
of Table~1.

Following similar reasoning, the coefficient of $\ln(1-z)/\eps$
in the amplitude matches the one in the splitting function
$$
\cg S^{[n]}_{i+1} \Split^{\rm tree}
 \Bigl(\discOne(s_{i,i+1})
+ \mu^{2\eps} F_{n:i+3}^{1{\rm m}}\Bigr|_{k_i\parallel k_{i+1}} \Bigr) \, ,
\eqn\SecondContrib
$$
where
$$
F_{n:i+3}^{1{\rm m}}\ \mathop{\longrightarrow}^{k_i\parallel k_{i+1}}\
-{1\over \eps^2} \Bigl( -(1-z)s_{i,i+1} \Bigr)^{-\eps}
    - \Li_2 \left(z\right) + \Ord(\eps) \,.
\anoneqn
$$
The contribution (\use\SecondContrib) corresponds to the the second
entry in the third column of Table~2.

The third and fourth rows in Table~3 contain singular terms which are
nowhere to be found in $n$-point amplitudes or in the factorizing
portion, so the coefficients must vanish, explaining the absence
of such contributions in Table~1.

Since we have collected all $\ln(-s_{i,i+1})/\eps$ singularities
together by moving such terms from the factorizing diagrams
to the non-factorizing contributions,
the coefficient of the $\discOne(s_{i,i+1})$ contribution is simply
$\Soft_i^{[n]}$ to match the same singularity in the amplitude, as
given in the third row of Table~1.  Finally, the coefficient of
$\discThree(s_{i,i+1})$, which is the only contribution to the
splitting function containing a $1/\eps$ singularity (without a
logarithm), is determined by the difference of the coefficients
$\Collinear^{[n]}$ for the $n$-point amplitude and
$\Collinear^{[n-1]}$ for the $(n-1)$-point amplitude.

Now consider factorization in a multi-particle channel
$t_i^{[r]} \rightarrow 0$.  The procedure for adjusting the
coefficients of potential contributions to the factorization functions
is similar to the two-particle collinear case.  Once again
the non-factorizing contributions must be proportional
to a product of factorized trees, because the singularities in $\eps$
are necessarily of this form.  Following the same logic the reader
may step through the rows of Table~4 to obtain the entries in Table~2.

\vskip .5 cm
\centerline{
\def\tend{\cr \noalign{\hrule}}
\vbox{\offinterlineskip
\hrule
\halign{
        \vrule#
        &\quad\hfil#\hfil\quad\vrule
        &\quad\hfil\strut#\hfil\quad\vrule
        \cr
height13pt & {\bf Potential Contribution}  &{\bf Selected Singular Part} \tend
height17pt & $F^{2{\rm m} h}_{n:r-1;i+1}$
& $\disp {1\over \eps}\ln(-s_{i-1,i})$ \tend
height17pt & $F^{{\rm 2m}\, h}_{n:n-r-1;i+r+1}$
& $\disp {1\over\eps} \ln(-s_{i+r-1,i+r})$ \tend
height17pt
& $\discTwo(t_i^{[r]}; t_{i+r}^{[r']}, t_i^{[r+r']})$
& $\disp -{1\over2\eps}\ln(-t_{i+r}^{[r']}/t_i^{[r+r']})$ \tend
height17pt &
$\discOne(t_i^{[r]})$
& $\disp {1\over\eps}\ln(-t_i^{[r]})$ \tend
height17pt &
$\discThree(t_i^{[r]})$
& $\disp{1\over\eps}$ \tend
\cr
}
}
}
\vskip .1in

\nobreak
{\baselineskip 13 pt
\narrower\smallskip\noindent\ninerm
{\ninebf Table 4:} Selected singular terms associated with integrals
with potential non-factorizing contribution in the multi-particle
$t_i^{[r]}$ channel.
\smallskip}

Thus we have fixed the coefficients of all non-factorizing
contributions in eqs.~(\use\loopsplit) and (\use\loopfact) for
$\Split^{\rm loop}$ and $\Fact_n$ and established the rules in
Tables~1~and~2.  The discussion we have presented here is valid to any
order in $\eps$; the only modification being that the box functions
appearing in these tables be kept to higher order.

\section{Discontinuity Functions as a Tool for Evaluating Integrals}
\tagsection\DiscontinuitySection

In this section we illustrate the use of discontinuity functions
as a tool for obtaining infrared divergent box integrals from
known infrared finite ones.

The infrared finite scalar box integrals of ref.~[\use\FourMassBox]
were evaluated in four dimensions.  In ref.~[\use\IntegralsLong] a separate
analysis for the infrared divergent integrals was performed
due to the need for dimensionally regularized expressions.
Since the discontinuity functions summarize the transition between
infrared finite and infrared divergent integrals this suggests a
procedure for obtaining infrared divergent scalar box integrals
directly from infrared finite ones. After subtracting off the
discontinuity function one simply takes the appropriate mass or
kinematic invariants to vanish to obtain the desired integrals.
This procedure is made practical by eq.~(\use\FourToThree) which
gives the necessary box discontinuity functions from the much simpler
triangle functions.  Thus, using known infrared finite box integrals,
any infrared divergent box integral is reduced
to an evaluation of simpler triangle integrals.

To illustrate the method we reproduce all the infrared divergent box
integrals (depicted in \fig\BoxDiscFigure) necessary for computations
in massless theories given in ref.~[\use\IntegralsLong].
Starting with the four-mass box one takes the limit of a
vanishing external kinematic invariant (or mass) and subtracts the
appropriate discontinuity function. For example, to obtain the
three-external-mass integral function (fig.~\use\BoxDiscFigure{b})
from the four-external-mass integral (fig.~\use\BoxDiscFigure{a}) given
in ref.~[\use\FourMassBox] we have from the first row of Table~5,
$$
\hskip -.3 cm
\lim_{K^2_1 \rightarrow 0}
\biggl[F^{4 \rm m}(K_1, K_2, K_3, K_4)
-\discTwo(K^2_1; K^2_2, (K_1 + K_2)^2)
-\discTwo(K^2_1; K^2_4, (K_4 +K_1)^2) \biggr] =
F^{3 \rm m}(k_1, K_2, K_3, K_4) \,.
\anoneqn
$$
This result agrees with the explicitly calculated result in
ref.~[\use\IntegralsLong].  The reader may verify that all other
infrared divergent box functions derived in ref.~[\use\IntegralsLong]
may be obtained in this way, by using the discontinuity functions
collected in table~5.%
\footnote{$^\star$}{In ref.~[\use\IntegralsLong] sign errors
were inadvertently introduced in the four-external-mass box integral obtained
from ref.~[\use\FourMassBox]; these signs are corrected in
ref.~[\use\SusyOne].}

This method for obtaining box integrals may be used for any of the
infrared divergent one-loop box integrals.  In particular, box integrals with
mixed massive and massless internal and/or external legs may also be
obtained efficiently in this way.  This method might also be useful
for higher-loop integrals, but one would need a method for obtaining
discontinuity functions that were obtained at one loop through
eq.~(\use\FourToThree).

\vskip -.8 cm
\LoadFigure\BoxDiscFigure
{\baselineskip 13 pt
\noindent\narrower\ninerm The box functions appearing in Table~5; the
upper-case $K_i$ represent off-shell ($K_i^2 \not = 0$) and lower-case
$k_i$ represent on-shell momenta.}
{\epsfysize 2.7 truein}{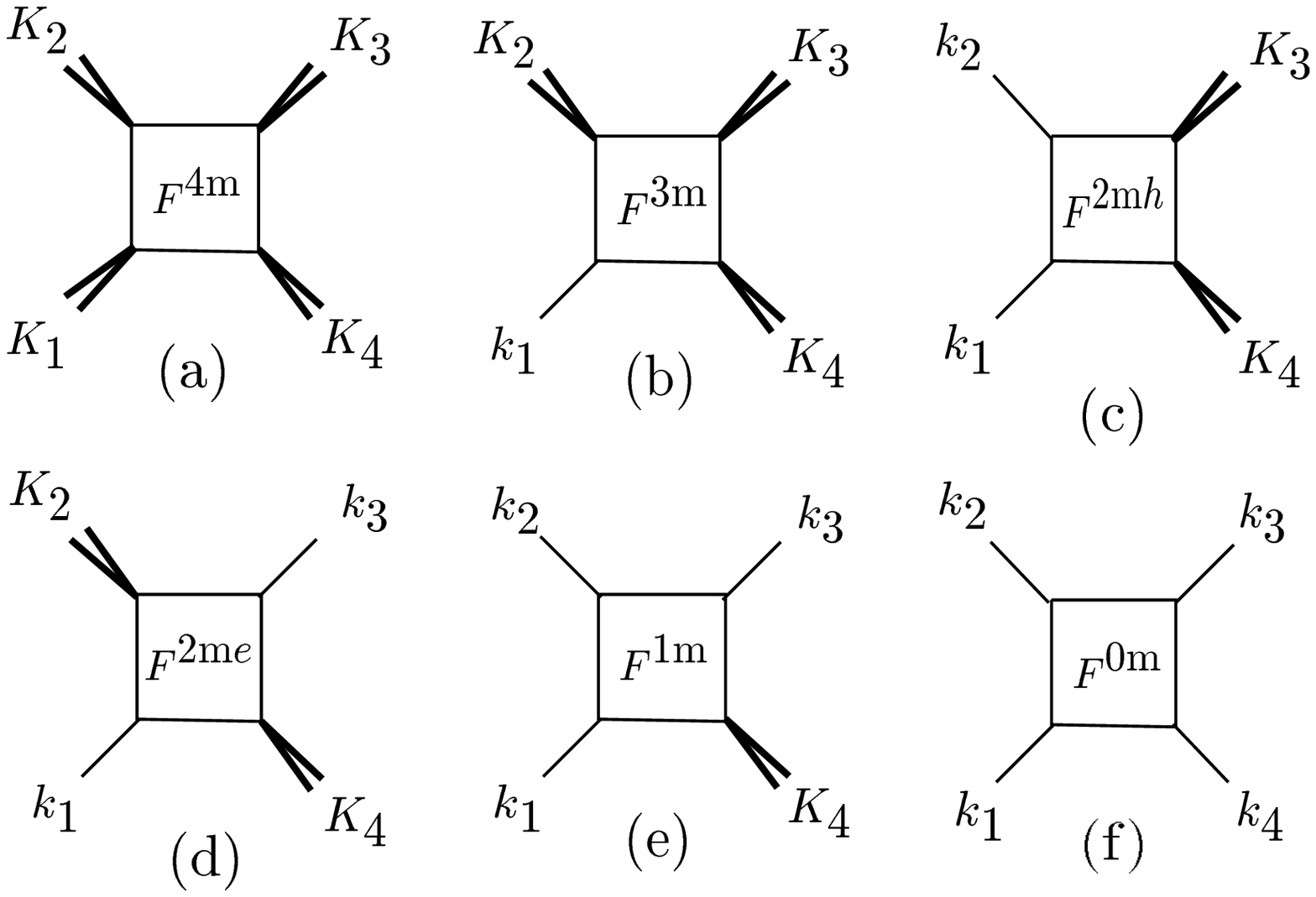}{}

\vskip .5 cm
\centerline{
\def\tend{\cr \noalign{\hrule}}

\vbox{\offinterlineskip
\hrule
\halign{
        \vrule#
        &\strut\quad\hfill#\hfill\quad\vrule
        &\strut\quad\hfill#\hfill\quad\vrule
        &\strut\quad\hfill#\hfill\quad\vrule
        &\strut\quad\hfill#\hfill\quad\vrule
        \cr
height15pt &{\bf Integral Functions}& {\bf Limit} & {\bf Discontinuity } \tend
height15pt &$F^{4 \rm m}\rightarrow F^{3 \rm m} $
& $K^2_1 \rightarrow 0$ &
$\disp\discTwo\bigl(K^2_1; K^2_2, (K_1 +K_2)^2\bigr) +
 \discTwo\bigl(K^2_1; K^2_4, (K_4+K_1)^2\bigr)$ \tend
height15pt & $F^{3 \rm m} \rightarrow F^{2 {\rm m} e}$ &
$K^2_3 \rightarrow 0$ &
$\disp \discTwo\bigl(K^2_3; K^2_2, (K_2+K_3)^2\bigr)
 + \discTwo\bigl(K^2_3; K^2_4, (K_3+K_4)^2\bigr)$ \tend
height15pt & $F^{3 \rm m} \rightarrow F^{2 {\rm m} h}$
& $K^2_2\rightarrow 0$ &
$\disp \half\discOne\bigl(K^2_2\bigr)
     + \discTwo\bigl(K^2_2; K^2_3, (K_2+K_3)^2\bigr)$ \tend
height15pt & $F^{2 {\rm m} e} \rightarrow F^{1 \rm m}$
& $K^2_2 \rightarrow 0$ &
$\disp \discOne\bigl(K^2_2\bigr)$ \tend
height15pt & $F^{2 {\rm m} h} \rightarrow F^{ {\rm 1 m}}$
& $K^2_3 \rightarrow 0$ &
$\disp\half\discOne\bigl(K^2_3\bigr)
+ \discTwo\bigl(K^2_3; K^2_4, (K_3+K_4)^2\bigr)$ \tend
height15pt & $F^{1 {\rm m}} \rightarrow F^{0 {\rm m}}$
& $K^2_4 \rightarrow 0$ &
$\disp \discOne\bigl(K^2_4\bigr)$ \tend
\cr
}
}
}
\vskip .2 cm
\nobreak
{\baselineskip 13 pt
\narrower\smallskip\noindent\ninerm
{\ninebf Table 5:} The discontinuity functions for all scalar box
integrals encountered in computations in massless gauge theories.
(See fig.~\use\BoxDiscFigure.)
\smallskip}

\vskip -.4 truecm
\section{Conclusions}

In this paper we have provided a general proof and discussion of
factorization in massless amplitudes. Infrared divergences associated
with massless theories significantly complicate matters as they induce
contributions which are not interpreted directly in terms of a naive sum of
products of amplitudes with a tree propagator.  The loop integrals
themselves may contain kinematic poles which cannot be interpreted in
any simple way in terms of tree propagators.  Furthermore, if the
amplitudes contain logarithms in the kinematic variable in which the
factorization is being performed, new infrared divergences develop.

Nevertheless, we provided a proof that factorization of massless
one-loop gauge theory amplitudes are described by a set of universal
functions linked to the known
[\use\KunsztSoper,\use\GG,\KunsztSingular] infrared divergences. In
massless (or high-energy)
QCD these singularities have been tabulated in previous papers
[\use\KunsztSoper,\use\GG,\use\KunsztSingular].  The proof presented
here was based on general properties of gauge theory amplitudes and
not on a specific diagrammatic analysis.  In particular, we made use
of the reduction of any one-loop amplitude in terms of a basis of scalar
box, triangle and bubble functions
[\use\Reduction,\use\PV,\use\VNV,\use\OtherInt,\use\IntegralsShort,%
\use\IntegralsLong]. By
identifying all possible sources of poles from the loop integrals, a
proof was presented which links the coefficients of all such kinematic
poles to the known infrared divergences appearing in the amplitudes.
The non-factorization of amplitudes is described by a limited set of
discontinuity and box integral functions, which were given explicitly.
The non-trivial aspect of our proof is that all finite (in $\eps$)
contributions to non-factorization are fixed in this way.

In particular, we proved the conjecture in ref.~[\use\SusyFour] that
the collinear splitting functions (or amplitudes) determined from
one-loop four- [\use\Long] and five-point
[\use\FiveGluon,\use\Fermion] amplitudes are universal functions for
an arbitrary number of external legs.  We also presented a procedure
for directly obtaining the splitting functions by calculating only
two- and three-point diagrams.  Although these diagrams are gauge
dependent, after combining these diagrams with discontinuity functions
whose coefficients are fixed by the infrared divergences, the gauge
invariant splitting functions are obtained.  An explicit sample
calculation of the $g \rightarrow gg$ splitting function in two
different gauges was provided.  We also calculated the contributions
of massive fermions and scalars to these loop splitting functions.

We have also shown how amplitudes factorize on multi-particle poles.
Although there are contributions which cannot be interpreted in terms
of simple factorization, these are given by a small number of
functions whose coefficients are again fixed by infrared divergences.
For calculations of six- and higher-point amplitudes multi-particle
factorization provides a powerful constraint on the form of the
amplitude.  An explicit six-point example of the constraints imposed
by multi-particle factorization was given.

We have also outlined a procedure for obtaining the splitting
functions to higher order in $\eps$.  The higher order terms
would be useful for performing phases space integrals at
next-to-next-to-leading order using the formalism of ref.~[\use\GG].

Although we have not explicitly discussed factorization as the momenta
of external legs become soft, the same type of analysis performed here
may be extended to that case.  By a general argument presented in
section~\ProofSection {.2}, any kinematic poles in the reduction
coefficients are proportional to the discontinuity functions.

A spin-off from our analysis is an efficient method for generating
dimensionally regularized infrared divergent box integral functions
from the known finite massive box integrals.  Indeed all box integral
functions contained in refs.~[\use\IntegralsLong] may be obtained by
taking the appropriate massless limits of the massive integrals after
subtracting off an appropriate set of discontinuity functions
determined by a recursion relation between between box and triangle
functions.  To extend this result to two-loops one would need a way of
determining the discontinuities.

The results presented here prove that universal factorization
properties hold for any gauge theory one-loop amplitude with
an arbitrary number of external legs.  We expect factorization to
continue to be a powerful tool in the calculation of gauge theory
amplitudes.

\vskip .2 cm
We thank L. Dixon, D.C. Dunbar, D.A. Kosower and A.G. Morgan for
helpful discussions and suggestions.  Research supported in part by
the US Department of Energy under grant DE-FG03-91ER40662, in part by
the National Science Foundation under grant PHY 9218990, and in part
by the Alfred P. Sloan Foundation under grant BR-3222.


\vskip -.4 truecm
\appendix{Reduction of One-Loop Integrals}
\tagappendix\ReductionAppendix

In this appendix we review and slightly modify the reduction procedure
[\use\Reduction,\use\PV,\use\VNV,\use\OtherInt,\use\IntegralsShort,%
\use\IntegralsLong] that rewrites $n$-point integrals in terms of
lower-point ones for use in constructing our proof
in section~\use\ProofSection.  This procedure
allows any Feynman diagram to be expressed as a linear combination of
$m\leq4$-point integrals.  The reduction procedure for tensor
integrals is a bit different than for scalar integrals so we discuss
these cases separately.

\subappendix{Tensor Integrals}
\tagappendix\TensorIntegralsAppendix

First we review the reduction of Feynman integrals with a loop
momentum dependent numerator and with five or more external legs.
Following this, the analogous decomposition of tensor box, triangle,
and bubbles is presented.

Consider then the reduction of any tensor integral function with more than
four external legs $(n> 4)$.  Denote by $I_n [l^{\alpha_1} \ldots
l^{\alpha_j} ]$ the tensor $n$-point integral with $j$ powers of loop
momenta in the numerator of the integrand:
$$
I_n [l^{\alpha_1} \ldots l^{\alpha_j} ]  \equiv i (-1)^{n+1}
(4\pi)^{2-\eps}
\int {d^{4-2\e}l\over
\L2\pi\R^{4-2\e}}\; {l^{\alpha_1} \cdots l^{\alpha_j} \over
( l^2 - m^2 ) ( \L l-p_1\R^2 - m^2 )
\cdots ( ( l-p_{n-1})^2 - m^2 ) } \,,
\eqn\tensorloopint
$$
where the momentum routing is taken to be $p_i = \sum_{j=1}^{i} K_j$,
and the $K_j$ are adjacent sums of the external momenta of the $n$-point
amplitude.
In gauge theory, in a Feynman-like gauge, the maximum number of powers
of loop momentum in the numerator is $n$ for an $n$-point function so
$j\leq n$.  Although we are interested in amplitudes with vanishing
internal masses, we have inserted a uniform mass $m$ in all internal
propagators in order to display the difference in the corresponding
reduction coefficients.  We will show that the pole structure is
independent of the mass, a fact we use in section~\use\ProofSection\
to explain why the poles in the massless reduction coefficients are
proportional to discontinuity functions.

The integrals in eq.~(\use\tensorloopint) may be reduced by projecting
the first component $\alpha_1$ of the tensor onto a basis of four
independent momenta; a simple choice of basis are the four vectors
$p_1,\ p_2,\ p_3,\ p_4$.  The final results for the reduction of the
amplitudes are independent of any particular choice of basis.
Projecting the loop integral onto the four vectors yields
$$
I_n [ l^{\alpha_1} l^{\alpha_2} \ldots l^{\alpha_j} ]
= p_1^{\alpha_1} A_1 + p_2^{\alpha_1} A_2
+ p_3^{\alpha_1} A_3 + p_4^{\alpha_1} A_4 \,,
\eqn\expand
$$
where we have suppressed the indices $\alpha_2 \ldots \alpha_j$ on
the right-hand side of eq.~(\use\expand) in $A_j$.  The functions
$A_i$ are found by first contracting eq.~(\use\expand) with the
momenta $p_{i}$, generating the four linearly independent equations,
$$
2I_n [ l\c p_i \ l^{\alpha_2} \ldots l^{\alpha_j} ] =
\sum_{k=1}^4 t_{ik} A_k \, , \hskip 1 cm  (i=1,2,3,4) \, ,
\eqn\matrixeqn
$$
where
$$
t_{ik}\equiv 2 p_{i}\c p_{k} \, .
\eqn\tmatrix
$$

Since the products $l\c p_i$ within the $n$-point integral on
the left-hand side of eq.~(\use\matrixeqn) may be expressed as
$$
2l\c p_i = -[(l-p_i)^2 - m^2 ] + [l^2 - m^2 ]
+ p_i^2 \,,
\eqn\massrewrite
$$
one can eliminate one of the Feynman denominators with each of
the first two terms in eq.~(\use\massrewrite),
which gives the integrand for an $(n-1)$-point
integral.  The third term in eq.~(\use\massrewrite) is independent
of the loop momenta and gives an $n$-point integral with only $j-1$
powers of loop momentum in the integrand.

Thus by solving the linear equations (\use\matrixeqn) for the $A_i$,
we may express the $n$-point integral with $j$ powers
of loop momentum as a linear combination of $n$- and $(n-1)$-point
integrals with $j-1$ powers of loop momentum.
By inverting $t$ we have
$$
\eqalign{
A_i & = \sum_{k=1}^{4} {[t]_{ki}\over \Delta} \
I_n [2 l\c p_{k}\ l^{\alpha_2} \ldots l^{\alpha_j} ]  \cr
    & = \sum_{k=1}^{4} {[t]_{ki}\over \Delta} \
\Bigl( I_{n-1}^{(k+1)}[l^{\alpha_2} l^{\alpha_3} \cdots l^{\alpha_j}]
- I_{n-1}^{(1)}[l^{\alpha_2} l^{\alpha_3} \cdots l^{\alpha_j}]
  + p_k^2 I_n[l^{\alpha_2} l^{\alpha_3} \cdots l^{\alpha_j}] \Bigr) \,,\cr}
\eqn\Asolution
$$
where the elements $[t]_{ki}$ in (\use\Asolution) are the cofactors of
$t$. The denominator $\Delta$ is the Gram determinant of the basis
of four vectors,
$$
\Delta = {\rm det}(t_{ij}) \, .
\eqn\GramDef
$$
(Note the relative signs, between $I_n$ and $I_{n-1}$, in the
definition of the integral functions (\use\tensorloopint).)

For dimensionally regularized expressions there is a technicality
which must be addressed.  Since the expansion on the right-hand side
of eq.~(\use\Asolution)
is in terms of a basis of four-dimensional external momenta while the
loop momentum on the left-hand side are $(4-2\eps)$-dimensional, one
might worry that this could lead to an error.  Consider, for
example, an $n$-point integral with two powers of loop momentum in the
numerator.  The full $(4-2\eps)$ expansion is
$$
I_n [ l^{\alpha} l^{\beta}]
=  \sum_{i,j=1}^4 p_i^\alpha p_j^\beta A_{ij}
    + \delta_{[-2\eps]}^{\alpha\beta} A_0 \,,
\eqn\dimreg
$$
where  $\delta_{[-2\eps]}^{\alpha\beta}$ is a metric
which is non-zero only in the $[-2\eps]$-dimensions.
By contracting $\delta_{[-2\eps]}^{\alpha\beta}$ into eq.~(\use\dimreg)
we obtain
$$
I_n[-l_\eps^2] = -2\eps A_0 \, ,
\eqn\EpsIntegral
$$
where $-l_\eps^2$ is the $[-2\eps]$ component of $l^2$.  After Feynman
parametrizing the loop integral (\use\EpsIntegral)
and following the discussion in
refs.~[\use\Mahlon], we break up the momentum integral as
$$
\int {d^{4-2\eps} l \over (2\pi)^{4-2\eps}}
= \int {d^4 l_4 \, d^{-2\eps} l_\eps \over (2\pi)^{4-2\eps}}
= - \eps {(4\pi)^\eps\over \Gamma(1-\eps)}
          \int {d^4 l_4 \over (2\pi)^{4}} \int_0^\infty dl_\eps^2 \,
          (l_\eps^2)^{-1-\eps} \,.
\anoneqn
$$
Performing the integrals one finds
$$
I_n[l_\eps^2] = -\eps I_n^{D=6-2\eps}[1] \,,
\eqn\HigherDimInt
$$
where $I_n^{D=6-2\eps}[1]$ is the $n$-point scalar integral in
$D=6-2\eps$ dimensions (given in appendix~\use\IntegralsAppendix{.4}).
Note that $I_n^{D=6-2\eps}$ is completely finite for $n>3$, since
there are no infrared or ultraviolet divergences in six dimensions for
$n>3$.  Thus $\delta_{[-2\eps]}^{\alpha\beta} A_0$ in
eq.~(\use\dimreg) is of $\Ord(\eps)$ and therefore does not enter into
the reduction procedure through $\Ord(\eps^0)$.  One can easily
generalize this argument to show that the difference between a
reduction in a four-dimensional basis and a $(4-2\eps)$-dimensional
basis is of $\Ord(\eps)$ as long as no ultra-violet divergences occur
in the integrals.  In renormalizable gauge theories (for a
Feynman-like gauge) all diagrams with five or more legs are
ultra-violet finite, so that the reduction coefficients for $n\ge 5$
do not depend on $\eps$.  (If one were to keep terms of $\Ord(\eps)$
or higher, then one would need to keep the higher dimension integrals
for $n\ge 5$, but these do not possess infrared divergences,
discontinuities as an external mass vanishes, or massless poles.
Following the discussion in section~\use\ProofSection,
these do not contribute to non-factorization.)

For $n\leq 4$ the reduction procedure is similar except that the
integral does not contain a complete set of external momenta that
spans four dimensions, so we must include the metric tensor.  The
metric tensor must be taken to be $4 - 2\eps$ dimensional since
ultra-violet divergences are encountered: boxes with four powers,
triangles with two or more powers of momentum, and bubble integrals
are all ultraviolet divergent.  The reduction coefficients of such
integrals can be expected to contain $\eps$-dependence.  (In the case
of supersymmetric theories, one can arrange all $n\ge 3$ integrals to
have at least two powers less of loop momentum in the numerator so
that all integrals with $n\ge 3$ are ultra-violet finite; this means
that the reduction coefficients are independent of $\eps$ in
supersymmetric theories [\use\SusyOne].)

Consider the standard expansion of the tensor integrals for $n\leq 4$,
\defeqn\IntExpansions
$$
\eqalignno{
& I_n[l^\alpha] = A_{1;i} p_i^\alpha \, ,
& ({\rm \IntExpansions{a}}) \cr
& I_n[l^{\alpha_1} l^{\alpha_2}] =
               A_{2;0} \delta_{[4-2\eps]}^{\alpha_1 \alpha_2} +
               A_{2;i,j} p_i^{\alpha_1} p_j^{\alpha_2} \, ,
& ({\rm \IntExpansions{b}}) \cr
& I_n[l^{\alpha_1} l^{\alpha_2} l^{\alpha_3}] =
     A_{3;i} \Bigl[ \delta_{[4-2\eps]}^{\alpha_1 \alpha_2} p_i^{\alpha_3} +
                    \delta_{[4-2\eps]}^{\alpha_1 \alpha_3}p_i^{\alpha_2} +
                    \delta_{[4-2\eps]}^{\alpha_2 \alpha_3}p_i^{\alpha_1}\Bigr]
            +  A_{3;i,j,k} p_i^{\alpha_1} p_j^{\alpha_2} p_k^{\alpha_3} \, ,
& ({\rm \IntExpansions{c}}) \cr
& I_n[l^{\alpha_1} l^{\alpha_2} l^{\alpha_3} l^{\alpha_4}] =
A_{4;0} \Bigl[
 \delta_{[4-2\eps]}^{\alpha_1 \alpha_2}\delta_{[4-2\eps]}^{\alpha_3\alpha_4} +
 \delta_{[4-2\eps]}^{\alpha_1 \alpha_3}\delta_{[4-2\eps]}^{\alpha_2\alpha_4} +
 \delta_{[4-2\eps]}^{\alpha_1 \alpha_4}\delta_{[4-2\eps]}^{\alpha_2\alpha_3}
         \Bigr] &  \cr
& \hskip 3 cm
 +  A_{4;ij} \Bigl[
    \delta_{[4-2\eps]}^{\alpha_1 \alpha_2} p_i^{\alpha_3} p_j^{\alpha_4} +
   {\rm sym.} \Bigr]
 + A_{4;i,j,k,l} p_i^{\alpha_1} p_j^{\alpha_2} p_k^{\alpha_3} p_l^{\alpha_4}\,,
& ({\rm \IntExpansions{d}})
}
$$
where the repeated indices are implicitly summed over and some of the
coefficients are related by symmetry.  By dotting the
momenta $p_i$ into these equations and contracting various indices one
obtains a complete set of equations which may be solved for the
coefficients $A$.  The equations obtained by contracting indices
contain explicit $\eps$-dependence through
$\delta_{[4-2\eps]}^{\alpha\beta}\;
\delta^{[4-2\eps]}_{\alpha\beta} = 4-2\eps$.

We now outline a modified procedure which eliminates explicit
$\eps$-dependence from the equations obtained by contracting indices.
Consider the first integral (\IntExpansions{a}); in this case the
metric tensor does not appear and the $A_{1;i}$ may be solved by
dotting external momenta into eq.~(\use\IntExpansions{a}); no
$\eps$-dependence enters in these equations.

Next we have the integral (\use\IntExpansions{b}), which has a
$\delta_{[4-2\eps]}^{\alpha_1 \alpha_2}$ in the tensor expansion.  To
solve for the $A_{2;0}$ we perform instead the contractions with the
tensor $\delta_{[-2\eps]}^{\alpha_1\alpha_2}$ which is non-zero only
for the $[-2\eps]$ dimensions.  Performing the contraction of
eq.~(\use\IntExpansions{b}) with
$\delta_{[-2\eps]}^{\alpha_1\alpha_2}$ we obtain
$$
I_n[l_\eps^2] = -\eps I_n^{D=6-2\eps}[1] =  2\eps A_{2;0}\,,
\eqn\EpsLoopInt
$$
so that
$$
A_{2;0} = -{1\over 2} I_n^{D=6-2\eps}[1]  \,.
\anoneqn
$$
The coefficient of $I_n^{D=6-2\eps}[1]$ does not contain $\eps$-dependence.
The remaining equations for $A_{2;i,j}$ obtained by dotting into
external momenta are manifestly free of $\eps$.

For $I_n[l^{\alpha_1} l^{\alpha_2} l^{\alpha_3}]$ we may contract
eq.(\use\IntExpansions{c}) with $2\delta_{[-2\eps]}^{\alpha_1
\alpha_2} p_m^{\alpha_3}$ to obtain
$$
2I_n[l_\eps^2 l\cdot p_j] = 2\eps \sum_{i=1}^{n-1} A_{3;i} t_{j i} \,.
\eqn\Third
$$
Using the expansion $
2 l\cdot p_j  = -[(l-p_j)^2 -  m^2] + [l^2 - m^2] + p_j^2
$
and eq.~(\use\HigherDimInt) we obtain the expanded form
$$
2I_n[l_\eps^2 l\cdot p_j] =
-\eps \Bigl( I_{n-1}^{(j+1), D=6-2\eps} - I_{n-1}^{(1), D=6-2\eps}
       + p_j^2 I_n^{D=6-2\eps} \Bigr) \,.
\anoneqn
$$
By inverting eq.~(\use\Third) we then have
$$
A_{3;i} = -{1\over 2} \sum_{j=1}^{n-1} t_{ij}^{-1}
\Bigl[ I_{n-1}^{(j+1), D=6-2\eps}[1] - I_{n-1}^{(1), D=6-2\eps}[1]
       + p_j^2 I_n^{D=6-2\eps}[1] \Bigr] \,.
\anoneqn
$$
Note again that there is no $\eps$-dependence in the coefficients of the
integrals.  Further equations which are obtained by only dotting
external momenta into eq.~(\use\IntExpansions{c}) are manifestly
independent of $\eps$.

Finally, consider the case of four powers of loop momentum in the numerator.
If we contract eq.~(\use\IntExpansions{d}) with
$\delta_{[-2\eps]}^{\alpha_1 \alpha_2} \delta_{[-2\eps]}^{\alpha_3\alpha_4}$
we obtain
$$
I_n[l_\eps^4] = -\eps(1-\eps) I_n^{D=8-2\eps}
           = (4\eps^2 - 4\eps) A_{4;0} \,,
\anoneqn
$$
so that
$$
A_{4;0} ={1\over 4} I_n^{D=8-2\eps} \,.
\anoneqn
$$
Once again all $\eps$-dependence cancels from the coefficients.
The other coefficients may be obtained by continuing the reduction process
and are also free of $\eps$.

Thus, by introducing the $D=6-2\eps$ and $D=8-2\eps$ dimension scalar
bubble, triangle and box functions (given in
appendix~\use\IntegralsAppendix{.4}) we may avoid $\eps$-dependence in the
reduction coefficients; this dependence has been pushed into these
integrals.   (In the dimensional reduction [\use\Siegel] or
FDH [\use\Long] schemes there are no other sources of $\eps$
dependence, since the numbers of particle states are the four
dimensional ones.)
Upon rewriting the higher-dimensional integral functions in
terms of $D=4-2\eps$ ones, the conventional reduction
(through $\Ord(\eps^0)$) is regained.

As an explicit example of the modified reduction procedure consider
the tensor bubble integral in a massless theory
$$
\eqalign{
I_2[l^\mu l^\nu]  & =  - i (4 \pi)^{2-\eps}
 \int {d^{4 -2 \eps} l \over (2\pi)^{4-2\eps} } \;
{ l^\mu l^\nu  \over l^2 (l-K)^2 } \cr
& \equiv B_1 \delta^{\mu\nu}_{[4-2\eps]} K^2  + B_2 K^\mu K^\nu \, .\cr}
\eqn\BubbleIntegral
$$
First consider a conventional reduction procedure.
By either tracing over the indices with $\delta_{[4-2\eps]}^{\mu\nu}$
or dotting into the $K^\mu K^\nu$ we obtain the
two equations
$$
(4- 2\eps) B_1 + B_2 = 0\, ,   \hskip 2 cm
 (B_1 + B_2) = {1\over 4} I_2[1]\, ,
\eqn\BubbleEqs
$$
where we used $2 K\cdot l = -(l-K)^2 + l^2 + K^2$ and also dropped
the tadpole diagrams, which vanish in dimensional regularization.
(For higher-point functions it is more convenient to dot only one
momentum at a time into the expansion for the integrals, followed by
iterating the equations for each power of loop momentum.)
Solving the two equations we have
$$
B_1 = -{1\over 4} {1\over 3 - 2\eps} I_2[1] \, , \hskip 2 cm
B_2 = {1\over 2} {2 - \eps \over 3 - 2\eps} I_2[1] \,,
\eqn\SolutionOne
$$
which contain explicit $\eps$-dependence in the coefficient of
the scalar bubble integral; it is this $\eps$-dependence which
we wish to avoid by using the modified procedure.

Now consider the modified procedure, where we extend our basis of
integral functions to include the six-dimensional bubble.
We obtain the first equation by tracing over the $[-2\eps]$ dimensions
in eq.~(\use\BubbleIntegral) to yield
$$
\eqalign{
I_2[l_\eps^2] & = 2\eps B_1 K^2 \cr
& =  - \eps I_2^{D=6-2\eps}[1]\,,  \cr}
\anoneqn
$$
which we may use to solve for $B_1$.
The second equation is obtained by dotting $K^\mu K^\nu$ into
eq.~(\use\BubbleIntegral)
$$
B_1 + B_2 = {1\over 4} I_2[1] \,,
\anoneqn
$$
so that
$$
B_1 = -{1\over 2 K^2} I_2^{D=6-2\eps}\, , \hskip 2 cm
B_2 = {1\over 4} I_2[1] + {1\over 2 K^2} I_2^{D=6-2\eps} \,.
\eqn\SolutionTwo
$$
Note that there is no $\eps$-dependence in the coefficients of the integral
functions in eq.~(\use\SolutionTwo).
It is easy to verify that this solution is identical to the one in
eq.~(\use\SolutionOne) after using eq.~(\use\BubbleSix) to
express the  $D=6-2\eps$ integral functions in terms of $D=4-2\eps$
ones.

\subappendix{Scalar Integrals}
\tagappendix\scalarintegrals

Consider now $n$-point scalar integrals with $n>5$.
Denote by $I_n [1]$ the scalar $n$-gon integral with no powers
of loop momenta in the numerator of the integrand:
$$
I_n[1] \equiv i (-1)^{n+1}
(4\pi)^{2-\eps} \int {d^{4-2\e}l\over (2\pi)^{4-2\e}}
\; { 1 \over ( l^2 - m^2 ) ( (l-p_1)^2 - m^2 )
   \cdots ((l-p_{n-1})^2 - m^2 ) }  \,.
\eqn\scalarintegral
$$
A convenient method for reducing the integral in
eq.~(\use\scalarintegral) is given in ref.~[\use\OtherInt].  This
method is based on the observation that for four-dimensional
momenta defined by $p_i = \mathop{\sum}_{j=1}^i K_j$, and for $n>5$ point
integrals a solution to
$$
\sum_{i=1}^{6} b_i p_{i}^{\alpha} = 0\, , \hskip 2 cm
\sum_{i=1}^{6} b_i = 0 \, ,
\eqn\scalarexpand
$$
can be found for some constants $b_i$.  At least six $b_i$ are
required for a non-trivial solution since five or more vectors are
linearly dependent in four-dimensions and the second equation provides
an additional constraint.  For $n= 5$ this solution breaks down
because there are only four independent momenta present within the
integral; another technique [\use\VNV,\use\IntegralsShort,\use\IntegralsLong]
may be used to reduce scalar pentagons whose results we quote below.

The reduction for $n>5$ proceeds by first multiplying the
integrand of eq.~(\use\scalarintegral) by unity in the form,
$$
1 = {\sum_i b_i \L p_{i}^2 - m^2 \R \over \sum_i b_i \L
p_{i}^2 - m^2 \R } \,,
\eqn\multiplybyone
$$
and using the properties of $b_i$ (i.e., $\sum_i b_i l^2 = 0$ and
$\sum_i b_i l\c p_i = 0$) so that we may express the
numerator in eq.~(\use\multiplybyone) as
$$
\mathop{\sum}_{i=1}^{6} b_i ( p_{i}^2 - m^2 ) =
\mathop{\sum}_{i=1}^{6} b_i \bigl( (l-p_{i})^2 - m^2 \bigr) \, ,
\anoneqn
$$
which corresponds to a sum over factors in the denominator of
the integrand in eq.~(\use\scalarintegral).
The mass dependence in the denominator of eq.~(\use\multiplybyone)
also drops out via $\sum_{i=1}^6 b_i m^2 = 0$, and the original
integral in eq.~(\use\scalarintegral) may then be expanded as,
$$
\eqalign{
I_n[1] & = {1\over \sum_{j=1}^{6} b_j p_{j}^2} \ \sum_{i=1}^{6} b_i
\ I_n [(l-p_{i})^2 - m^2] \cr
& = -
{1\over \sum_{j=1}^6 b_j p_j^2} \sum_{i=1}^6 b_i I_{n-1}^{(i+1)}[1] \,,\cr}
\eqn\Nscalarreduct
$$
where $I_{n-1}^{(i)}$ is the $(n-1)$-point integral obtained
from the $n$-point integral by removing the propagator in
eq.~(\use\scalarintegral) between legs $i-1$ and $i$.

The solution of eqs.~(\scalarexpand) are ratios of kinematic
determinants, given in detail by Melrose [\use\OtherInt]
The result in solving for the $b_i$ is that the coefficients
in front of the $ I_{n-1}^{(i+1)}$ in eq.~(\use\Nscalarreduct) are
inversely proportional to
$$
\det \Bigl[-{1\over 2} (p_{i-1} -p_{j-1})^2 \Bigr] \, .
\eqn\singularsource
$$
Any poles in the scalar integral reduction coefficients (for $n\ge
6$), for the case of a uniform or zero internal mass, must come from
this determinant.  Since the mass dependence $m$ in the integral reduction
coefficients drops out for the case of a uniform mass around the loop,
the poles for $m=0$ or $m\neq 0$ are the same.

This leaves us with the special case of $n=5$, that is, the reduction
of scalar pentagon integrals down to boxes.  For this case, we quote
the results of ref.~[\use\IntegralsShort].  The scalar pentagon with a
uniform internal mass and external momenta $K_i$ may be expressed as
$$
I_5 = \half \mathop{\sum}_{i=1}^5 c_i I_{4}^{(i)} +
 \eps c_0 \, I_5^{D=6-2\eps} \,,
\eqn\pentagonreduceB
$$
where
$$
c_i = \mathop{\sum}_{j=1}^5 S_{ij}^{-1} \, , \hskip 2 cm
c_0 = \sum_{i=1}^5 c_i\,, \hskip 2 cm
S_{ij} = m^2 - \half (p_{i-1} - p_{j-1})^2 \,.
\eqn\fivepointreduce
$$

In summary, by iterating the above tensor and scalar reduction
procedures from $n$- to $(n-1)$-point integrals any $D=4-2\eps$ one-loop
amplitude may be expressed as a linear combination of scalar bubble,
triangle, and box integral functions.
 By using a slightly modified Passarino-Veltman reduction of tensor
integrals we avoid all $\eps$-dependence in the reduction coefficients
at the cost of introducing higher-dimensional scalar integrals.  The
explicit forms of scalar functions which may appear in massless gauge
theories is summarized in appendix~\use\IntegralsAppendix.

\vskip -.4 truecm
\appendix{Kinematic Poles in Reduction Coefficients}
\tagappendix\PoleAppendix

In this appendix we find the potential kinematic poles in a particular
channel within the integral reduction coefficients, reviewed in the
previous appendix.  We will systematically step through and discuss
the various kinematic denominators of the coefficients to exhibit
their structure.  Furthermore, we find that one can side-step the
poles in a given channel for the reduction of integrals down to $n=5$
and $n=6$ for tensor and scalar integrals, respectively.  We also track
the difference in the kinematic denominators when there is a uniform
internal mass in the loop and when there is no mass, which is used in
section~\use\ProofSection\ to understand why these poles are
proportional to discontinuity functions.

\subappendix{Poles from $n\ge 6$ integral reductions}

The coefficients which may appear, in each step of the reduction of
integrals with six or more external legs, are proportional to the
inverses of the two kinematic determinants given in
eqs.~(\use\GramDef) and (\use\singularsource),
$$
\eqalign{
\hbox{(a)} &  \quad {\det [2 p_{\sigma_i} \c p_{\sigma_j}} ] \, ,
   \hskip 1.4 cm (i,j=1, \ldots,4)  \, ,  \cr
\hbox{(b)} &  \quad {\det[ -{1\over 2} (p_{\sigma_i} - p_{\sigma_j})^2]} \, ,
   \hskip 1 cm (i,j=1, \ldots,6)  \, , \cr}
\anoneqn
$$
where the $p_{\sigma_j}$ (with $j=1\ldots n-1$) are any of the $p_i =
\sum_j K_j$. The $K_j$ are external momenta of the integral
function being reduced and may be on- or off-shell.  The final result
for the amplitudes is independent of any particular choice of the
bases $\{ p_{\sigma_j} \}$ at each step of the procedure. (This
clearly holds for tensor reductions [\use\Reduction,\use\VNV] and has
been explicitly shown by Melrose [\use\OtherInt] for the case of
scalar reductions.)  The first determinant (a) is the Gram determinant
arising from tensor reductions, while the second (b) is a modified
Cayley determinant coming from the reduction of scalar integrals.

Whenever either of these determinants has zeros the reduction
coefficients can have poles.  The Gram determinant will vanish
whenever the momentum basis of the reduction collapses, which may
happen if two (even non-adjacent) massless external momenta are
collinear or one external momentum is soft.  Note that as a
multi-particle kinematic variable vanishes, the two determinants
considered as general analytic functions will not vanish (except for
special kinematics, such as when all kinematic variables are
time-like); thus we only need consider collinear (and soft) poles.

In any channel where the momenta of given pair of external legs become
collinear one can avoid zeros in the determinants (a) and (b) by a
judicious choice of the momentum basis.  For example, by not including
either $p_1$ or $p_2$ in the basis, as the momenta of legs 1 and 2
become collinear the Gram determinant (a) does not vanish. (For $n\ge
6$ there are always at least five momenta from which to choose the
four needed in the Gram determinant.)  In this way we may avoid
collinear poles in any given channel coming from the tensor reduction
coefficients for $n\ge6$.  The same type of reasoning allows one to
avoid zeros in the second determinant (b) for $n>6$.  For the case of
$n=6$ one may explicitly check that there are zeros.  Thus by making
use of arbitrariness in the momentum basis choice of the reduction one
can avoid the poles in the coefficients in any channel of the $n$ goes
to $n-1$ reduction, down to scalar hexagons ($n=6$).  The poles in the
scalar hexagon reduction coefficient do not, however, depend on the
uniform internal mass.

In another channel one would choose a different momentum basis to
again avoid poles in $n> 6$ reduction coefficients.  The consistency
of this procedure follows from the fact that all final results for
amplitudes are independent of the choice of basis in each step; a
change of momentum basis only shifts the poles to different steps of
the reduction.  Our choice is to always push the collinear pole in any
given channel to the scalar hexagon and the pentagon reduction,
discussed below.

\subappendix{Poles from $n\leq 5$ integral reductions}

Now consider the potential kinematic poles that arise from the
reduction of $n\leq 5$ point integrals.  As discussed
in the previous appendix, the only denominators which may appear are
$$
\eqalign{
\hbox{(a)} & \quad \Delta =  {\det [2 p_i \c p_j] } = \det [2 K_i \c K_j] \, ,
   \hskip 1 cm (1\leq i,j \leq n-1 \leq 4) \,, \cr
\hbox{(b)} & \quad \det S =  {\det\Bigl[m^2 - {1\over 2}(p_{i-1} - p_{j-1})^2
                  \Bigr]} \, ,
   \hskip 1.4 cm (i,j=1, \ldots,5)  \,, \cr}
\eqn\detlist
$$
where $m$ is a uniform mass in the loop.  The first one comes from
tensor reductions with $n\le 5$ and the second from reducing scalar
pentagons as given in eq.~(\use\pentagonreduceB).  There are no
multi-particle zeros in either determinant for $n>2$.  For $n=2$ the
multi-particle pole appearing in the reduction coefficient of the
tensor bubble (\use\SolutionTwo) multiplies $I_2^{D=6-2\eps}$, given
in eq.~(\use\BubbleSix), which then cancels the pole.  One cannot avoid
collinear poles for $n\le 5$ since there is only one basis choice for
the reduction of these integrals; the poles which have been
side-stepped above for $n \ge 6$ appear here.  For example, as $s_{12}
\rightarrow 0$ the diagram depicted in fig.~\PentagonPoleFigure, with
on-shell legs $k_1$ and $k_2$, is the only type of pentagon which has
a pole in this channel.

The Gram determinant (a), which has no dependence on the uniform
internal mass, contains zeros whenever any pair of on-shell momenta
$K_i=k_i$ become collinear. (Recall that lower case momenta denote on-shell
external momenta of the amplitude.)

For the case of $m=0$, it is straightforward to investigate zeros of
the second determinant (b) in multi-particle and collinear channels.
As a multi-particle kinematic variable $t_i^{[r]}$ $(r>2$) vanishes
$\det S$ will not vanish (except for special kinematic
configurations).  There will however be zeros in $\det S$ for
collinear massless momenta.  For example, the kinematic configuration
in fig.~\PentagonPoleFigure\ gives
$$
\eqalign{
\det S & =
      -{ 1\over 16} s_{1 2} \Bigl(s_{2 3}\, s_{3 4} \,s_{4 5} \,s_{5 1}
           - s_{2 3}\, s_{1 2}\, s_{5 1}\, K_4^2
           + s_{5 1}\, K_3^2\, K_5^2\, s_{2 3}
           - s_{3 4}\, K_5^2\, s_{2 3}^2
           - K_3^2\, s_{4 5}\, s_{5 1}^2 \Bigr)  \, ,
            \cr}
\anoneqn
$$
which clearly vanishes in the limit $s_{12} \rightarrow 0$.  Thus,
poles of the form $1/s_{i, i+1}$ may develop in the
reduction of scalar pentagons to scalar boxes.

For the case of $m\not=0$, the second determinant (b) contains
explicit dependence on the uniform internal mass. However, this
dependence is relatively simple since
$$
\det S \Bigr|_{m \not =0} = \det S \Bigr|_{m=0} + {m \over 16} \Delta \; .
\anoneqn
$$
The Gram determinant has zeros whenever $\det S |_{m=0}$ has one,
so that $\det S|_{m \not =0}$ will have the same zeros as $\det S|_{m=0}$.

In summary, in any given multi-particle or collinear channel the
integral reduction may be performed so that there are no poles in the
coefficients for $n> 6$.  For $n\leq 6$ poles may appear as momenta
become collinear (or soft).  However, one finds the same apparent
poles in the reduction of loop integrals containing a uniform internal
mass as for the massless case.


\vskip -.4 truecm
\appendix{Higher Dimension Integrals: Special Cases}
\tagappendix\HigherDimAppendix

In this appendix we show that the two special cases, the higher
dimension integrals $I_2^{D=6-2\eps}$ and $I_3^{ {\rm 1m}, \,
D=6-2\eps}$ in eq.~(\use\BubbleTrouble), do not contribute as
discontinuity functions.  Both are proportional (with $\eps$ dependent
coefficients) to the $D=4-2\eps$ scalar bubble in
eq.~(\use\BubbleInt), which becomes a discontinuity function as the
single external mass vanishes.  The potential appearance of these
integrals as discontinuity functions might introduce contributions to
the splitting and factorization functions not linked to the
singularities in $\eps$.  In this appendix we show that such potential
contributions do not occur.

For both the bubble $I_2^{D=6-2\eps}(K^2)$ and triangle $I_3^{{\rm
1m},\, D=6-2\eps}(K^2)$ we will examine the appearance of poles in
the $K^2$ channel.  Poles may in principle come from either tree
propagators, as discussed in section~\use\ProofSection{.1} or from
reduction coefficients, as in section~\use\ProofSection{.2}.  (These
two higher-dimensional integral functions themselves do not have
poles.)

First consider the appearance of $I_2^{D=6-2\eps}$. As discussed in
appendix~\use\ReductionAppendix, the Passarino-Veltman reduction takes
an $n$-point integral function with $m$ powers of loop momentum and
reduces it to a combination of $n$ and $n-1$ point integral functions
with $m-1$ powers of loop momentum.  Thus the only way to obtain a
tensor bubble (two-point) integral function with two powers of loop
momentum, which generate the $D=6-2\eps$ scalar bubbles via equations
(\use\BubbleIntegral) and (\use\SolutionTwo), is to start with an
$m$-point integral with $m$ powers of loop momentum in the numerator.
Thus, to obtain $D=6-2\eps$ bubble functions we must consider
(maximal) tensor integral functions of the form
$$
I_m [l^{\alpha_1} \ldots l^{\alpha_m} ]  \equiv i (-1)^{n+1}
(4\pi)^{2-\eps} \int {d^{4-2\e}l\over
\L2\pi\R^{4-2\e}}\; {l^{\alpha_1} \ldots l^{\alpha_m} \over
  l^2 \L l-K_1\R^2  \cdots  \L l-K_m\R^2 } \,,
\eqn\TensorLoopMax
$$
In ordinary or background field Feynman gauge this maximum occurs for
diagrams where all legs attached to the loop are gauge boson legs
[\use\SusyOne]; diagrams where fermion lines are entering or exiting
the loop will have one less power of loop momentum.  Thus we need
consider loops with only gluons attached.

First consider the case where the kinematic pole arises from a
reduction coefficient.  As discussed in appendix~\use\PoleAppendix, no
multi-particle poles arise from reduction coefficients.  For the case
of two-particle (collinear) poles one can side-step the poles from
tensor integrals down to
pentagons, as discussed in appendix~\use\PoleAppendix{.1}.  Furthermore,
for the pentagon and below, using the formulas of
appendix~\use\ReductionAppendix, it is straightforward to check that
the coefficients of the $D=6-2\eps$ bubbles contain at most a single
pole in any given channel. Since $I_2^{D=6-2\eps}[1](K^2)$ is
proportional to $(K^2)^{1-\eps}$ the pole is spurious since it is
canceled.  Alternatively, one may investigate the $m$-point loop
integrals with $m$ powers of loop momentum directly; this type of
analysis has already been discussed in
ref.~[\use\AllPlus,\use\GordonConf] with the result that they do not
contribute a pole to the amplitude.

Now consider the bubble function $I_2^{D=6-2\eps}$ from diagrams with
a tree propagator as on the right-hand-side of
fig.~\use\MultiFactFigure.  The bubble functions can arise either
directly from diagrams with two-point loops or from the integral
reduction of higher-point loop diagrams.  The diagrams with a bubble
loop are part of the `factorizing' contribution and therefore do not
concern us in this appendix.  For the bubble functions that arise from
higher-point diagrams we show below that all integrals with a maximum
number of powers of loop momentum suppress the pole from the
intermediate leg, since they are proportional to either
$$
K^2 \delta^{\alpha_1 \alpha_i}_{[4-2\eps]}\hskip 1 cm  \hbox{or}
\hskip 1 cm K^{\alpha_1}  \,,
\eqn\disctensor
$$
where the index $\alpha_1$ dots into the gluon
(or gauge boson) line of the intermediate factorized leg and
$K_1=K$ in eq.~(\use\TensorLoopMax).  The integrals cannot produce
any further poles in external mass $K^2$, as can be seen from the fact
that such poles do not exist in the reduction coefficients, discussed
in appendix~\use\PoleAppendix, or in the basis of integral functions,
discussed in appendix~\use\IntegralsAppendix.  (The $D=4-2\eps$ single
mass triangles are an exception, but such integrals with more than
four-point kinematics do not occur in the reduction of diagrams with
tree poles that we are considering.)  The first type of term is
clearly subdominant in the factorization since the factor of $K^2$
cancels the pole from the tree propagator.  The second type of term is
also subdominant because the gluon contracts against a tree amplitude
which makes up a conserved current; the longitudinal term vanishes
sufficiently fast to be irrelevant.

First we contract eq.~(\use\TensorLoopMax) with the external momentum
$K^{\alpha_1}_1$ and rewrite the scalar product as $ 2l\c K_1 = -
(l-K_1)^2 + l^2 + K_1^2$.  This produces three new integrals,
$$
2 I_m [l \cdot K_1 \, l^{\alpha_2} \ldots l^{\alpha_m} ]
=
 I_{m-1}^{(2)} [l^{\alpha_2} \ldots l^{\alpha_m}]
                - I_{m-1}^{(1)} [l^{\alpha_2} \ldots l^{\alpha_m}]
                + K_1^2 I_m  [l^{\alpha_2} \ldots l^{\alpha_m}]  \,,
\anoneqn
$$
where $I_{m-1}^{(j)}$ is the loop integral with $m-1$ legs obtained from
(\use\TensorLoopMax) by removing the $j$th propagator.

The first two integrals are functions of $K_1+K_2$ and $K_1+K_m$ and
will never reduce further into a bubble integral with momentum
$K^{\alpha}_1$ flowing through.  The third integral does not possess a
$K_1^2$ pole within the integral function and is then suppressed by
a factor of $K_1^2$ in front.  Thus in the factorizing limit
there is no unsuppressed contribution.

Dotting the integral (\use\TensorLoopMax) into $K_1^\alpha$
leads only to potential discontinuities suppressed by $K_1^2$; this
necessarily means that the integral (\use\TensorLoopMax) must be
proportional to either of the two tensors in eq.~(\use\disctensor),
which as discussed above are suppressed in the factorization limit.

The single-external-mass triangle $I_3^{{\rm 1m},\, D=6-2\eps}$
(\use\SixDimTri) also requires special care as it is discontinuous in
the limit that the external mass vanishes.  The argument of these
integrals are the $s_{i, i+1}$ where $i$ and $i+1$ label the momenta
of two massless legs of the triangle.  Such triangles are only
discontinuous for the collinear or soft limits of $k_i$ and $k_{i+1}$.
In the integral reduction, this triangle function can appear from two
places.  The first is from the factorizing diagrams displayed in
fig.~\use\MasterSplitLoopFigure.  The integrals from these diagrams
are directly taken into account in our analysis since we calculate
them explicitly; indeed, a combination of $D=4-2\eps$ and $D=6-2\eps$
integrals from these diagrams is what makes $\Split^{\rm fact}$
non-zero in eq.~(\factorizing).

The second place $I_3^{{\rm 1m}, \,D=6-2\eps}$ may arise is in the
reduction of integrals of the form in fig.~\use\IntPoleFigure, which
is given for the $s_{12}$ channel (all remaining $s_{i,i+1}$ channels
are similar); in this figure $I_3^{{\rm 1m},\, D=6-2\eps}$ would arise
in the reduction term where all loop propagators between $k_3$ and
$k_n$ (following the clockwise ordering) are removed; the single
external mass in this case is $s_{12}$.  As discussed in
appendix~\PoleAppendix, down to pentagon integrals we can avoid poles
from tensor integrals in any given channel by choosing an appropriate
momentum basis in which to perform the reduction.  For the pentagon to
box, and box to triangle reductions it is straightforward to verify,
using the formulas of appendix~\use\ReductionAppendix, that the
coefficient of $I_3^{{\rm 1m},\, D=6-2\eps}(s_{12})$ does not have a
pole as $s_{12}\rightarrow 0$.  Thus $I_3^{{\rm 1m},
\,D=6-2\eps}(s_{12})$ does not contribute to the non-factorizing parts
of the splitting and factorization functions.


\vskip -.4 truecm
\appendix{The Basis of Integral Functions}
\tagappendix\IntegralsAppendix

In this appendix we collect the integral functions useful for the
discussions in the text; these integral functions were obtained from
ref.~[\use\IntegralsShort,\use\IntegralsLong].  The $n$-point scalar
one-loop integral in $4-2\eps$ dimensions is
$$
I_n = (-1)^{n+1}\, i\, \L4\pi\R^{2-\eps}
 \int{d^{4-2\eps}p \over (2\pi)^{4 -2\eps} } {1 \over
p^2 (p-K_1)^2 (p-K_1 - K_2)^2 \cdots (p-K_1-K_2 - \cdots - K_{n-1})^2}
\, ,
\eqn\NPointLoopIntegral
$$
where $K_i$, $i=1,\ldots,n$ are the external momenta, which may be
either on- or off-shell.  As discussed in appendix~\ReductionAppendix\
any gauge theory one-loop amplitude can be reduced to a linear
combination of (a) $D=4-2\eps$ scalar box, triangle and bubble
integrals, (b) $D=6-2\eps$ scalar triangle and bubble integrals and
(c) $D=8-2\eps$ box integrals.  The higher dimension scalar
integrals are defined by replacing the $4-2\eps$ in
eq.~(\use\NPointLoopIntegral) with the appropriate dimension $D$.
Following the conventions of ref.~[\use\IntegralsLong] integrals
without a dimension label are taken to be in $D=4-2\eps$.
We first present the $D=4-2\eps$ integrals and subsequently give the higher
dimension ones.

\subappendix{Bubble functions}

The $D=4-2\eps$ two-point integral function is
$$
\eqalign{
& I_2(K^2) = {\rg\over \eps (1-2\eps)} (-K^2)^{-\eps} \, ,\cr }
\eqn\BubbleInt
$$
where
$$
\rg =  {\Gamma(1+\e)\Gamma^2(1-\e)\over\Gamma(1-2\e)} \,.
\eqn\RGdef
$$

\subappendix{Triangle functions}

In fig.~\TriDiscExFigure\ the three types of triangle integral functions
that may appear in massless gauge theories are given.
The three $D=4-2\eps$ scalar triangle functions are
\defeqn\Triangles
$$
\eqalignno{
& I_{3}^{3 \rm m} (K^2_1, K^2_2, K^2_3)
=\ {i\over \sqrt{\Delta_3}}  \sum_{j=1}^3
  \left[ \Li_2\left(-\left({1+i\delta_j \over 1-i\delta_j}\right)\right)
       - \Li_2\left(-\left({1-i\delta_j \over 1+i\delta_j}\right)\right)
  \right] \,,
& {\rm (\use\Triangles{a})} \cr
& I_{3}^{2 \rm m}(K^2_1,K^2_2) = {\rg\over\e^2}
{(-K^2_1)^{-\eps}-(-K^2_2)^{-\eps}\over (-K^2_1)-(-K^2_2) }\,,
& {\rm (\use\Triangles{b})} \cr
&I_{3}^{1\rm m}(K_1^2) = {\rg\over\e^2} (-K_1^2)^{-1-\eps} \, ,
& {\rm (\use\Triangles{c})} \cr}
$$
where
$$
\eqalign{
\delta_1 & = { K^2_1 - K^2_2 - K^2_3 \over
\sqrt{\Delta_3}} \, ,  \hskip 1 cm
\delta_2 = {-K^2_1 + K^2_2 - K^2_3 \over
\sqrt{\Delta_3}} \, ,  \hskip 1 cm
\delta_3 = {-K^2_1 - K^2_2 + K^2_3 \over
\sqrt{\Delta_3}}\, ,  \cr}
\anoneqn$$
and
$$
\Delta_3\equiv -(K^2_1)^2 - (K_2^2)^2 - (K^2_3)^2
+ 2 K^2_1 K^2_2 + 2 K^2_2 K^2_3 + 2 K^2_3 K^2_1\, .
\anoneqn
$$

For our purposes it is a bit more convenient to deal with
integral functions where the denominators have been scaled out
so we define
\defeqn\RescaledTri
$$
\eqalignno{
& I_{3}^{3 \rm m}(K^2_1, K^2_2, K^2_3) =
  i {\rg \over \sqrt{\Delta_3}} T_{3}^{3 \rm m}(K^2_1, K^2_2, K^2_3) \,,
& {\rm (\use\RescaledTri{a})} \cr
& I_{3}^{2 \rm m}(K^2_1, K^2_2) =
 {\rg \over K^2_1 - K^2_2}  T_{3}^{2 \rm m}(K^2_1, K^2_2) \,,
& {\rm (\use\RescaledTri{b})} \cr
& I_{3}^{1\rm m} (K_1^2) = -{\rg \over K^2_1} T_3^{1\rm m} (K_1^2)
& {\rm (\use\RescaledTri{c})} \,.\cr}
$$

\subappendix{Box Functions}

The scalar box (four-point) functions appearing in computations with
massless internal lines have already been extensively discussed in
refs.~[\use\FourMassBox,\use\IntegralsLong].
The reader is referred to these papers for
further details.  It is convenient to define this
function as
$$
F(K_1,K_2,K_3,K_4) = -{2\sqrt{\mathop{\rm det} S}\over\rg} \,I_4 \,,
\eqn\GeneralFdefn$$
where the symmetric $4\times4$
matrix $S$ has components ($i$, $j$ are $\mod 4$)
$$
S_{ij} = -{1\over2}\L K_i+\cdots+ K_{j-1}\R^2, \quad i\neq j;
\hskip 1truecm S_{ii} = 0 \, .
\anoneqn
$$
The external momentum arguments $K_{1\ldots4}$ in equation~(\use\GeneralFdefn)
are sums
of external momenta $k_i$ that are the arguments of
the $n$-point amplitude.   From ref.~[\use\IntegralsLong], through
$\Ord(\eps^0)$, we have
(after correcting a sign in the first box function)
\def\hs{\hskip 2.5 cm}
\defeqn\Fboxes
$$
\eqalignno{
& F^{4{\rm m}} (K_1, K_2, K_3, K_4) =
{1\over 2}
\biggl\{ - \Li_2\left(\hf(1-\lambda_1+\lambda_2+\rho)\right)
   + \Li_2\left(\hf(1-\lambda_1+\lambda_2-\rho)\right) \cr
 & \hs
 - \Li_2\left(
   \textstyle-{1\over2\lambda_1}(1-\lambda_1-\lambda_2-\rho)\right)
  + \Li_2\left(\textstyle-{1\over2\lambda_1}(1-\lambda_1-
    \lambda_2+\rho)\right) \cr
  &\hs - {1\over2}\ln\left({\lambda_1\over\lambda_2^2}\right)
   \ln\left({ 1+\lambda_1-\lambda_2+\rho \over 1+\lambda_1
        -\lambda_2-\rho }\right) \biggr\} \,,
  & \hbox{(\Fboxes{a})} \cr
&  F^{3{\rm m}}(k_1, K_2, K_3, K_4) =
  -{1\over\e^2} \Bigl[ (-s)^{-\e} + (-t)^{-\e}
     - (-K_2^2)^{-\e}- (-K_3^2)^{-\e} - (-K_4^2)^{-\e} \Bigr] \cr
  &\hs    - {1\over2\e^2}
    \biggl( { (-K_2^2)(-K_3^2) \over (-t) } \biggr)^{-\e}
     - {1\over2\e^2}
    \biggl( { (-K_3^2)(-K_4^2) \over (-s) } \biggr)^{-\e}\cr
  &\hs + \Li_2\left(1-{K_2^2\over s}\right)
   + \Li_2\left(1-{K_4^2\over t}\right)
   -  \Li_2\left(1-{K_2^2K_4^2\over st}\right)
       + {1\over 2} \ln^2\left({s\over t}\right) ,
 & \hbox{(\Fboxes{b})} \cr
&  F^{2{\rm m} \, h}(k_1, k_2, K_3, K_4) =
  -{1\over\e^2} \Bigl[ (-s)^{-\e} + (-t)^{-\e}
              - (-K_3^2)^{-\e} - (-K_4^2)^{-\e} \Bigr]
    - {1\over2\e^2}
    \biggl( { (-K_3^2)(-K_4^2) \over (-s) } \biggr)^{-\e} \cr
 &\hs  + \Li_2\left(1-{K_3^2\over t}\right)
   + \Li_2\left(1-{K_4^2\over t}\right)
   + {1\over 2} \ln^2\left({s\over t}\right) ,
  & \hbox{(\Fboxes{c})} \cr
&  F^{2{\rm m}\, e}(k_1, K_2, k_3, K_4) =
  -{1\over\e^2} \Bigl[ (-s)^{-\e} + (-t)^{-\e}
              - (-K_2^2)^{-\e} - (-K_4^2)^{-\e} \Bigr] \cr
  &\hs  + \Li_2\left(1-{K_2^2\over s}\right)
    + \Li_2\left(1-{K_2^2\over t}\right)
    + \Li_2\left(1-{K_4^2\over s}\right)
    + \Li_2\left(1-{K_4^2\over t}\right) \cr
  &\hs  - \Li_2\left(1-{K_2^2K_4^2\over st}\right)
    + {1\over2} \ln^2\left({s\over t}\right) ,
 &  \hbox{(\Fboxes{d})} \cr
&  F^{1{\rm m}}(k_1, k_2, k_3, K_4) =
  -{1\over\e^2} \Bigl[ (-s)^{-\e} + (-t)^{-\e} - (-K_4^2)^{-\e} \Bigr] \cr
 &\hs  + \Li_2\left(1-{K_4^2\over s}\right)
   + \Li_2\left(1-{K_4^2\over t}\right)
   + {1\over 2} \ln^2\left({s\over t}\right) + {\pi^2\over6} \, ,
 & \hbox{(\Fboxes{e})} \cr
&  F^{0{\rm m}}(k_1,k_2,k_3,k_4)  =
 - {1\over\e^2} \Bigl[ (-s)^{-\e} + (-t)^{-\e} \Bigr]
  + {1\over 2} \ln^2\left({s\over t}\right) + {\pi^2\over 2} \,,
  & \hbox{(\Fboxes{f})} \cr}
$$
where the $k_i$ denote on-shell momenta and the $K_i$ off-shell momenta.
The kinematic variables are
$$
s = (k_1 + k_2)^2 \, , \hs t = (k_2 + k_3)^2\,,
\anoneqn
$$
or with $k$ relabeled as $K$ for off-shell (massive) legs
and the functions appearing in $F_4^{4 \rm m}$ are
$$
 \rho\ \equiv\ \sqrt{1 - 2\lambda_1 - 2\lambda_2
+ \lambda_1^2 - 2\lambda_1\lambda_2 + \lambda_2^2}\ ,
\eqn\rdefinition
$$
and
$$
\lambda_1 = {K_1^2 \, K_3^2
\over (K_1 + K_2)^2 \, (K_2 + K_3)^2 } \; , \hskip 1.5 cm
\lambda_2 = {K_2^2  \, K_4^2 \over
 (K_1 + K_2)^2 \, (K_2 + K_3)^2  } \; .
\anoneqn
$$

When dealing with the kinematics of an $n$-point amplitude it is convenient
to label the integral functions in terms of the kinematic variables
which may appear. Thus, we define
\defeqn\NotationChangeBoxes
$$
\eqalignno{
& \Ffour{r, r', r'';i} = F^{\rm 4m}
         (k_i + \cdots + k_{i+r-1}, k_{i+r} + \cdots + k_{i+r+r'-1},
          k_{i+r+r'} + \cdots + k_{i+r+r'+r''-1},  & \null \cr
& \hskip 5 cm
          k_{i+r+r'+r''} + \cdots + k_{i-1}) \, ,
& ({\rm \NotationChangeBoxes a})\cr
&\Fthree{r,r';i} = F^{3{\rm m}}(k_{i-1}, k_i + \cdots k_{i+r-1}, k_{i+r}
+ \cdots + k_{i+r+r'-1} , k_{i+r+r'} + \cdots + k_{i-2}) \, ,
& ({\rm \NotationChangeBoxes b})\cr
&\Fhard{r;i} =  F^{2{\rm m}\, h} (k_{i-2}, k_{i-1}, k_i + \cdots + k_{i+r-1},
                                k_{i+r} + \cdots +  k_{i-3}) \, ,
& ({\rm \NotationChangeBoxes c})\cr
&\Feasy{r;i} = F^{2{\rm m}\, e} (k_{i-1}, k_i + \cdots + k_{i+r-1}, k_{i+r},
                                k_{i+r+1} + \cdots +  k_{i-2}) \, ,
& ({\rm \NotationChangeBoxes d})\cr
&\Fone{i} =  F^{1{\rm m}} (k_{i-3}, k_{i-2}, k_{i-1},
                                k_i + k_{i+1} + \cdots +  k_{i-4}) \, ,
&  ({\rm \NotationChangeBoxes e})\cr
& F_4^{\rm 0m} \equiv  F^{0{\rm m}}(k_1,k_2,k_3,k_4) \, ,
&  ({\rm \NotationChangeBoxes f})\cr}
$$
corresponding to the kinematics depicted in \fig\BoxesFigure.  The labels
on the momenta are defined mod $n$.

\vskip -.8 cm
\LoadFigure\BoxesFigure
{\baselineskip 13 pt
\noindent\narrower\ninerm  The kinematics of the box functions defined
in eq.~(\use\NotationChangeBoxes).}
{\epsfysize 2.6 truein}{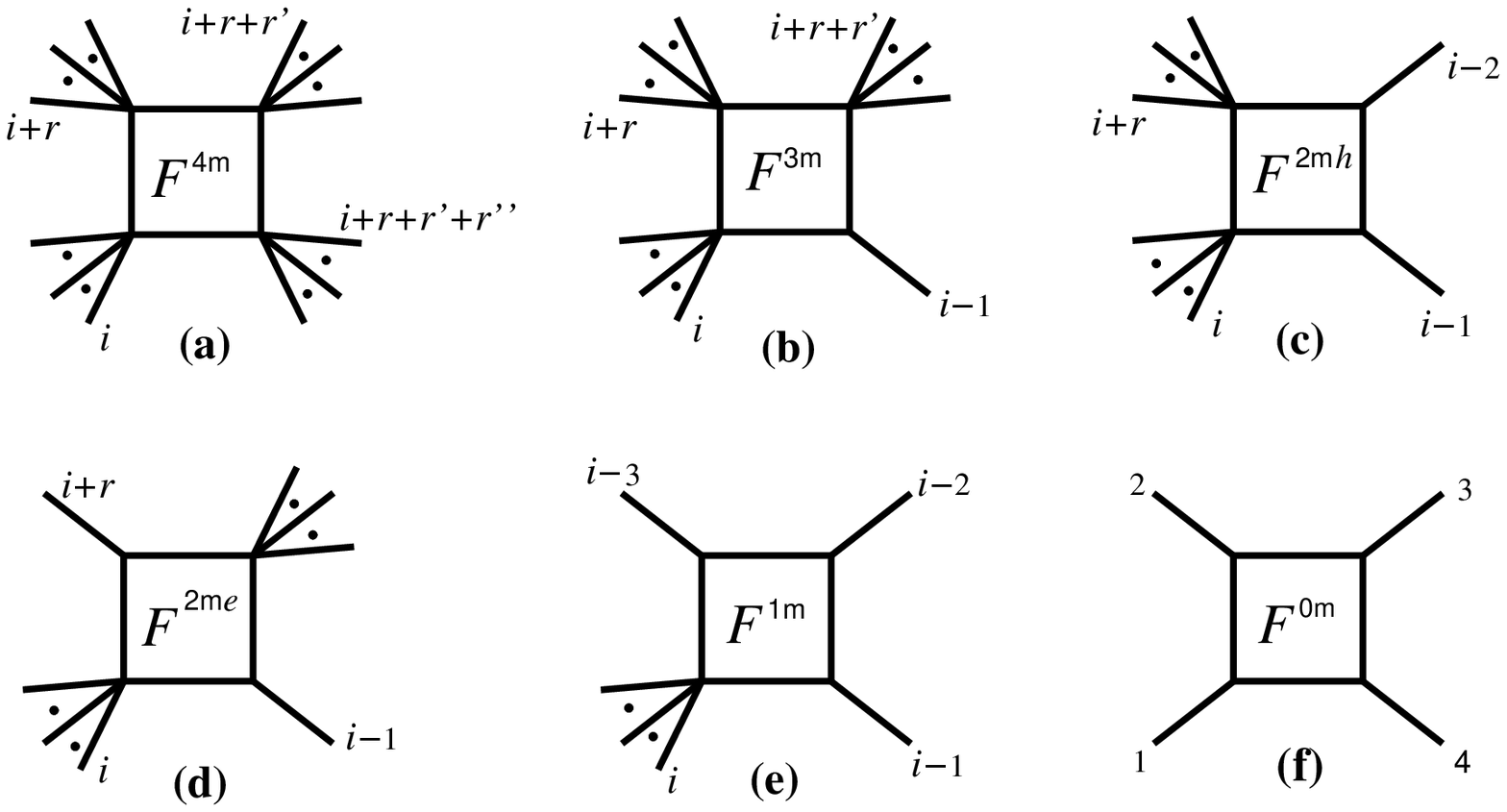}{}

For the purposes of this paper it is useful to exhibit the poles in
eq.~(\use\GeneralFdefn) contained in the integral functions with
$n$-point kinematics,
\defeqn\RescaledBoxes
$$
\eqalignno{
I_{4: r, r', r'', i}^{4{\rm m}} & =
-2 \rg {\Ffour{r, r', r'';i}\over t_i^{[r+ r']}\; t_{i+r}^{[r'+r'']}\;\rho}\,,
 & ({\rm\RescaledBoxes{a}})   \cr
  I_{4:r,r',i}^{3{\rm m}}
&=\ -2 \rg {\Fthree{r,r';i}
  \over t^{[r+1]}_{i-1}t^{[r+r']}_i - t^{[r]}_{i}t^{[n-r-r'-1]}_{i+r+r'} }\,,
 & ({\rm\RescaledBoxes{b}}) \cr
  I_{4:r;i}^{2{\rm m}h}
&=\ -2 \rg {\Fhard{r;i} \over t^{[2]}_{i-2} t^{[r+1]}_{i-1} } \,, &
 ({\rm\RescaledBoxes{c}}) \cr
 I_{4:r;i}^{2{\rm m}e}
&=\ -2 \rg {\Feasy{r;i}
   \over t^{[r+1]}_{i-1} t^{[r+1]}_{i} - t^{[r]}_{i} t^{[n-r-2]}_{i+r+1} }\,, &
     ({\rm\RescaledBoxes{d}}) \cr
  I_{4:i}^{1{\rm m}} &=\ -2 \rg {\Fone{i} \over t^{[2]}_{i-3} t^{[2]}_{i-2} }
        \, , \hskip 4 cm & ({\rm\RescaledBoxes{e}}) \cr}
$$
where the dimensionful prefactors have been extracted.

\subappendix{Higher dimension scalar integrals}

Due to the modified integral reduction procedure discussed in
appendix~\use\ReductionAppendix, higher-dimensional bubble, triangle
and box functions may also appear in the basis of functions in terms
of which amplitudes are expressed.
These integrals may be obtained from the
$D=4-2\eps$ integrals using the recursion formulas
$$
I_n^{D=6-2\eps} =
{1\over (n-5+2\eps)\, c_0} \biggl( 2 I_n -
\sum_{i=1}^n c_i\ I_{n-1}^{(i)} \biggr)\, ,
\eqn\SixDimInt
$$
$$
I_n^{D=8-2\eps} =
{1\over (n-7+2\eps)\, c_0} \biggl( 2 I_n^{D=6-2\eps} -
\sum_{i=1}^n c_i\ I_{n-1}^{(i),D=6-2\eps} \biggr) \, ,
\eqn\EightDimInt
$$
where
$$
  c_i = \sum_{j=1}^n S^{-1}_{ij}\, ,
\hskip 2 cm
  c_0 = \sum_{i=1}^n c_i\ =\ \sum_{i,j=1}^n S^{-1}_{ij}\, ,
\anoneqn
$$
and
$$
S_{ij} = - \half (p_{i-1} - p_{j-1})^2 \, ,
\anoneqn
$$
is a symmetric matrix. As defined previously the $(n-1)$-point
integral $I_{n-1}^{(i)}$ is obtained by removing the internal
propagator between external lines $(i-1)$ and $i$ from the original
$n$-point (scalar) integral.  The $D=6-2\eps$ bubble function is of
particular interest and is
$$
I_2^{D=6-2\eps}(K^2) = - {\rg\over 2 \eps (1-2\eps)(3-2\eps)} \,
                       (-K^2)^{1-\eps} \, .
\eqn\BubbleSix
$$
For one- and two-external mass triangles eq.~(\use\SixDimInt) is ill-defined
because the $S_{ij}$ matrix is not invertible; in this case the
$D=6-2\eps$ integrals may be obtained by direct integration
[\use\IntegralsLong] yielding
\defeqn\SixDimTri
$$
\eqalignno{
I_3^{{\rm 1m},\, D= 6-2\eps}(K_1^2) & =
                   {\rg\over 2\eps (1-\eps) (1-2\eps)}\,
(-K_1^2)^{-\eps} \, ,
 & {\rm (\use\SixDimTri{a})} \cr
I_3^{{\rm 2m}, D= 6-2\eps}(K_1^2, K_2^2) & =
{\rg\over 2\eps (1-\eps) (1-2\eps)}\,
         {(-K_1^2)^{1-\eps} - (-K_2^2)^{1-\eps} \over K_2^2 - K_1^2} \, .
& {\rm (\use\SixDimTri{b})} \cr}
$$
Observe that all higher dimension scalar integrals are related to
$D=4-2\eps$ dimensional scalar integrals by coefficients which depend
on $\eps$. For the integrals which may be obtained from the recursion
formulas (\use\SixDimInt) and (\use\EightDimInt), the
$\eps$-dependence appears in the overall factor.  Note also that the
six-dimension triangles (\use\SixDimTri) may be expressed in terms of
the four-dimensional bubble function (\use\BubbleInt) with rational
coefficients containing explicit $\eps$-dependence.

As discussed in appendix~\ReductionAppendix\ all $\eps$-dependence in
integral reduction coefficients may be eliminated by including the
$D=6-2\eps$ triangle functions and $D=8-2\eps$ box functions in the
basis of integral functions in which amplitudes are expressed; the
$\eps$-dependence is absorbed into the enlarged set of integral
functions.  Observe that the $D=6-2\eps$ scalar box is both infrared
and ultraviolet finite we may drop the $\eps$ in the overall
coefficient in eq.~(\use\SixDimInt) with no effect through
$\Ord(\eps^0)$; thus $I_4^{D=6-2\eps}$ may be written in terms of
$D=4-2\eps$ integrals with no $\eps$ appearing in the coefficients.
There is thus no need to include $I_4^{D=6-2\eps}$ in the basis of integral
functions.

\vfill

\baselineskip 14 pt
\listrefs
\end